\PassOptionsToPackage{colorlinks=true,allcolors=blue}{hyperref}
\documentclass[aps,prc,reprint,nofootinbib,superscriptaddress]{revtex4-2}

\usepackage{graphicx} 
\usepackage{amsmath,amssymb}
\usepackage{siunitx}
\usepackage{array}
\usepackage{multirow}
\usepackage{dcolumn}
\usepackage{bm}
\usepackage{braket}
\usepackage{rotating}
\usepackage{lineno}
\usepackage{adjustbox}
\usepackage{xcolor}
\usepackage{colortbl}
\usepackage{capt-of}  

\usepackage[T1]{fontenc}
\usepackage[utf8]{inputenc}
\DeclareUnicodeCharacter{2212}{-} 

\usepackage[colorlinks=true,allcolors=blue]{hyperref}
\usepackage{orcidlink}

\begin{document}
	
\title{\Large Realizing the Scientific Program with Polarized Ion Beams at EIC}
\collaboration{The EPIOS Scientific Consortium}
	
\author{Grigor Atoian}
\affiliation{Brookhaven National Laboratory, Upton, NY, USA}
	
\author{Nigel Buttimore\,\orcidlink{0000-0003-2945-8843}}
\affiliation{School of Mathematics, Trinity College Dublin, Dublin, Ireland}

\author{Giuseppe Ciullo\,\orcidlink{0000-0001-8297-2206}}
\affiliation{University of Ferrara and Istituto Nazionale di Fisica Nucleare, 44100 Ferrara, Italy}

\author{Ian Cloet}
\affiliation{Argonne National Laboratory, Lemont, IL, USA}

\author{Marco Contalbrigo\,\orcidlink{0000-0002-8612-7998}}
\affiliation{University of Ferrara and Istituto Nazionale di Fisica Nucleare, 44100 Ferrara, Italy}

\author{Jaydeep\,Datta\,\orcidlink{0000-0002-4444-2922}}
\affiliation{Stony Brook University, Stony Brook, NY, USA}
	
\author{Abhay Deshpande\,\orcidlink{0000-0003-3724-4749}}
\affiliation{Stony Brook University, Stony Brook, NY, USA}
	
\author{Shubham Dutta,\orcidlink{0000-0001-9650-8121}}
\affiliation{Saha Institute of Nuclear Physics, Kolkata, India}
	
\author{Oleg Eyser}
\affiliation{Brookhaven National Laboratory, Upton, NY, USA}
	
\author{Muhammad Farooq}
\affiliation{University of New Hampshire, Durham, NH, USA}
	
\author{Renee\,Fatemi}
\affiliation{University of Kentucky, Lexington, KY, USA}
	
\author{Ishara\,Fernando}
\affiliation{University of Virginia, Charlottesville, VA, USA}
	
\author{Michael Finger}
\affiliation{Charles University, Prague, Czech Republic}
	
\author{Wolfram Fischer}
\affiliation{Brookhaven National Laboratory, Upton, NY, USA}
	
\author{Dave Gaskell}
\affiliation{Thomas Jefferson National Accelerator Facility, Newport News, VA, USA}
	
\author{Prakash Gautam}
\affiliation{University of Virginia, Charlottesville, VA, USA}
\affiliation{Thomas Jefferson National Accelerator Facility, Newport News, VA, USA}
	
\author{Ralf\,Gebel\,\orcidlink{0000-0001-5290-4237}}
\affiliation{GSI Helmholtz Centre for Heavy Ion Research, Darmstadt, Germany}
\affiliation{Forschungszentrum Jülich, Jülich, Germany}
	
\author{Boxing\,Gou\,\orcidlink{0000-0002-8918-3514}}
\affiliation{Institute of Modern Physics, Chinese Academy of Sciences, Lanzhou, China}
	
\author{Daoning Gu}
\affiliation{GSI Helmholtz Centre for Heavy Ion Research, Darmstadt, Germany}
\affiliation{III. Physikalisches Institut B, RWTH Aachen University, Aachen, Germany}
	
\author{Yoshitaka Hatta}
\affiliation{Brookhaven National Laboratory, Upton, NY, USA}
	
\author{Mohammad Hattawy}
\affiliation{Old Dominion University, Norfolk, VA, USA}
	
\author{Volker\,Hejny\,\orcidlink{0000-0003-0713-5859}}
\affiliation{Forschungszentrum Jülich, Jülich, Germany}
	
\author{Kiel Hock\,\orcidlink{0000-0002-1886-0661}}
\affiliation{Brookhaven National Laboratory, Upton, NY, USA}
	
\author{Georg\,Hoffstaetter}
\affiliation{Cornell University, Ithaca, NY, USA}
\affiliation{Brookhaven National Laboratory, Upton, NY, USA} 

\author{Haixin Huang\,\orcidlink{0000-0003-1189-4874}}
\affiliation{Brookhaven National Laboratory, Upton, NY, USA}
	
\author{Christopher Ianzano}
\affiliation{Laboratory for Nuclear Science, Massachusetts Institute of Technology, Cambridge, MA, USA}
	
\author{Jiangyong Jia\,\orcidlink{0000-0002-5725-3397}}
\affiliation{Stony Brook University, Stony Brook, NY, USA}
	
\author{Andro\,Kacharava\,\orcidlink{ 0000-0003-4864-8087}}
\affiliation{Forschungszentrum Jülich, Jülich, Germany}
	
\author{Maggie Kerr}
\affiliation{Laboratory for Nuclear Science, Massachusetts Institute of Technology, Cambridge, MA, USA}
	
\author{Wolfgang\,Korsch\,\orcidlink{0000-0002-6670-3011}}
\affiliation{University of Kentucky, Lexington, KY, USA}
	
\author{Dario Lattuada}
\affiliation{Università Kore di Enna \& INFN-LNS, Enna \& Catania, Italy}
	
\author{Andreas Lehrach}
\affiliation{III. Physikalisches Institut B, RWTH Aachen University, Aachen, Germany}
\affiliation{Forschungszentrum Jülich, Jülich, Germany}
	
\author{Minxiang\,Li\,\orcidlink{0009-0001-7794-1664}}
\affiliation{Institute of Modern Physics, Chinese Academy of Sciences, Lanzhou, China}
	
\author{Xiaqing\,Li}
\affiliation{Shandong University, Jinan, China}

\author{Paolo\,Lenisa\,\orcidlink{0000-0003-3509-1240}}
\affiliation{University of Ferrara and Istituto Nazionale di Fisica Nucleare, 44100 Ferrara, Italy}

\author{Win Lin}
\affiliation{Stony Brook University, Stony Brook, NY, USA}
	
\author{James Maxwell}
\affiliation{Thomas Jefferson National Accelerator Facility, Newport News, VA, USA}
	
\author{Aleksei Melnikov\orcidlink{0009-0002-3339-7326}}
\affiliation{Institute for Nuclear Research, Russian Academy of Sciences, Moscow, Russia}
	
\author{Zein-Eddine\,Meziani\,\orcidlink{0000-0001-9450-2914}}
\affiliation{Argonne National Laboratory, Lemont, IL, USA}
	
\author{Richard\,G.\,Milner\,\orcidlink{0000-0002-0031-1963}}
\affiliation{Laboratory for Nuclear Science, Massachusetts Institute of Technology, Cambridge, MA, USA}
	
\author{William R. Milner\,\orcidlink{0000-0003-0510-0775}}
\affiliation{Research Laboratory for Electronics, Massachusetts Institute of Technology, Cambridge, MA, USA}
	
\author{Iurii Mitrankov}
\affiliation{Stony Brook University, Stony Brook, NY, USA}
	
\author{Hamlet Mkrtchyan\orcidlink{0000-0003-4358-9380}}
\affiliation{A. I. Alikhanyan National Science Laboratory, Yerevan Physics Institute, Yerevan 0036, Armenia}
	
\author{Prajwal MohanMurthy\,\orcidlink{0000-0002-7573-7010}}
\affiliation{Laboratory for Nuclear Science, Massachusetts Institute of Technology, Cambridge, MA, USA}
	
\author{Christoph\,Montag}
\affiliation{Brookhaven National Laboratory, Upton, NY, USA}
	
\author{Sergei Nagaitsev}
\affiliation{Brookhaven National Laboratory, Upton, NY, USA}
	
\author{Charles-Joseph Naim\,\orcidlink{0000-0001-5586-9027}}
\affiliation{Stony Brook University, Stony Brook, NY, USA}
	
\author{Alexander Nass\,\orcidlink{0000-0003-2929-9109}}
\affiliation{Forschungszentrum Jülich, Jülich, Germany}
	
\author{Dien Nguyen\,\orcidlink{0000-0002-6964-6761}}
\affiliation{University of Tennessee, Knoxville, TN, USA}
	
\author{Nikolai\,Nikolaev\,\orcidlink{0000-0001-9362-4813}}
\affiliation{L.D. Landau Institute for Theoretical Physics, Chernogolovka, Russia}
\affiliation{Moscow Institute of Physics and Technology, National Research University, Dolgoprudny, Moscow Region 141701, Russian Federation}
\affiliation{Bogoliubov Laboratory of Theoretical Physics, International 	Intergovernmental Scientific Research Organization, Joint Institute for Nuclear Research, Dubna, Moscow Region,  141980 Russia}

\author{Luciano Pappalardo\,\orcidlink{0000-0002-0876-3163}}
\affiliation{University of Ferrara and Istituto Nazionale di Fisica Nucleare, 44100 Ferrara, Italy}

\author{Chao Peng}
\affiliation{Argonne National Laboratory, Lemont, IL, USA}

\author{Anna Piccoli\,\orcidlink{0009-0008-3313-7413}}
\affiliation{University of Ferrara and Istituto Nazionale di Fisica Nucleare, 44100 Ferrara, Italy}

\author{Andrei\,Poblaguev}
\affiliation{Brookhaven National Laboratory, Upton, NY, USA}
	
\author{Deepak\,Raparia\,\orcidlink{0000-0002-5149-6363}}
\affiliation{Brookhaven National Laboratory, Upton, NY, USA}
	
\author{Frank Rathmann\,\orcidlink{0000-0003-0824-2103}}
\affiliation{Brookhaven National Laboratory, Upton, NY, USA}
	
\author{Thomas Roser}
\affiliation{Brookhaven National Laboratory, Upton, NY, USA}
	
\author{Premkumar Saganti}
\affiliation{Prairie View A\&M University, Prairie View, TX, USA}
	
\author{ Andrew\,Sandorfi\,\orcidlink{0000-0001-8309-8581}}
\affiliation{University of Virginia, Charlottesville, VA, USA}
	
\author{Medani\,Sangroula}
\affiliation{Brookhaven National Laboratory, Upton, NY, USA}
	
\author{Vincent Schoefer}
\affiliation{Brookhaven National Laboratory, Upton, NY, USA}
	
\author{Yousif Shabaan El-Feky\,\orcidlink{0009-0009-7830-6134}}
\affiliation{American University in Cairo, New Cairo 11835, Egypt}

\author{Rahul Shankar\,\orcidlink{0009-0006-7626-5824}}
\affiliation{University of Ferrara and Istituto Nazionale di Fisica Nucleare, 44100 Ferrara, Italy}

\author{Vera Shmakova\,\orcidlink{0000-0003-4035-4949}}
\affiliation{Brookhaven National Laboratory, Upton, NY, USA}
	
\author{Evgeny\,Shulga}
\affiliation{Brookhaven National Laboratory, Upton, NY, USA}
	
\author{Jamal Slim\,\orcidlink{0000-0002-9418-8459}}
\affiliation{Deutsches Elektronen-Synchrotron, Hamburg, Germany}
	
\author{Dannie Steski}
\affiliation{Brookhaven National Laboratory, Upton, NY, USA}
	
\author{Bernd Surrow}
\affiliation{Temple University, Philadelphia, PA, USA}
	
\author{Noah Wuerfel\,\orcidlink{0000-0001-9872-5330}}
\affiliation{Laboratory for Nuclear Science, Massachusetts Institute of Technology, Cambridge, MA, USA}
	
\author{Yaojie Zhai\,\orcidlink{0000-0001-9250-7623}}
\affiliation{Institute of Modern Physics, Chinese Academy of Sciences, Lanzhou, China}
	
\date{\today}

\begin{abstract}
Polarized ion beams at the Electron Ion Collider are essential to address some of the most important open questions at the twenty-first century frontiers of understanding of the fundamental structure of matter.   Here, we summarize the science case and identify polarized $^2$H, $^3$He, $^6$Li and $^7$Li ion beams as critical technology that will enable experiments which address the most important science.  Further, we discuss the required ion polarimetry and spin manipulation in EIC. The current EIC accelerator design is presented. We identify a significant R\&D effort involving both national laboratories and universities that is required over about a decade to realize the polarized ion beams and estimate (based on previous experience) that it will require about 20 FTE over 10 years (or a total of about 200 FTE-years) of personnel, including graduate students, postdoctoral researchers, technicians and engineers. Attracting, educating and training a new generation of physicists in experimental spin techniques will be essential for successful realization.  AI/ML is seen as having significant potential for both acceleration of R\&D and amplification of discovery in optimal realization of this unique quantum technology on a cutting-edge collider. The R\&D effort is synergistic with research in atomic physics and fusion energy science.
\end{abstract}

\maketitle

\tableofcontents

\section{Introduction}
The science case for the Electron-Ion Collider has been developed over more than two decades.  It was initiated and matured through three US Nuclear Physics Long Range Planning Exercises culminating in the clear recommendation in 2015:

\medskip\noindent
{\bf We recommend a high-energy high-luminosity polarized EIC as the highest priority for new facility construction following the completion of FRIB.}
\medskip\noindent

Further, in 2018, an independent committee assembled by the U.S. National Academy of Sciences and Engineering strongly validated the science case and found that 

\medskip\noindent
{\bf An EIC with highly polarized electrons and ions, with sufficient high luminosity and sufficient variable center-of-mass energy, can uniquely address three profound, high-priority science questions about neutrons and protons and how they are assembled to form the nuclei of atoms.}
\medskip\noindent

In early 2020, the US Department of Energy granted CD-0 and made the decision to realize EIC at the existing Relativistic Heavy Ion Collider at Brookhaven National Laboratory at Upton, New York.
A call to consider EIC detector proposals led to the unanimous recommendation in March 2022 by the EIC Detector Proposal Advisory Panel to select the ECCE reference design, on which the ePIC detector is based enhanced with a novel barrel imaging calorimeter.  Subsequently, the EIC Project received CD-1 approval in mid-2021 and CD-3A in March 2024.  CD-4 approval is now estimated to occur in 2034.

\medskip
In the 2030s, it is anticipated that the U.S.-based Electron-Ion Collider (EIC) will be the premier facility worldwide for hadronic and nuclear physics~\cite{AbdulKhalek:2021gbh,NAP25171,Accardi:2012qut}. In particular, EIC is motivated by the desire to comprehensively understand the spin structure of the proton and neutron in terms of fundamental quarks and gluons within the framework of Quantum Chromodynamics (QCD). Further, over the years, polarized beams have provided a unique handle to probe dynamical phenomena that might be inaccessible with unpolarized beams. Indeed, polarized beams have enabled an enhanced sensitivity to physics observables critical to probe novel phenomena and spin dependent phenomena. The latter encompasses a broad spectrum of scientific inquiries, from enabling 3-dimensional spatial and momentum partonic structure of nucleons to tests of the standard model of particle physics. Using highly polarized beams at the EIC has always been an integral part of its overall design including polarized electrons colliding with polarized light ions, namely polarized protons (spin 1/2), deuterons (spin 1) and helions (spin 1/2). A desirable addition of lithium-6 (spin 1) and lithium-7 (spin 3/2) ions would also enhance the scientific program. 

One of the important questions in nuclear physics is whether one can infer the intrinsic properties of a nucleus from its fundamental QCD degrees of freedom, namely quarks and gluons. Traditionally nuclear physics has focused on providing  {\it ab initio} calculations of the properties of light nuclei using nucleon-meson degrees of freedom. More recently the focus has shifted to using effective field theory to include two-body and three-body forces consistently. These calculations represent an important reference once the investigation of the direct role quarks and gluons play in nuclei is underway. Physics examples that are enabled with light nuclei are the investigation of the polarized EMC effect and its $A$ dependence, hidden color configurations, six quarks contributions to the deuteron (free or embedded in \(^6\)Li) through measurements of the \(b_1\) structure function and the gluon transversity distribution not yet accessed anywhere else. The ultimate goal is to unravel the role of quarks and gluons in nuclei beyond the nucleon-meson picture as well as search for exotic phenomena using the nucleus and its spin as a QCD laboratory, an example is given in~\cite{Jaffe:1989, Maxwell2018}.

Here, we summarize the main
conclusions from the workshop {\href{https://indico.cfnssbu.physics.sunysb.edu/event/343/}{\it Polarized Ion Sources and Beams at EIC}} 
that took place at Stony Brook University, New York on March 10-12, 2025.
The goals of the workshop were: 
\begin{itemize}
    \item {} to raise the visibility in the EIC and spin communities of the exciting scientific case for spin measurements; 
    \item {} to assess, in the context of the considerable scientific motivation, the status of ion sources development for EIC;
    \item {} to identify critical path R\&D essential for a successful polarized EIC scientific program that can be implemented on day-1;
    \item{} to motivate the education and training of a new generation of young physicists with expertise in spin polarization technology.  This will be essential for realization of the EIC polarized program.
\end{itemize}

A program of R\&D to develop the necessary spin technology is presented and the necessary personnel and resources to successfully realize the polarized beams at EIC demanded by the science case is outlined. 

\section{Executive Summary}

Over the course of two days, the workshop featured an excellent  series of presentations and stimulating, insightful discussions from participants, both in-person and remote. There was a clear sense of urgency that a significant R\&D effort directed at realization of polarized ion beams at EIC must get underway soon.  Thus, the EPIOS ({\bf E}IC {\bf P}olarized {\bf IO}n {\bf S}ource) scientific consortium was formed to drive and coordinate this essential effort.  This paper is a summary of the EPIOS perspective arising from the meeting in March 2025. 

The major points arising from the meeting were:
\begin{itemize}
    \item It must be realized that the adaptation of a well-understood polarization technique into a reliable ion source operating at maximum performance injecting with high reliability into an accelerator requires a sustained R\&D effort by a critical mass of suitably skilled personnel for about a decade.
    \item If we consider the major polarization experimental efforts in nuclear physics over the last half century, university-based research groups played an essential role in developing the technical capabilities and in attracting and training the generations of physicists who carried out the research. 
    \item{} The EIC science demands the widest available range of polarized ions and innovative source technologies must be pursued.
    \item{} The EIC accelerator design team must make it a priority to develop ion spin control and manipulation from source to collisions of the deuteron $^3$He, $^6$Li and $^7$Li nuclei.
    \item {} AI/ML can serve as both accelerators of R\&D (through fast modeling, optimization, and control) 
    and amplifiers of discovery (through advanced data analysis and anomaly detection).  Their integration will be key to meeting the ambitious polarization goals of the EIC.    \item {} It is recommended that a specific amount of funding be set aside to target R\&D associated with the realization of polarized ion beams at EIC.  This would support education and training of young physicists with the necessary expertise.   
    \item {} It is recommended that a focused multi-week program on the science case for polarized ion beams at EIC being organized at the Institute for Nuclear Theory in the next year.
    \item {} It is recommended that an annual summer school for young physicists on the science and technical realization of polarized ion beams at EIC be established.  Existing summer schools and the U.S. Particle Accelerator School can be leveraged to include relevant lectures and classes.
    \item It is recommended that the BNL Tandem be utilized to secure the future with $\vec d$ and $\vec{^6\text{Li}}$ ions, as there is not sufficient space available for these sources at BNL EBIS.    
    \item {} We identify the AGS as a very valuable platform to carry out beam studies of polarized sources, polarized beams and spin manipulators in the era when RHIC is dark.  We recommend that EPIOS and the BNL-CAD together consider the possibilities and develop a plan that takes advantage of the AGS. 
    \item {} Finally, we point out that the polarized atom sources required for the production of EIC polarized ion beams can also be utilized to feed windowless gas targets internal to a charged particle storage ring.  If desired, this would make possible a program of fixed target physics at one of the storage rings in the EIC accelerator complex.
\end{itemize}


\section{Science Case for Polarized Ion Beams at EIC}
Spin is the origin of order and structure for the visible matter in our universe.  Particles with half-integer spins, such as electrons, quarks, neutrinos and even composite particles like protons or neutrons, are {\it fermions}.  Particles with integer spins, such as photons, Ws, Zs, gravitons and gluons and even nuclei like the deuteron, are {\it bosons}.  Over the past 100 years, physicists have discovered and described how fermions interact via the exchange of bosons, mediators of Nature's four fundamental forces.  The Standard Model, together with Einstein's theory of gravity, successfully explains all laboratory experiments to date.  

No two identical spin-1/2 particles can be in the same quantum state. This is why atomic electrons organize themselves into {\it shells}.  These arrangements form the basis for the Periodic Table, which groups chemical elements according to their electronic structure and chemical properties.  A similar spin-based ordering determines the arrangement of protons and neutrons in atomic nuclei.  Without spin, there would be no atoms and hence no observable universe as we know it.

This century-long journey to deeper understanding of the fundamental structure of matter has brought us to confronting in the twenty-first century how the fundamental spin-1/2 quarks and spin-1 gluons of QCD form the proton and neutron, the elementary building blocks of nuclei. Spin plays a central role in this exciting scientific quest.  Consequently, polarized electron and ion beams are absolutely essential to realize the EIC scientific program. In particular, polarized ion beams are required to carry out some of the main scientific research thrusts:

\medskip
\begin{itemize}

\item{} to provide polarized neutron targets;
\item{} to study how the nucleon's quark and gluon spin distributions can be modified in the nucleus;
\item{} to access the new parton and transverse momentum distributions that allow full three-dimensional imaging of the proton, neutron and nuclei;
\item{} to carry out pioneering searches for spin-1 and higher components, e.g. {\it exotic gluons}, in the nucleus. 
\end{itemize}

Nuclear spin results from the spins of valence protons and neutrons occupying shell model orbits around an unpolarized core.  For scattering from polarized nuclei with increasing atomic mass number, the interactions with the unpolarized core will dominate and dilute the scattering asymmetry.  Thus, polarized ion beams are of greatest scientific interest at the EIC for nuclei up to atomic mass number (A) of about 20. However, polarized medium light or heavy nuclei with A greater than 20 can be considered in tandem with the detection of the recoiling unpolarized core to mitigate the dilution of the scattering asymmetry.

3D imaging of nuclei represents the natural extension of a fundamental scientific pursuit: the desire to see inside complex systems to reveal their inner workings with precision. Just as imaging technologies have revolutionized our understanding of biological structures, materials, and the cosmos, 3D imaging of nuclei allows us to explore the deep structure of matter itself—mapping the spatial and momentum distributions of quarks and gluons to uncover the dynamics that govern nuclear structure at the most fundamental level. The role of polarization of nuclei in this endeavor is of paramount importance.

\medskip
\noindent
Experimentally, the collider offers unique advantages for spin physics:
\begin{itemize}
    \item {} realization of pure electron-nuclear scattering, unlike current fixed-target experiments, where the polarized nucleus is embedded in a significant amount of extraneous material;
    \item{} straightforward control of the ion spin direction from transverse to longitudinal with respect to the beam direction;
    \item {} detection of particles scattered close to the beamline;
    \item {} efficient tagging of spectator systems, which travel in the laboratory with high energy close to the beam direction.  
\end{itemize}
For all the reasons outlined above, the collider will provide access to a significantly broader phase space for a wide range of scattering processes—including inclusive, semi-inclusive, and exclusive measurements. A key advantage lies in the ability to tag recoil spectators, which allows for precise identification of the underlying reaction mechanism and greatly enhances the figure of merit for each measurement. For instance, in order to accurately determine the polarization carried by a neutron in the outer shell of a nucleus, it is essential to measure its momentum distribution in the nuclear ground state—something that can be effectively achieved through exclusive reactions with full detection of the final states of the reaction.

\subsection{Quark and Gluon Structure of Nucleon Spin}
\subsubsection{Longitudinal Structure}
The advent of highly polarized beams and targets, nearly four decades ago, opened a frontier in experimental hadronic physics that transformed our understanding of the spin structure of the nucleon and the underlying theory of Quantum Chromodynamics. These fixed-target experiments operated as ``matter microscopes", effectively using the scattering of high-energy lepton beams from nuclear matter to probe the nature of the interactions between the quarks confined inside protons and neutrons. This type of microscope provides resolutions ranging from $Q = 1 - 10$ GeV ($\sim 0.2 - 0.02$ fm) and allows for flavor-separated extractions of the momentum and spin degrees of freedom for quarks carrying 1-90$\%$ ($0.01 < x < 0.9$) of the nucleon's momentum. 

\begin{figure}[htb]
\centering
\includegraphics[width=0.95\columnwidth]{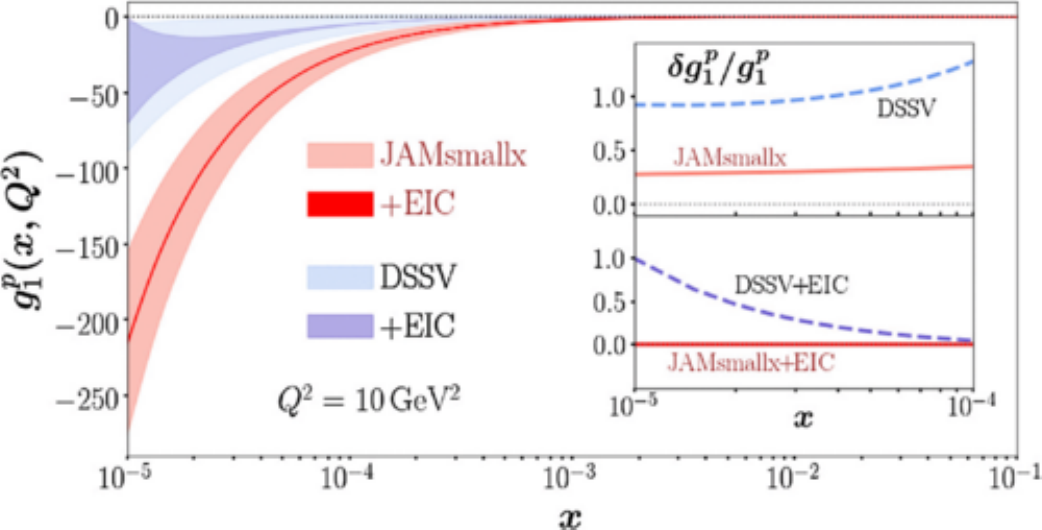}
\caption{The gluon helicity distribution is determined from the $Q^2$ evolution of the $g_1^p$ structure function. The curves above are extracted from the DSSV and JAMlowx global analysis~\protect\cite{Adamiak:2021ppq} and show the associated errors with (darker) and without (lighter) EIC data. The curves diverge for $x < 0.001$ due to the application of different theoretical evolution formalisms.}
\label{fig:DSSV_JAM_lowx}
\end{figure}

EIC will usher in a new era of hadronic physics by replacing nuclear targets with high-energy ion and polarized proton beams. The move to a collider configuration will extend the resolution range to 100 GeV ($\sim 0.002$ fm) and, for the first time, provide a glimpse of the sea of quarks, anti-quarks and gluons that exist at extremely low nucleon momentum fractions ($0.0001 < x < 0.1$). Existing measurements indicate that the gluons contribute to $40\%$ of the total nucleon momentum and could also contribute up to $40\%$ of the total spin of the proton. Unfortunately, this latter number has a large uncertainty ($\pm20\%$) due to the lack of data for $x < 0.01$. The kinematics of the collisions at the EIC will push these measurements to much lower momentum fractions, providing critical new constraints on both the extraction and the low-$x$ evolution of the gluon helicity distribution. 

\begin{figure*}[t]
\vspace{-0.4cm}
\includegraphics[width=0.85\textwidth]{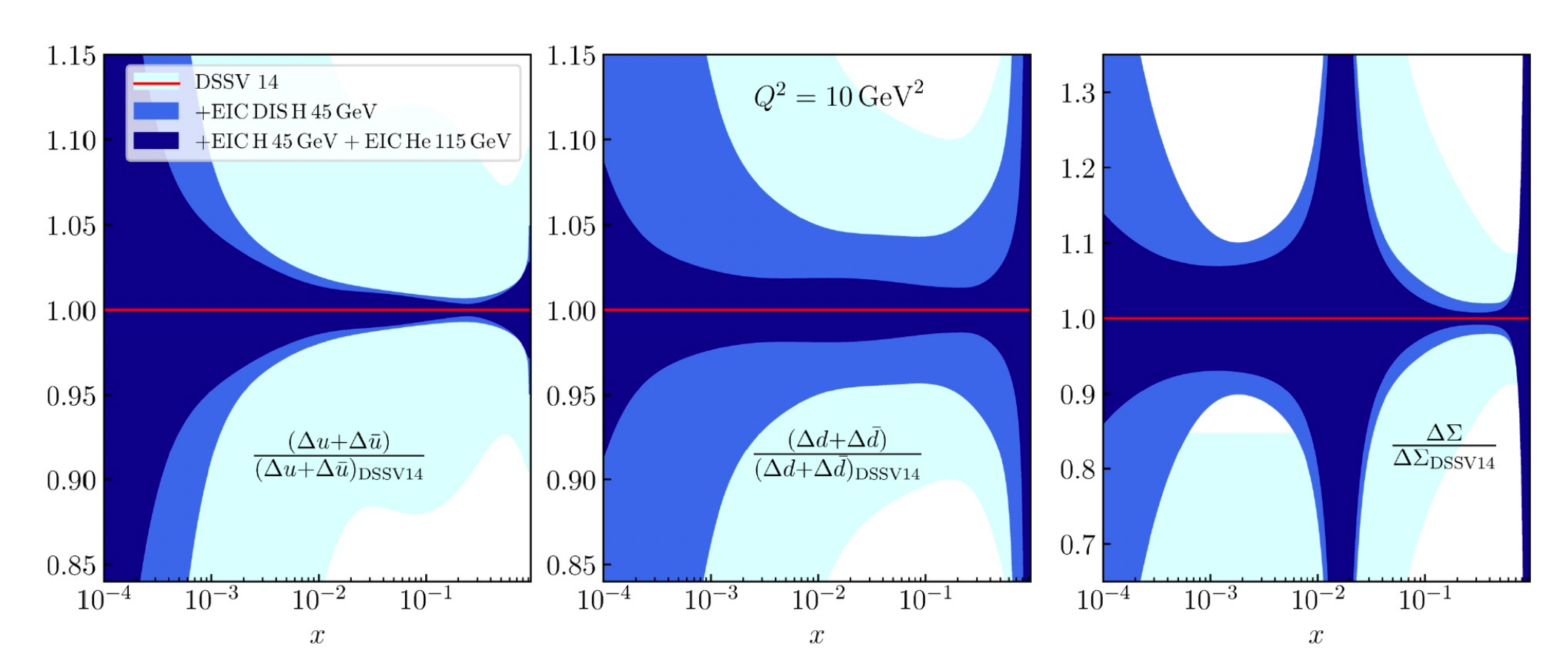}
\caption{The impact of projected inclusive scattering data off polarized proton and $^3$He beams at the EIC on the relative uncertainties of the extracted up, down and quark singlet helicity distributions.}
\label{fig:EIC_PDFs_I}
\end{figure*} 
\par\medskip
As we embark on the EIC era it is imperative to pursue the development of highly polarized light ion beams, such as deuterium ($^2$H), helium-3 ($^3$He) and lithium-6 ($^6$Li). These ion beams serve as proxies for free neutrons, providing additional information and constraints that cannot be provided by polarized proton beams alone. A striking example is the flavor separation of the quark helicity distributions, which have traditionally been constrained in semi-inclusive deep-inelastic-scattering (SIDIS) fixed-target experiments by using detected final-state hadrons to tag the flavor of the interacting quark. This requires existing knowledge, or the parallel extraction, of fragmentation functions and ultimately results in additional associated systematic errors. In contrast, it is possible to use inclusive scattering data, which is typically higher precision than SIDIS channels since there are no requirements on the final state, from both proton and ``neutron" beams. The first two panels of Fig. \ref{fig:EIC_PDFs_I} demonstrate the significant constraints~\cite{PhysRevD.102.094018} that can be made using EIC data compared to existing global analysis extractions, especially for the down quarks. The third panel shows the reduced uncertainties on the flavor singlet at low $x$. In addition to flavor separation, inclusive measurements made over a wide range of $x$ and $Q^2$ will facilitate the determination of the Bjorken Sum Rule, $\int_0^1[g_1^p(x,Q^2)$-$g_1^n(x,Q^2)]dx$. The $Q^2$ evolution of the sum rule can be exploited to make a precision extraction of $\alpha_s$, the strong coupling constant \cite{PhysRevD.110.074004}.
\par\medskip

The utility of polarized light-ion beams is not limited to inclusive channels, it cuts across all aspects of the EIC physics program. A flagship effort at the future EIC will be the extraction of the Generalized Parton Distributions (GPDs), functions that map the spatial dependence of the quark and gluon spin-momentum correlations in the plane transverse to the longitudinal momentum of the interacting parton. The neutron GPDs can be accessed rather cleanly via measurements of exclusive final states in Deeply Virtual Compton Scattering (DVCS) and Deeply Virtual Meson Production (DVMP) from deuterium beams. Studies of the coherent DVCS channel, where the recoiling nucleus is reconstructed, will explore the gluon GPDs and their potential modification in a nuclear medium. It will also be possible to access information about the nuclear energy-momentum tensor and therefore the pressure and shear forces inside the nucleus. 
\par\medskip
The analysis of $e+^3$He data relies on effective neutron polarizations that are extracted from non-relativistic nuclear structure models. Nuclear modifications arise from $\Delta$ isobars in the $^3$He nucleus at high $x$ and from spin-dependent nuclear shadowing at lower momentum fractions. Uncertainties from this method can be avoided by measuring the spin asymmetry of the spectator proton as a function of neutron virtuality, with the free neutron polarization extracted in the limit that the virtuality goes to zero. This method requires the tagging and momentum reconstruction of the spectator proton in the far-forward region. Additional information on the detector requirements for forward proton detection in tagged DIS, as well as the general physics motivation for polarized light-ion beams at the EIC, may be found in the EIC Yellow Report~\cite{AbdulKhalek:2021gbh}.

\subsubsection{Mapping the Full 3D Structure of Nucleon and Nuclei}
One of the key paradigm shift in the studies of the nucleon spin structure that emerged during the late 1990s early 2000s~\cite{Ji:1996nm,Mueller:1998fv,Radyushkin:1996nd,Collins:2003fm} was the realization that a complete understanding of the nucleon structure requires more than just one-dimensional momentum distributions.
To fully characterize how quarks and gluons (partons) are distributed inside the nucleon in both position and momentum space -- a more complete, three-dimensional picture is necessary. This 3D structure is being explored through two complementary frameworks: Generalized Parton Distributions (GPDs) and Transverse Momentum Dependent distributions (TMDs). GPDs provide spatial imaging of partons in the transverse plane as a function of their longitudinal momentum, effectively offering a “tomographic” view of the nucleon. TMDs, by contrast, describe the distribution of partons in transverse momentum space and capture key features such as spin-momentum correlations and orbital angular momentum dynamics. As an example, 
the leading twist TMDs of a spin-one target are illustrated in Fig.~\ref{fig:spin_one_tmds}, which are in addition to the TMDs that appear in a spin-half target like the nucleon, there are 3 additional $T$-even and 7 additional $T$-odd quark TMDs associated with tensor polarization
\begin{figure*}[t]
   \centering
    \includegraphics[width=0.7\textwidth]{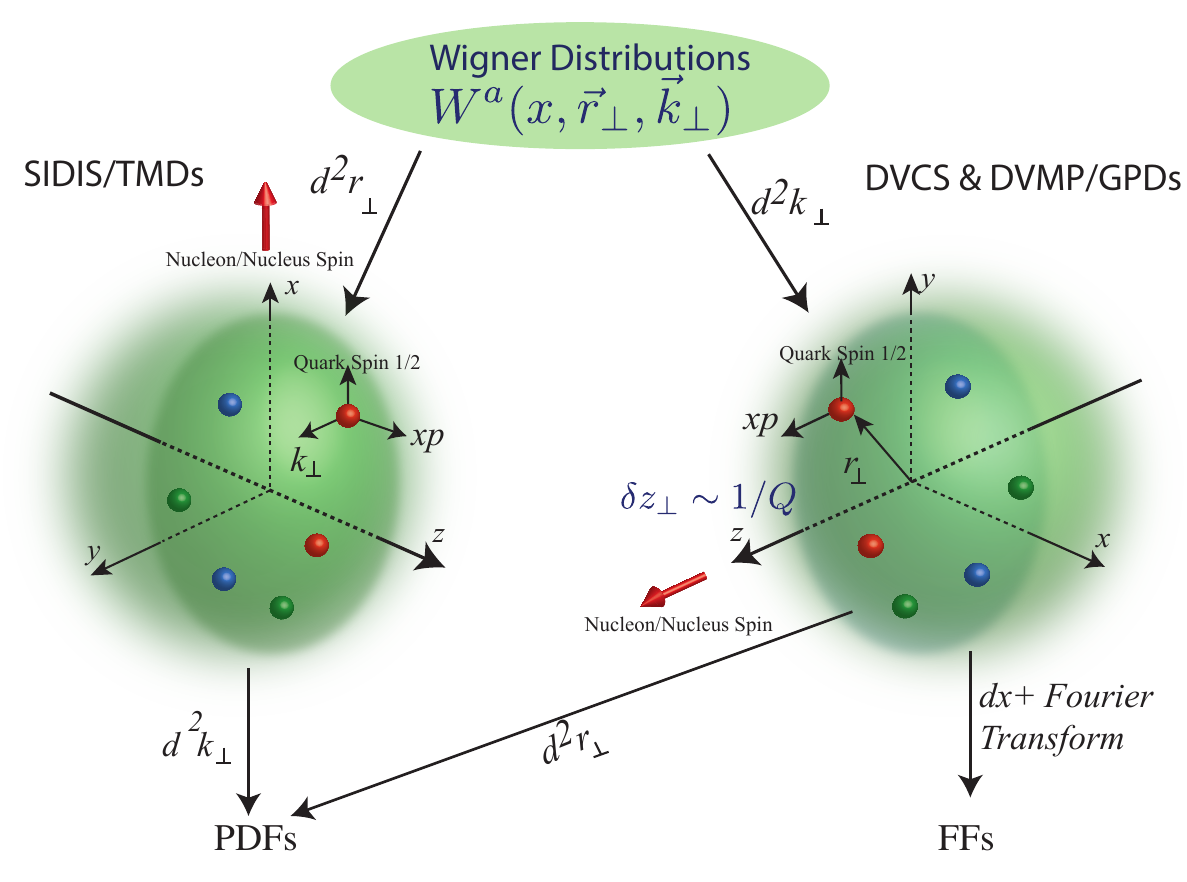}
    \caption{Schematic representation of Wigner distributions in the lightcone frame for each parton species $a$ and their relation to the daughter distributions, GPDs and TMDs, and how these further connect to the one dimensional parton distribution functions (PDFs) and form factors (FF)~\protect\cite{Dudek:2012vr}. }
\label{fig:wigner}
\end{figure*}

Together, GPDs and TMDs underpin a broader quantum phase-space description known as Wigner distributions~\cite{Ji:2003ak} -- a multidimensional formalism that, while not directly observable, provides a unifying conceptual framework for nucleon structure, see Fig.\,\ref{fig:wigner} depicting these relations.

While ongoing experimental efforts—including DVCS and DVMP, SIDIS, and polarized Drell-Yan processes—are steadily enriching our picture of the nucleon, the upcoming Electron-Ion Collider (EIC) will be the definitive facility for mapping GPDs and TMDs and completing the 3D imaging of the nucleon~\cite{Accardi:2012qut}. With its luminosity, wide kinematic reach and polarized beams, the EIC will enable comprehensive studies of nucleon structure in the regime dominated by gluons and sea quarks.

This evolving 3D understanding is not only revolutionizing our view of the nucleon—it is essential to addressing some of the most fundamental questions in quantum chromodynamics (QCD), including the origin of nucleon spin and mass, the dynamics of confinement, and the role of orbital motion in hadronic matter. Furthermore, this approach can be extended to nuclei, providing a unified framework that connects their conventional hadronic description, mediated by pions, with their underlying partonic structure~\cite{Hen:2016kwk,Cloet:2019mql}.

In this new framework,important questions about the structure of the nucleon can be posed. The experimentally accessible observables and their sensitivity to the different components of the partonic structure of the nucleon are described in Appendix A.

One example of a central open question we pose is the role of orbital angular momentum (OAM) in resolving the proton spin decomposition. While quark and gluon helicities have been constrained through polarized DIS at SLAC, DESY, CERN and Jefferson Lab and, RHIC polarized proton-proton collisions data, the spin sum rule remains incomplete without OAM. 

Two nucleon spin decompositions sum rules are available to explore the structure of the nucleon the Ji sum rule and the Jaffe--Manohar sum rule.
The Ji sum rule~\cite{Ji:1996ek} offers a gauge-invariant decomposition, relating the total angular momentum of quarks and gluons to the second moment of GPDs: 
\[
J_{q,g} = \tfrac{1}{2} \int_0^1 dx \, x \big[ H_{q,g}(x,\xi,t=0) + E_{q,g}(x,\xi,t=0) \big].
\]
This relation highlights the importance of accessing the GPD combination $H+E$, which directly encodes quark and gluon angular momentum. Deeply virtual Compton scattering (DVCS) provides the primary experimental channel to constrain $H$ and especially the elusive $E$ GPD. Measurements at HERMES~\cite{Airapetian:2008aa}, COMPASS~\cite{Adolph:2012qw}, and Jefferson Lab 12~GeV~\cite{Defurne:2017paw} have yielded the first quantitative estimates of quark angular momentum in the valence region, offering partial closure of the Ji sum rule. However, the gluon and sea-quark contributions remain essentially unconstrained. In QCD, OAM is not a conserved quantity; even if absent at one renormalization scale, it reappears at another through evolution. This makes OAM a dynamical, scale-dependent component of the proton spin structure. Theoretically, the gauge-invariant completion of the Jaffe--Manohar sum rule has clarified the distinction between canonical and kinetic OAM~\cite{Jaffe:1989jz,Hatta:2011ku}, both of which can be expressed within the framework of \emph{Wigner distributions}. These five-dimensional phase-space distributions reduce to \emph{generalized transverse momentum dependent distributions} (GTMDs) upon integration~\cite{Meissner:2009ww,Lorce:2011kd}, providing a natural language to encode spin--orbit correlations and the connection of OAM to both GPDs and TMDs. At small-$x$, OAM is predicted to cancel helicity, leading to intriguing phenomena such as perfect spin--orbit anti-correlation and partonic entanglement in Bell-state--like configurations~\cite{Bhattacharya:2024sck}. Yet, OAM has never been directly measured. The Electron--Ion Collider provides a unique opportunity to access OAM, particularly through observables such as longitudinal double-spin asymmetries in diffractive dijet production~\cite{Hatta:2019ixj,Bhattacharya:2024sck}, potentially opening the way to map quark and gluon OAM via their GTMD/Wigner function representations and to finally close the proton spin sum rule.

In the TMD sector another example worth mentioning is the quest for the sign change between the Sivers asymmetry measured in SIDIS and Drell-Yan. The Sivers function encodes a correlation between parton transverse momentum and the nucleon’s transverse spin, and QCD with gauge links predicts a process-dependent sign: it enters with one sign in SIDIS (final-state interactions) and the opposite sign in Drell–Yan or $W/Z$ production (initial-state interactions)~\cite{Collins:2002kn,Belitsky:2002sm}. Over two decades, SIDIS measurements at HERMES/COMPASS/JLab have established sizable, flavor-dependent Sivers asymmetries and enabled global fits with TMD evolution~\cite{HERMES:2009lmz,Bury:2020vhj}. The decisive test is in polarized Drell–Yan (or $W/Z$) where the sign should flip. COMPASS has now reported final pion-induced polarized Drell–Yan results combining 2015+2018 data; the extracted Sivers asymmetry is \emph{consistent} with the predicted sign reversal~\cite{COMPASS:2017jbv,COMPASS:2023qvt}. At RHIC, STAR’s $W/Z$ transverse single-spin asymmetries remain consistent with zero within current precision but are compatible with (and mildly prefer) the sign-change hypothesis~\cite{STAR:2015vmv,STAR:2023jwh}. Taken together with global fits, the data support the QCD sign reversal, but the confirmation is not yet statistically definitive. The EIC  will deliver high-precision, multi-differential SIDIS over wide $(x,Q^2,P_{hT})$ with excellent PID, crucial to map the quark (including sea and strange) Sivers function and to test TMD evolution with unprecedented rigor, thereby tightening theory-driven predictions for Drell–Yan sign-change tests and enabling complementary access to the \emph{gluon} Sivers function in DIS channels (heavy flavor, dijets)~\cite{AbdulKhalek:2021gbh}.
 
\subsection{Is the Nucleon Spin Structure Modified in the Nucleus?}
The quest to understand how quarks and gluons bind together to form the vast bulk of visible matter in the universe (e.g., protons, neutrons, and nuclei) has been significantly accelerated by several important experimental discoveries that challenge our understanding of quark-gluon dynamics as governed by QCD. Two standout examples include: 1) The spin-puzzle~\cite{EuropeanMuon:1987isl}, where polarized DIS experiments in the 1980s implied that the fraction of the proton's spin carried by the quarks was much smaller than expected. At the time, this result was consistent with zero, which was in contrast with relativistic constituent quark model predictions which found a quark spin fraction of around two-thirds~\cite{Myhrer:2007cf}. 2) The EMC effect~\cite{EuropeanMuon:1983wih}, where unpolarized DIS experiments on iron in the 1980s found that nuclear structure functions are quenched with respect to the deuteron in the valence region ($0.3 \lesssim x \lesssim 0.7$). At the time the expectation was that the separation of scales between quark and nucleon degrees of freedom in a nucleus implied that nuclear structure functions would simply be given by the sum of proton and neutron structure functions (i.e. $F_{2A} \simeq Z\,F_{2p} + N\,F_{2n}$) in the valence region.

In the forty years since these discoveries, efforts to understand the spin-puzzle have led to a much deeper appreciation of the role of gluon spin and quark/gluon orbital angular momentum in the proton~\cite{Leader:2013jra}. Similarly, efforts to understand the EMC effect -- which implies that valence quarks in a nucleus carry less momentum than in a nucleon -- have revealed a shortcoming in traditional nuclear physics approaches based on nucleon degrees of freedom and a nucleon-nucleon potential, because these approaches have not been able to explain the EMC effect~\cite{Smith:2002ci}. Instead, the EMC effect points to modification of nucleon structure by the nuclear medium, with the mechanism likely being the nuclear scalar and vector mean-fields~\cite{Cloet:2009qs} and/or short-range correlations~\cite{Weinstein:2010rt,Arrington:2012ax}.

Given the tremendous impact of these two measurements on our understanding of the quark and gluon structure of nucleons and nuclei, it is perhaps surprising that in the preceding 40 years since these discoveries their combined impact has not been explored in any detail via DIS experiments of the spin structure of nuclei. Measurements of the spin structure of the deuteron and $^3$He exist, but these have primarily been used as effective neutron targets rather than for probing nuclear spin structure. In fact, nuclei offer a tremendously rich laboratory with which to study QCD with an abundance of new observables manifesting in targets with $J \geqslant 1$ such as the deuteron, $^6$Li, $^7$Li, $^9$Be, $^{10}$B, etc. For example, the leading-twist gluon transversity distribution, $\Delta_T g(x)$, can only appear in targets with $J \geqslant 1$ because of helicity conservation~\cite{Jaffe:1989xy,Maxwell:2018gci}. Observation of a gluon transversity distribution in the deuteron would be the first direct evidence for non-nucleonic components in nuclei, such as exotic glue, a $\Delta\Delta$ component, etc. Such an observation would be a major advance in the pursuit to understand the QCD origin of the nucleon-nucleon interaction. 

Among the rich possibilities provided by nuclei, studying the EMC effect for the spin structure functions, that is the {\it polarized EMC effect} for light nuclei is perhaps the most important next step. To explore the spin structure functions of a nucleus it is important to note that the spin of a nucleus is, to a very good approximation, carried by the valence nucleons, that is, those nucleons in the outermost shells. This implies that it is only these nucleons that contribute to spin structure functions, which contrasts with the unpolarized case where all nucleons contribute. Therefore, the spin structure functions are suppressed by $1/A$ relative to their unpolarized counterparts making light nuclei ideal systems for such studies. A natural starting point would be the $^3$H/p ratio, which could be measured at Jefferson Lab; however, radioactivity from the $^3$H source presents significant challenges. Other light nuclei of interest include $^3$He, $^{6,7}$Li, $^{7,9,10}$Be, $^{10,11}$B, and $^{14,15}$N. 
\begin{figure}[tbp]
\includegraphics[width=0.48\textwidth]{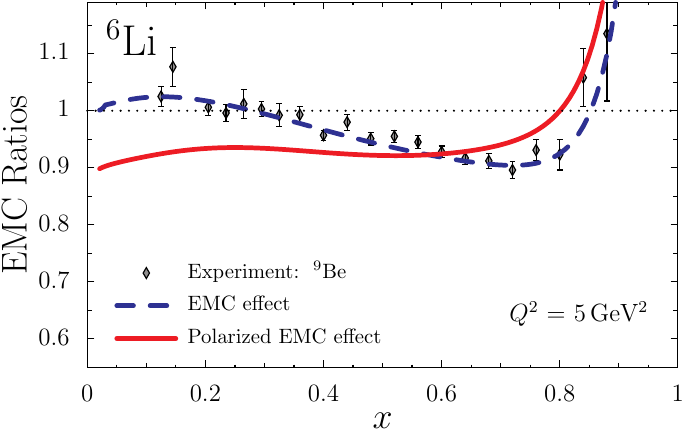} \hfill
\includegraphics[width=0.48\textwidth]{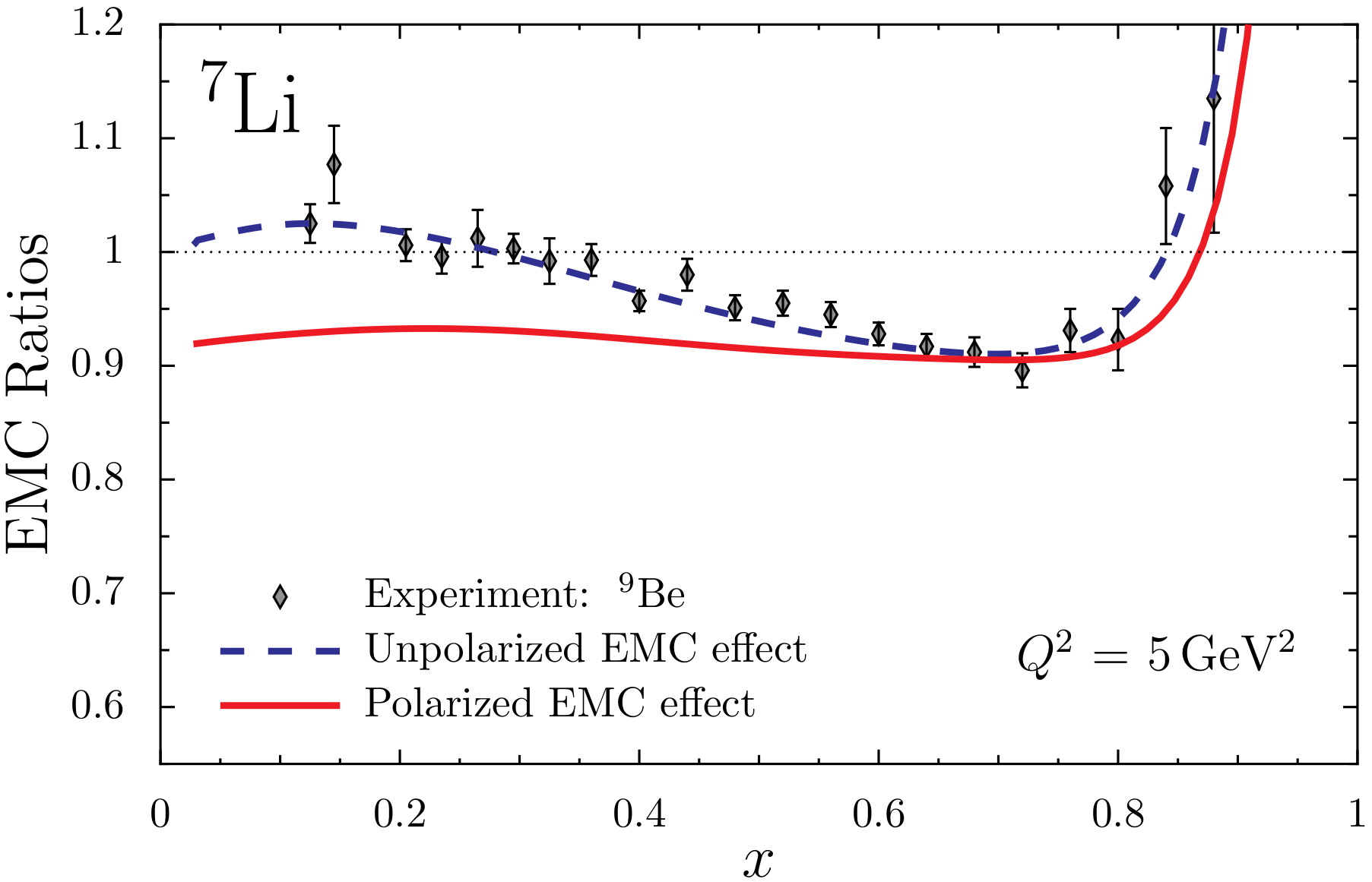}
\caption{Results for the EMC and polarized EMC effects in $^6$Li and $^7$Li, where the latter predictions are from Ref.~\protect\cite{Cloet:2006bq}.}
\label{fig:emc_effects}
\end{figure}

In the study of QCD effects in nuclei, it is useful to define quantities whose deviation from unity indicates some non-trivial nuclear effect, which can be achieved with ratios of the form $R= ({\rm observed})/({\rm naive~expectation})$. For the EMC effect this gives the definition:
\begin{eqnarray}
R_{\rm EMC}^A &:=& \frac{F_{2A}(x)}{Z\,F_{2p}(x) + N\,F_{2n}(x)} \nonumber \\
&\simeq& \frac{F_{2A}(x)\big|_{\rm ISO}}{F_{2D}(x)},
\label{eq:emc_effect}
\end{eqnarray}
where the latter approximation is often employed because there is no free neutron target to obtain $F_{2n}$, and instead the denominator is given by the deuteron $F_{2D}$ structure function and isoscalarity corrections are then applied to the numerator~\cite{Malace:2014uea}. In analogy with Eq.~\eqref{eq:emc_effect} 
the polarized EMC effect~\cite{Cloet:2005rt} can be defined as

\begin{equation}
\Delta R_{\rm EMC}^A := \frac{g_{1A}(x)}{P_p\,g_{1p}(x) + P_n\,g_{1n}(x)}.
\end{equation}
%
where $P_p$ and $P_n$ are the effective polarizations of the protons and neutrons in the nucleus of interest, which need to be determined using state-of-the-art methods in traditional nuclear physics. Several predictions exist for the polarized EMC effect~\cite{Cloet:2005rt,Smith:2005ra,Tronchin:2018mvu} for a polarized proton embedded in nuclear matter and all these studies find an effect that is at least as large as the EMC effect. If such predictions are confirmed it would provide significant insight into the quark and gluon spin structure of nuclei, and imply that the nuclear medium converts quark spin into orbital angular momentum. For light nuclei, $^7$Li and $^{11}$B are of particular interest because their spins are proton dominated, so knowledge of $g_{1n}(x)$ is less important, and $^6$Li, $^{14}$N which, together with the deuteron, are the only stable spin-one nuclei, making this trio a particularly compelling system for study. Predictions for the polarized EMC in $^6$Li and $^7$Li are given in Fig.~\ref{fig:emc_effects}, where the $^6$Li result does not yet account for configuration mixing in the p-shell which will be included in a future study.

With a polarized light ion beam at the EIC there are numerous other QCD effects in nuclei to explore. For example, the EMC and polarized EMC effect for the gluon PDFs can be studied~\cite{Wang:2021elw}. For spin-one systems there are several important new observables associated with the tensor polarization of these nuclei, including the new structure function $b_1(x)$ and three new leading twist TMDs~\cite{Ninomiya:2017ggn}. Similarly, connections between traditional nuclear structure and QCD can be made via the study of nuclear GPDs where phenomena such as deuteron and alpha clustering should manifest, together with the donut and dumbbell configurations in the deuteron. So it is clear that a polarized light-ion program at the EIC promises significant new insights into QCD and nuclei.

\subsection{Exotic Gluon States in the Nucleus}

Understanding the glue that binds nuclear matter is a key challenge of nuclear physics and a central goal of the EIC project. Despite the gluon's integral role in nuclear structure, the gluon does not couple to photons and can only be probed indirectly in electron scattering experiments. The observation of a nuclear glue effect, free from contributions of any nucleon, would provide an invaluable look into the workings of the strong force. In ``Nuclear Gluonometry''  \cite{Jaffe:1989}, Jaffe and Manohar identified just such a nuclear structure function---which they labelled ${\Delta}(x,Q^2)$---available via deep inelastic scattering from spin--1 or greater nuclei, which is sensitive \textbf{only} to gluonic states in the nucleus.

\begin{figure}[h!]
	\begin{center}
		\includegraphics[width=2in]{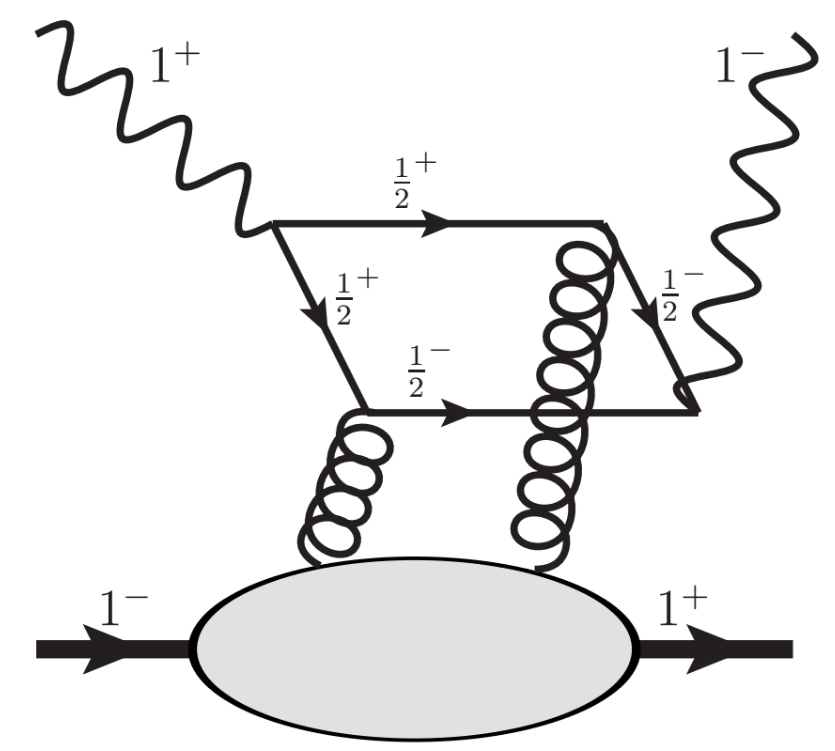}
	\end{center}
	\caption{Example of leading DIS process sensitive to ${\Delta}(x,Q^2)$ \protect\cite{Shanahan}.}
	\label{fig:nuglu}
\end{figure}

As we generalize from spin--$\frac{1}{2}$ nuclei to higher spin, new leading twist structure functions are necessary to describe the additional complexity, hinting that we are observing interactions between nucleons. For instance, the same quark and gluonic operators that lead to the spin--$\frac{1}{2}$ functions produce the new tensor structure functions $b_1$ and $b_2$ at spin--1. However, Jaffe and Manohar realized there was another tower of gluonic operators in the operator product expansion that give one more leading twist function. This structure function comes from helicity flip amplitudes $A_{+-}$ and $A_{-+}$, which correspond to a photon helicity flip of two in the interaction, as seen in Figure \ref{fig:nuglu}. Because bound nucleons or pions cannot contribute two units of helicity, the ${\Delta}(x,Q^2)$ double helicity-flip structure function is purely a gluonic observable.

In a hadronic, parton model interpretation, ${\Delta}(x,Q^2)$ can be expressed as 
 $$\Delta(x,Q^2) = - \frac{\alpha_s(Q^2)}{2\pi}\mathrm{Tr} \mathcal{Q}^2x^2 \int_{x}^{1} \frac{dy}{y^3}[g_{\hat{x}}(y,Q^2)-g_{\hat{y}}(y,Q^2)].$$
 
Here $g_{\hat{x},\hat{y}}(y,Q^2)$ represent the probability of finding a gluon with momentum fraction $y$ linearly polarized in $\hat{x}$,$\hat{y}$ direction in a target polarized in the $\hat{x}$ direction. In essence, this is an expression of how much more momentum of a transversely polarized particle is carried by gluons \textbf{aligned} rather than \textbf{perpendicular}, making $\Delta(x,Q^2)$ a signal of gluonic transversity. In the case of larger nuclei, $\Delta(x,Q^2)$ would show gluons not associated with individual nucleons in that nucleus: exotic gluonic states.

The double helicity-flip structure function has never been measured directly, but we have some clues to the magnitude of its lowest moment. Calculations from 1990 \cite{Sather}, shortly after the release of the original gluonometry paper, estimated the first moment of $\Delta(x,Q^2)$ would be quite small (-0.012$\alpha_s(Q^2)$). Renewed interest in a measurement of $\Delta$ at Jefferson Lab beginning in 2014 \cite{Maxwell} inspired initial Lattice QCD calculations on its first moment. These efforts produced a definitive signal using a spin--1 $\phi$ particle ($s\bar{s}$) with an exaggerated pion mass of 405 MeV  \cite{Shanahan}. The following year, the same group did calculations on a non-physical deuteron with pion mass of 806 MeV, and again saw a definitive signal \cite{winter_first_2017}. These results motivated further proposals for  $\Delta(x,Q^2)$ measurements at Fermilab and NICA.

A DIS measurement of $\Delta(x,Q^2)$ requires electron scattering from transversely polarized nuclei. Spin--1 or greater is needed, but as a multi-nucleonic effect, $\Delta$ is expected to be larger in compact nuclei with strong nuclear binding, much like the EMC effect. This might explain the significant signal seen from LQCD with heavier pion masses, making a more compact deuteron. So while the deuteron should be investigated, the best chance for discovery may come from larger nuclei. When considering available species that are known to be polarizable for targets or ion sources, obvious choices are spin--1 $^6$Li or $^{14}$N, or spin--$\frac{3}{2}$ $^7$Li or $^{23}$Na.

The polarized target or ion source technology necessary to provide polarized nuclei for such experiments remains a key challenge. Such polarization schemes---such as Heidelberg's polarized Li and Na atomic beam source \cite{steffens_source_1977}---have for the most part been demonstrated in the past, but they require the development and maintenance of expertise as much as technique.
The EIC offers an ideal place to search for a signal of $\Delta(x,Q^2)$, but a concerted effort is required to ensure the successful development of sources of polarized ions and means to keep them polarized in the EIC ring. 


\section{The Electron-Ion Collider}

\subsection{The Electron-Ion Collider}
\subsubsection{Introduction}
The Electron-Ion Collider (EIC), under development at Brookhaven National Laboratory, represents a significant advancement in our ability to explore the fundamental structure of matter. By enabling collisions between polarized electron beams and polarized protons or ions, the EIC will provide unprecedented insights into the role of gluons—the carriers of the strong force—in binding quarks within nucleons and nuclei. This facility is poised to address critical questions in Quantum Chromodynamics (QCD), thereby deepening our understanding of the strong interaction that governs atomic nuclei.

The scientific case for the EIC has been built and strengthened over more than two decades of community studies, culminating in the recommendation by the Nuclear Science Advisory Committee in 2023 to expeditiously complete the EIC as the highest priority for facility construction. According to the National Academy Report in 2018~\cite{NAP25171}, the compelling science case for the EIC rests on five pillars:
\begin{itemize}
\item How do quarks, gluons, and orbital angular momentum contribute to proton spin?
\item Does the mass of visible matter emerge from quark-gluon interactions?
\item How can we understand QCD dynamics and the relation to confinement?
\item How do quark-gluon interactions create nuclear binding?
\item Does gluon density in nuclei saturate at high energy?
\end{itemize}
These questions led to the DOE Mission-Need Statement for the Electron-Ion
Collider in late 2019 with the following requirements:
\begin{itemize}
\item A peak electron-proton luminosity of $10^{33}$ to
  $10^{34}~{\rm cm}^{-2}{\rm sec}^{-1};$
\item A variable electron-proton center-of-mass energy of 20 to 100~GeV, upgradeable to 140~GeV;
\item A high degree of polarization (70\%) for both electrons and light ions;
\item Availability of ion beams from deuterons to the heaviest stable nuclei;
\item Possibly more than one interaction region.
\end{itemize}
The EIC design~\cite{CDR} achieves all these requirements.

\subsubsection{Interaction Region}
Compared to the electron-proton collider HERA~\cite{hera}, which operated from 1992 to 2007, the EIC achieves a peak luminosity that is approximately 200 times higher, albeit at a lower center-of-mass energy. This luminosity increase is achieved by a larger number of bunches, a higher electron bunch charge, flat proton beam emittances, and tighter focusing at the interaction point (IP), as listed in Table~\ref{hera-eic-comparison}.
\begin{table}
  \caption{\label{hera-eic-comparison}Comparison of some key parameters of
    the electron-hadron colliders HERA and EIC}
  \begin{center}
  \begin{tabular}{lcc}
    \hline\hline
    & HERA & EIC\\
    \hline
    circumference [km] & 6.3 & 3.8\\
    center-of-mass energy [GeV] & 320 & 105\\
    number of bunches & 174 & 1160\\
    proton bunch charge [nC] & 11.7 & 11.0\\
    electron bunch charge [nC] & 5.3 & 28\\
    proton bunch length [cm] & 16 & 6\\
    electron $\beta^{\ast}$ (h/v) [cm] & 62/26 & 45/5.6\\
    proton $\beta^{\ast}$ (h/v) [cm] & 245/18 & 80/7.2\\
    RMS electron emittance (h/v) [nm] & 20/3 & 20/1.3\\
    RMS proton emittance (h/v) [nm] & 4/4 & 11/1\\
    luminosity [$10^{34}\,{\rm cm}^{-2}{\rm sec}^{-1}$] & 5.3 & 1000\\
    \hline
    \end{tabular}
    \end{center}
  \end{table}
The small $\beta$-functions at the IP require low-$\beta$ quadrupole magnets as close as possible to the interaction point. However, the detector itself occupies some $\pm 4.5$ to $5\,{\rm m}$ around the IP, thus pushing the closest focusing elements to a distance of about 5 to 5.5~m from the IP. To accomplish this for both the electron and the hadron beams, beams collide under a crossing angle of 25~mrad. This allows installation of electron and hadron quadrupoles side-by-side and avoids any separator dipoles which would result in a wide, hard synchrotron radiation fan that would have to be passed safely through the detector. The crossing angle is compensated by a set of crab cavities on either side of the IP that rotate the bunches around a vertical axis such they collide head-on at the IP.

\begin{figure*}
  \centering
  \includegraphics[width=0.7\linewidth]{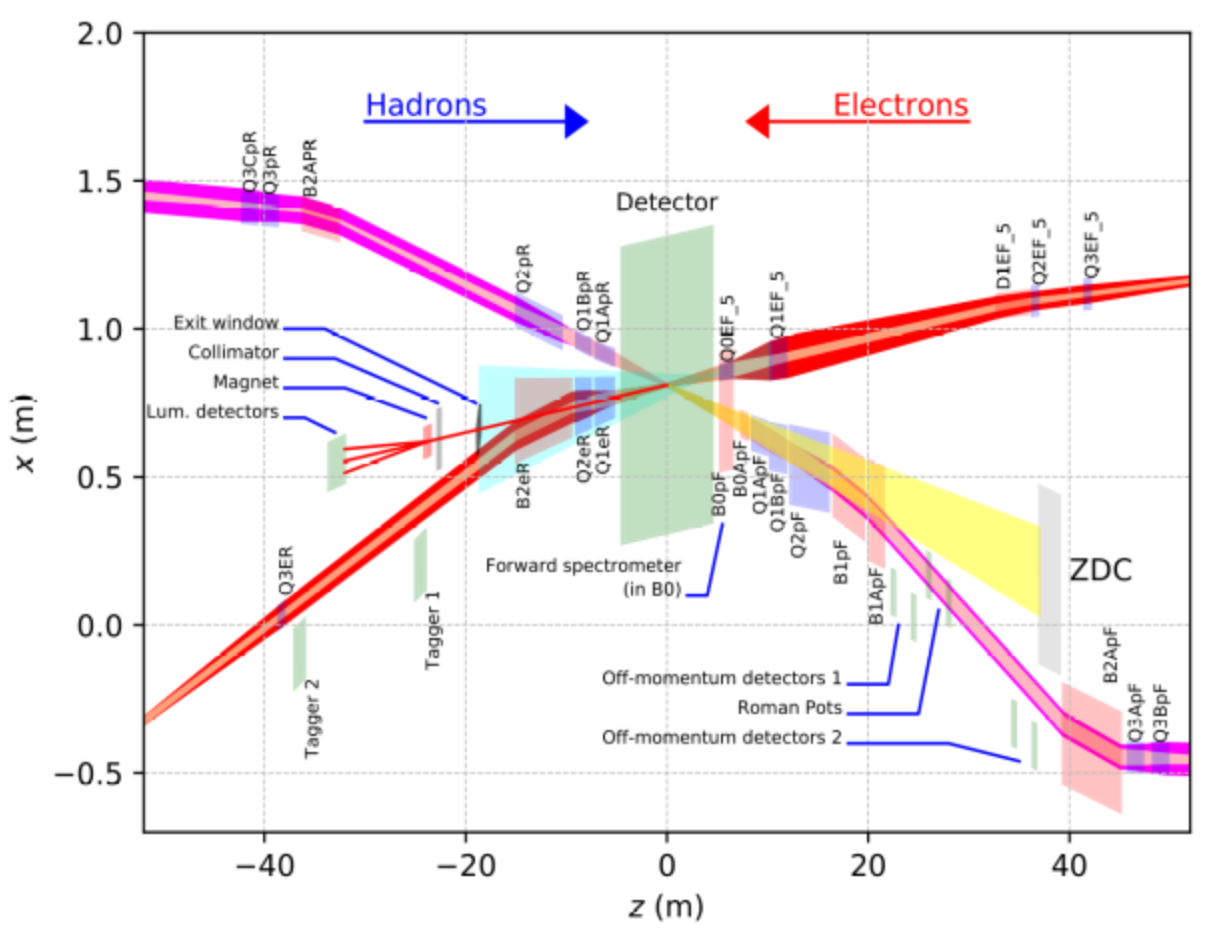}
  \caption{\label{inner-ir}Schematic view of the magnet and auxiliary detector arrangement in the interaction region. Magnet apertures are indicated by colored boxes.}
\end{figure*}

Besides focusing elements, dipole magnets are used to separate the hadron beam on the forward side of the detector from the $\pm 4\,{\rm mrad}$ neutron cone that is passed along to the Zero Degree Calorimeter (ZDC). The hadron magnets on the forward side need to have large apertures to accommodate not only the beam itself, but also the neutron cone as well as scattered particles with a transverse momentum of up to 1.3~GeV/c that are detected by ``Roman pots'' along the beamline. Electron magnet apertures on the rear side of the detector are determined by the need to safely pass the synchrotron radiation photons generated by the quadrupoles on the (``upstream'') forward side through the superconducting magnets. A dipole magnet deflects the electron beam away from the synchrotron radiation fan as well as from the much harder Bethe-Heitler photons that are used for luminosity measurement. Figure~\ref{inner-ir} shows the arrangement of magnets and auxiliary detectors in the interaction region.

Proton spin rotation from vertical polarization in the arcs to longitudinal polarization at the IP is accomplished by the same helical dipole rotators~\cite{helical-snakes} that are currently in use at RHIC. In the ESR, two sets of solenoid-based rotators will be installed. These solenoids rotate the vertical spins coming from the arcs by 90 degrees into the radial direction. The subsequent dipoles then provide the rotation into the longitudinal direction. This process is reversed downstream of the detector, so spins are vertical in the arcs. There are dedicated solenoids for 5 and for 18~GeV installed in the appropriate locations with respect to the orbit angles between solenoids and IP; for operation at 10~GeV all solenoids are powered.

\subsubsection{From RHIC to EIC}
\begin{figure*}
  \centering
    \includegraphics[width=0.7\linewidth]{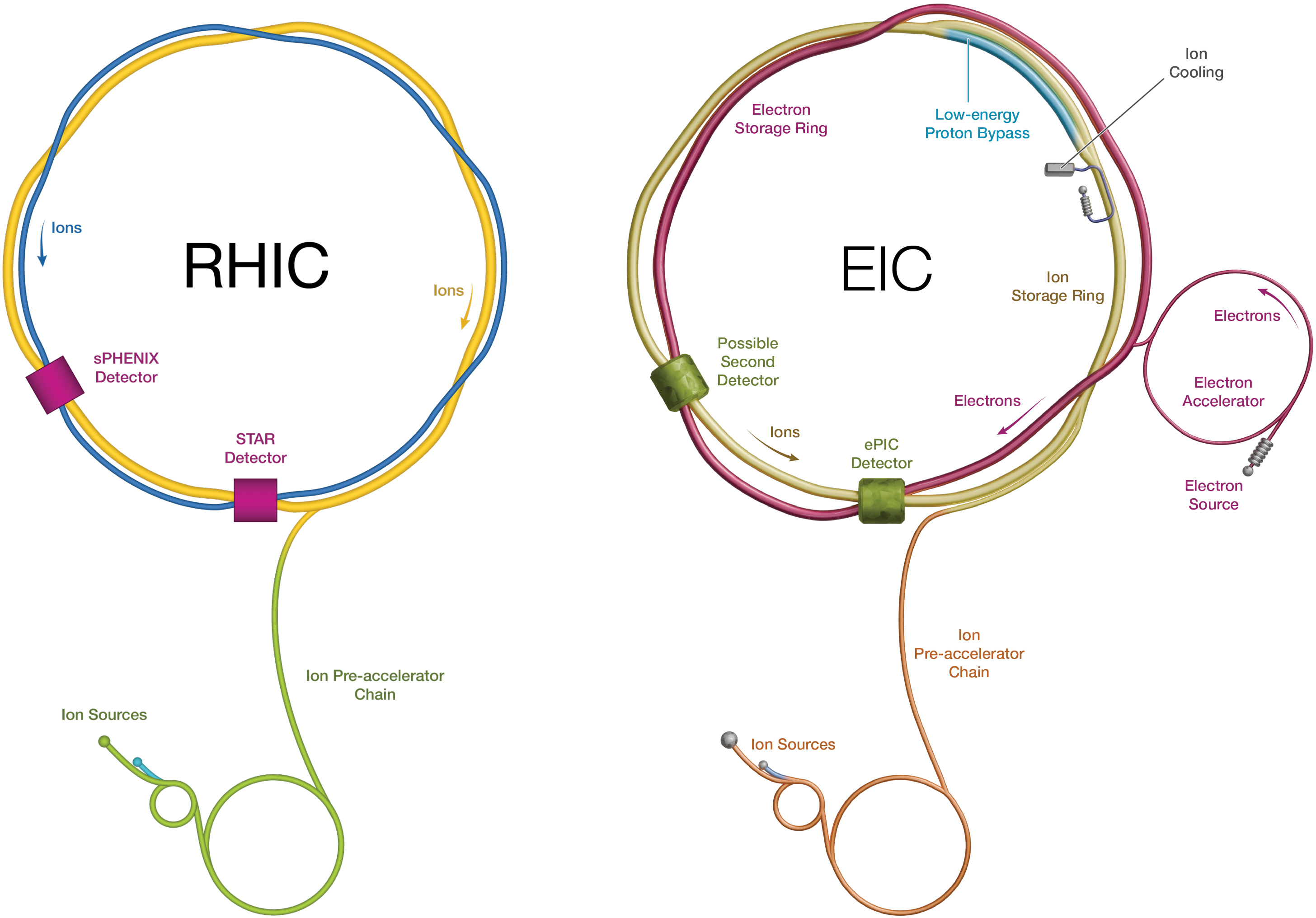}
  \caption{\label{RHIC-and-EIC}Schematic layouts of RHIC (left) and of the EIC (right) with their respective injectors.}
  \label{fig:EIC}
\end{figure*}

The EIC will be based on the existing RHIC facility, utilizing its existing tunnel, infrastructure, and injector chain. RHIC consists of two superconducting storage rings with a circumference of 3.8~km, ``Blue'' and ``Yellow''. The two rings intersect at six equally spaced locations around the ring circumference, two of which are currently equipped with the nuclear physics detectors ``STAR'' in IR6, and ``sPHENIX'' in IR8. Ion species from protons to uranium have been accelerated and collided in RHIC at beam energies up to 275~GeV for protons and 100~GeV/nucleon for heavy ions. Proton beams have reached polarization levels of approximately 60\% at 275~GeV, which makes RHIC the only polarized proton collider in the world. During its nearly 25 years of operation, RHIC has exceeded its original design luminosity by a factor 44. RHIC operation will end in December 2025, and construction of the EIC will begin. The ``Yellow'' RHIC ring will serve as the Hadron Storage Ring (HSR) of the EIC, while a new electron storage ring will be added in the collider tunnel. Figure~\ref{RHIC-and-EIC} shows the layouts of the RHIC collider with its injector chain and of the Electron-Ion Collider.
\subsubsection{Hadron Storage Ring}
Several modifications and upgrades to the ``Yellow'' RHIC ring are necessary to make it suitable as the Hadron Storage Ring for the EIC. To accommodate the much larger number of bunches with their shorter bunch length, actively cooled beam screens will be inserted in-situ in the cold bore of the superconducting magnets. These screens are copper-coated to reduce the electric resistivity compared to the bare stainless steel beam pipe, which reduces the heat load due to image currents. A secondary layer of amorphous carbon reduces the secondary electron yield (SEY) to prevent the build-up of electron clouds.

While each of the two present RHIC rings is equipped with two Siberian snakes for polarization preservation, four additional Siberian snakes will be installed in the EIC Hadron Storage Ring -- the former "Yellow" RHIC ring. Two of these devices will be directly transplanted from the "Blue" RHIC ring, while the other two will be constructed using the individual magnets of the "Blue" spin rotators.

The large center-of-mass energy range is realized by changing the two beam energies. While the velocity of the electron beam remains practically unchanged over the energy range from 5 to 18~GeV, the velocity of the ions changes appreciably. To keep the two beams in collision at the interaction point, their revolution frequencies have to be equal. This synchronization is accomplished by applying a radial shift of up to $\pm 20\,{\rm mm}$ in the arcs, which facilitates a range of the Lorentz factor of $118<\gamma<293.$ To allow for even lower ion energies, a ``Blue'' arc between IR12 and IR2 will be utilized as a bypass. The average radius of this arc is about 90~cm smaller than that of the corresponding ``Yellow'' arc, which reduces the circumference of the HSR by roughly 90~cm. The resulting circumference then corresponds to a Lorentz factor of $\gamma=43.5.$

Flat hadron beams with an emittance ratio of up to $\epsilon_x/\epsilon_y=11/1$ are created by electron cooling at injection and subsequent blow-up of the horizontal emittance. High-precision decoupling of the transverse planes ensures that this flatness is preserved during ramping and store. This injection energy cooler is based on a conventional electron cooling scheme, albeit using bunched electron beams. This concept has been operationally demonstrated at the Low Energy RHIC electron Cooler (LEReC).

\begin{figure}[htb]
  \centering
   \includegraphics[width=\columnwidth]{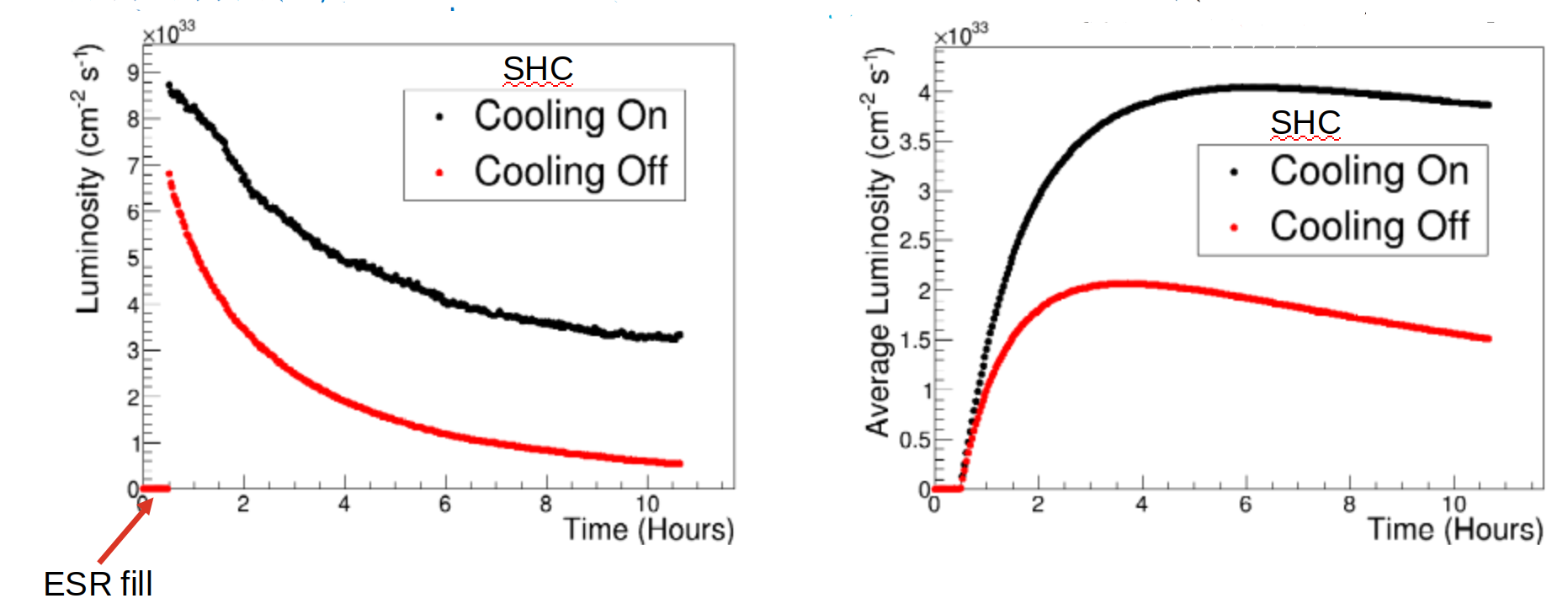}
  \caption{\label{luminositymodel}Luminosity evolution during a store (left) and average luminosity vs. store length (right) in the EIC, for collisions of 10~GeV electrons and 275~GeV protons. The black curves correspond to the case of collision energy cooling during stores in addition to injection energy cooling, while the red curves correspond to operation with injection energy cooling only. }
\end{figure}

At a later stage, cooling at collision energies could be added. Based on a detailed luminosity model for the EIC, this addition would increase the average luminosity by a factor of two if the cooling scheme is based on Coherent electron Cooling (CeC)~\cite{cec}, as shown in Figure~\ref{luminositymodel}. Since CeC is essentially a very high bandwidth stochastic cooling system, only particles within a rather small volume around the center of the hadron bunch are actually cooled, while particles at larger amplitudes are heated and therefore eventually lost from the beam. These losses are the dominant factor limiting the average luminosity.

\subsubsection{Electron Storage Ring}
The Electron Storage Ring (ESR) is based on a conventional FODO lattice. The bending sections in the arcs are realized as ``super-bends'' by splitting the dipoles into three segments, a short, 0.89~m long dipole centered between two 2.73~m long ones. For optimum luminosity over the entire center-of-mass range of the EIC, the electron beam emittance needs to be nearly constant at 20 to 24~nm. This is accomplished by a combination of two methods. At 18~GeV beam energy, the lattice is set to a betatron phase advance of 90 degrees per FODO cell, while at 10~GeV that phase advance is set to 60 degrees. At 5~GeV, the center dipoles of the super-bends are powered at reverse polarity. The resulting sharp, reverse bend increases the emittance to the desired value of 20~nm. In addition, it enhances radiation damping to support large beam-beam parameters even at the lowest electron beam energy.

\begin{figure}[htb]
  \centering
    \includegraphics[width=\columnwidth]{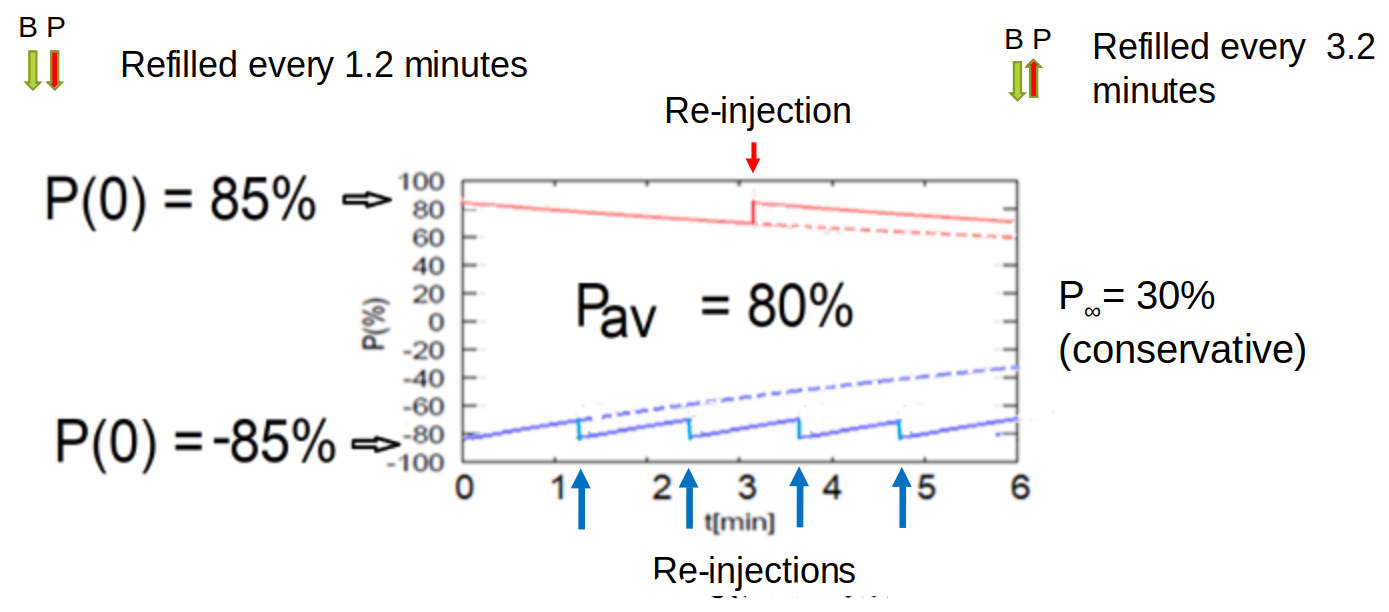}
  \caption{\label{polarization-sawtooth}Illustration of the polarization in a single ESR bucket vs. time, for a bunch with spins parallel (blue) and anti-parallel (red) to the main dipole field of the accelerator.}
  \end{figure}

High polarization in arbitrary spin patterns (spin ``up'' and ``down'') is achieved by injecting polarized bunches with the desired spin orientation at full intensity and energy. The polarization in the stored bunches will then decay due to Sokolov-Ternov self-polarization and spin diffusion towards an equilibrium. To counteract depolarization, bunches will be continuously replaced at a rate of one bunch per second. As a result, the polarization in every single bucket will exhibit a sawtooth pattern, as illustrated in Figure~\ref{polarization-sawtooth} for an injected polarization level of $\pm 85$~percent, an equilibrium polarization of 30~percent, and a target average polarization of 80~percent. As shown in the figure, bunches injected with negative polarization (spins parallel to the main dipole field) depolarize at a higher rate because the equilibrium polarization is positive. These bunches therefore have to be replaced more frequently than those with positive polarization to achieve the same average polarization.

\subsubsection{Electron Injector}
Longitudinally polarized bunches are generated in a high intensity polarized HVDC gun. This gun has been developed over several years by BNL and Stony Brook University and is capable of delivering bunches with up to 11~nC charge. A Wien filter is then used to rotate the spins into the vertical direction before acceleration in a 750~MeV normal-conducting S-band LINAC. Up to 28 single bunches of 1~nC each are accelerated at a rate of 30~Hz and injected into a Beam Accumulator Ring (BAR) for accumulation into one single 28~nC bunch per second. This bunch is then transferred to the Rapid Cycling Synchrotron (RCS) for further acceleration up to 18~GeV.

The RCS will be installed in a new, separate tunnel adjacent to the IR4 location of the collider. This choice was made to free up space in the highly congested collider tunnel. The highly periodic lattice of the 1.4~km circumference RCS ensures polarization preservation during the energy ramp. This is accomplished by a combination of high periodicity and high (integer) vertical tune, which drives the lowest-order intrinsic resonances to beam energies outside the operational energy range of the RCS~\cite{RCS-vahid}. As a consequence, the beam does not encounter any depolarizing intrinsic resonances during the energy ramp. This principle is routinely applied at the Brookhaven AGS, where proton beam depolarization is observed only at those energies where the resonance condition for intrinsic resonances is fullfilled. In the case of the AGS, where protons do encounter these depolarizing resonances during the energy ramp, these resonances are overcome by additional measures such as partial Siberian snakes and an AC dipole.

\subsubsection{Project Status}
The Electron-Ion Collider project was awarded CD-0 (``Mission Need'') in December 2019, and the site selection occurred shortly after in January 2020. CD-1 (``Alternative Selection and Cost Range'') was awarded in 2021, and the project is currently gearing up for CD-2 status (``Performance Baseline'') in 2026, potentially in conjunction with CD-3 (``Start of Construction''). Early procurement of long-lead items (CD-3A) was approved in early 2024, and a second round of early procurements (CD-3B) is currently awaiting approval. Project completion and start of operations (CD-4) are currently envisioned for 2034.

\subsection{Polarized electron beams at EIC}

\begin{figure*}[t]
\centering
\includegraphics[width=0.95\textwidth]{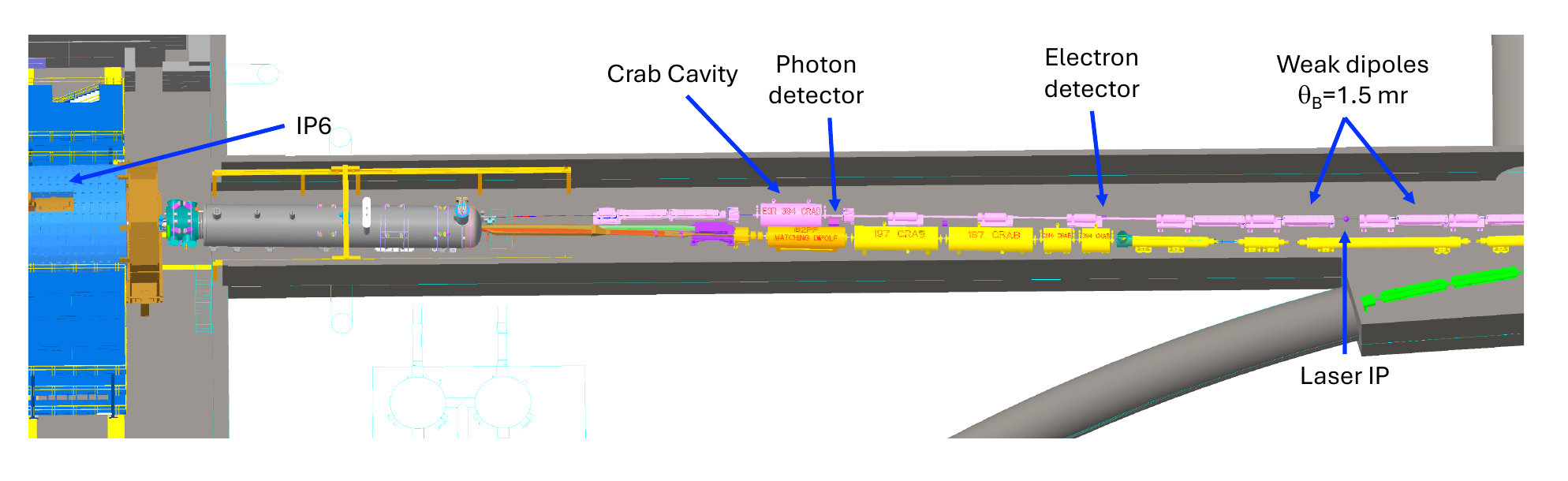}
\caption{\label{figure:compton_layout} Layout of the ESR Compton Polarimeter. Due to the dipoles between the Compton IP and the experiment, the electron spin direction will precess between the two locations.}
\end{figure*}

The polarized electron beam planned for the Electron Ion Collider poses unique challenges in the areas of polarized electron beam generation, beam transport, and electron polarimetry. The EIC electron beam complex is shown in Fig.~\ref{fig:EIC} (right panel). Electron bunches of $\approx$1~nC are generated fully polarized using a photocathode source.  An S-band normal-conducting linac accelerates the electron bunches to 750 C.  A beam-accumulator-ring (BAR) is then used to provide 7~nC or 28~nC bunches (depending on final beam energy) to the rapid-cycling-synchrotron (RCS).  The RCS accelerates the electron bunches from 750~MeV to 5, 10, or 18 GeV before they are injected into the electron storage ring (ESR).  The resulting electron beam in the ESR is required to have high polarization ($P>70\%$), with bunch spacings of only 10 ns (25 ns) at 5 GeV and 10 GeV (18 GeV). Average beam currents will be up to 2.5~A, resulting in significant synchrotron radiation.

Since polarized electron beams are still actively used both at Jefferson Lab and Mainz, there is significant existing expertise in this area. Nonetheless, there are certain challenges that should not be overlooked in order to successfully achieve the specifications required for EIC.

The first issue that must be addressed is the fabrication of photocathode material for polarized electron sources.  The primary source of polarized photocathodes no longer exists and the program at Jefferson Lab is presently running using its inventory of photocathodes from this source. In response to this need, there is an active R\&D program at Brookhaven National Lab, Jefferson Lab, Old Dominion University, and elsewhere to develop the expertise to provide reliable photocathode material~\cite{biswas, stutzman}.  While these projects are making good progress, there is still not a consensus choice of technique for photocathode fabrication, nor have any of these new photocathodes been deployed in a "production" environment.  It is also possible that different technologies may be necessary to meet the needs of the EIC program as opposed to the Jefferson Lab program (where the required bunch charges are smaller).

Electron polarimetry will also pose unique challenges at EIC. Electron polarimetry will be required in both the RCS and ESR. The RCS polarimeter will be used primarily for beam setup so needs only modest precision.  The RCS Compton polarimeter will operate in "multi-photon" mode with many (a few thousand) backscattered photons generated for every crossing of the laser and electron bunches.  This polarimeter will measure the transverse polarization of the electrons averaged over several bunches using an integrating technique, similar to the operating mode of the LEP Compton polarimeter (for example).  The ESR Compton (see Fig.\,\ref{figure:compton_layout}) will operate in single-photon mode and measure the polarization of each electron bunch.  This places challenging constraints on the detector response time (due to the 10 ns bunch spacing at 5 and 10~GeV) and measurement time (due to the $\approx2$~minute lifetime of an electron bunch in the ring at 18 GeV). In addition, since the electron spin will undergo some precession between the Compton interaction point and the experiment IP, the ESR Compton will need to measure both longitudinal and transverse electron polarization with high precision to meet the needs of the EIC experimental program.  Finally, it is worth noting that while significant polarimetry expertise exists at Jefferson Lab and Mainz, electron polarimetry in a storage ring has qualitatively different considerations than at a fixed-target facility.

We note that the polarization of a 669 MeV stored electron beam at the MIT-Bates South Hall Ring was flipped with high efficiency using an RF dipole magnet~\cite{Morozov2001}.

Finally, the last issue that poses a significant at challenge at EIC is the acceleration and transport of the highly polarized electrons from the injector to the storage ring.  The rapid acceleration of the electron bunches in the RCS is of concern (hence the need for a polarimeter in the RCS), and the polarization lifetime of the electron bunches at the highest energy requires significant attention.

\subsection{Plans for Polarized Beams at EIC-China (EicC)}
\label{sec:Overview}
\label{sec:HIAF_Introduction}

\begin{figure*}[htb]
\centering
\includegraphics[width=0.75\textwidth]{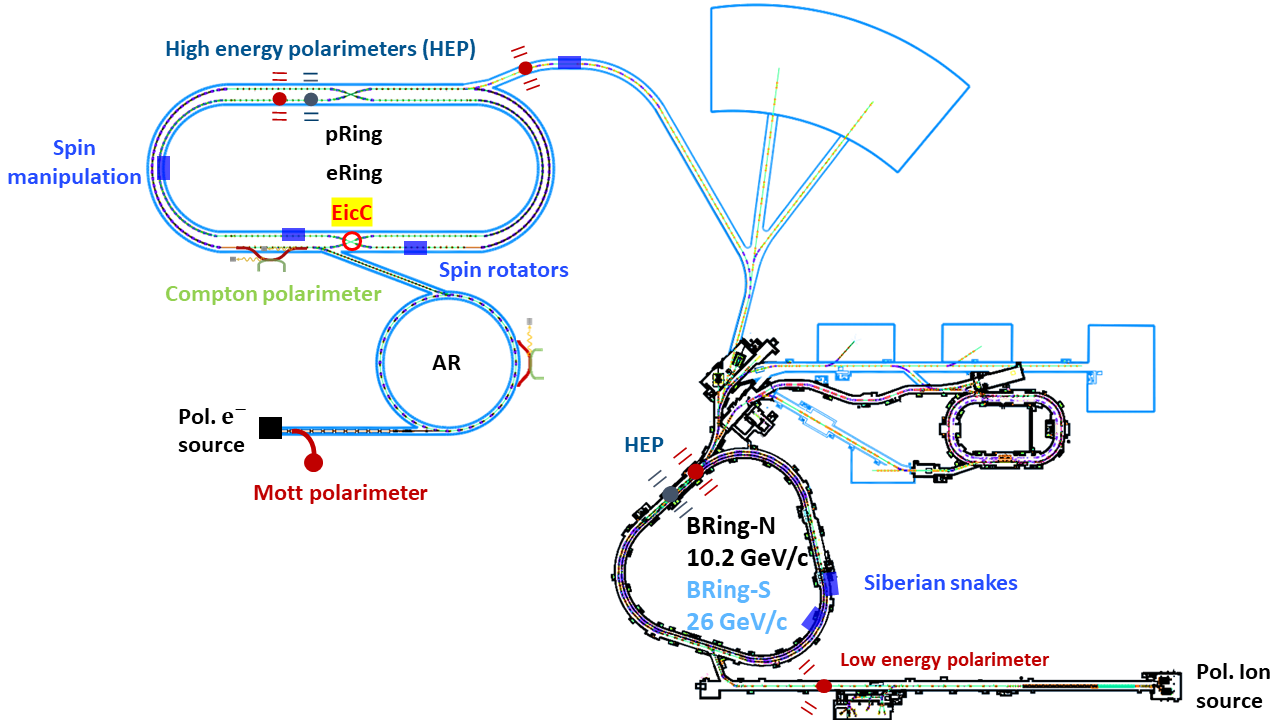}
\caption{The HIAF accelerator complex (black) and its future extension CNUF (cyan). Setups related to polarized beams, such as the polarized ion/electron sources, Siberian snakes and beam polarimeters, are indicated.}
\label{fig:HIAF_CNUF}
\end{figure*}

The \textbf{H}igh \textbf{I}ntensity heavy-ion \textbf{A}ccelerator \textbf{F}acility (HIAF)~\cite{yang2013high, zhou2022status} hosted by the Institute of Modern Physics, Chinese Academy of Sciences (IMP, CAS), is currently being constructed in Huizhou, China. As shown in the bottom right of Fig.~\ref{fig:HIAF_CNUF}, HIAF is composed of several ion sources, a cascade of accelerators, and several experimental stations. Low energy beams of nuclei ranging from hydrogen up to uranium provided by the ion sources will be accelerated in an RFQ and a superconducting ion linear accelerator (iLinac) to 17 - 48 MeV/u before being injected into the booster ring (BRing).  With its maximum rigidity of 34~Tm, the 569-meter-long triangular BRing can accelerate heavy-ion beams up to $0.835\sim2.6$~GeV/u and $1.6\sim6.0\times10^{11}$~ppp, depending on the ion species. For proton beams in BRing, a highest kinetic energy of 9.3~GeV and a maximum beam intensity of $2.0\times10^{12}$~ppp can be expected. By bombarding graphite or tantalum targets with BRing heavy-ion beams, a large amount of nuclear fragments will be produced and selected as secondary radioactive beams by the high-energy fragment separator (HFRS)~\cite{HFRS20201} to explore the territory far beyond the valley of $\beta$ stability on the nuclear chart. Secondary beams can be sent either to an external station or to the spectrometer ring (SRing) for nuclear mass measurements and internal-target experiments.
Besides, both proton and heavy-ion beams from BRing can be extracted to the high-energy experimental station for physics programs such as nuclear phase diagram, hypernuclei, and hadrons like hyperon and nucleon excited states.
HIAF is currently under construction and will be commissioned by the end of 2025. In parallel, as the future upgrade plan of HIAF, a group of next-generation accelerators and experiments under the name of \textbf{C}hina advanced \textbf{NU}clear physics research \textbf{F}acility (CNUF) are being conceived and proposed~\cite{CNUF_Zhao}. 
In Fig.~\ref{fig:HIAF_CNUF}, most parts of CNUF are drawn in cyan color, except the superconducting synchrotron (BRing-S) which shares the same tunnel with the current normal-conducting booster synchrotron (BRing-N).
BRing-S has a maximum rigidity of 86~Tm thus is able to increase the proton beam energy up to about 26~GeV/c. 
A major project of CNUF is the electron-ion collider in China (EicC)~\cite{anderle2021electron}. EicC is complementary to the EIC experiment at Brookhaven in the sense that it will explore the nucleon structure in the valence and sea quark region at lower CM energies between 15 and 20~GeV.
Hadron beams (p, d or $\mathrm{^3He}$) extracted from BRing-S will be stored in the pRing synchrotron and collide with electron beams ($2.8\sim5$~GeV) in the eRing synchrotron. 
Just as pRing, eRing serves only for beam storage, electrons are accelerated and accumulated in the AR synchrotron to the required energy and intensity before they are injected into eRing. 
Forseen setups related to polarized beams, such as the polarized ion/electron sources, Siberian snakes and beam polarimeters, are highlighted in Fig.~\ref{fig:HIAF_CNUF}.

\section{Polarized Ion Sources}
\label{chap:polionsources}
\subsection{Polarized Proton Beam}
\label{sec:OPPIS}

\begin{figure*}[htb]
    \centering
    \includegraphics[width=0.6\textwidth]{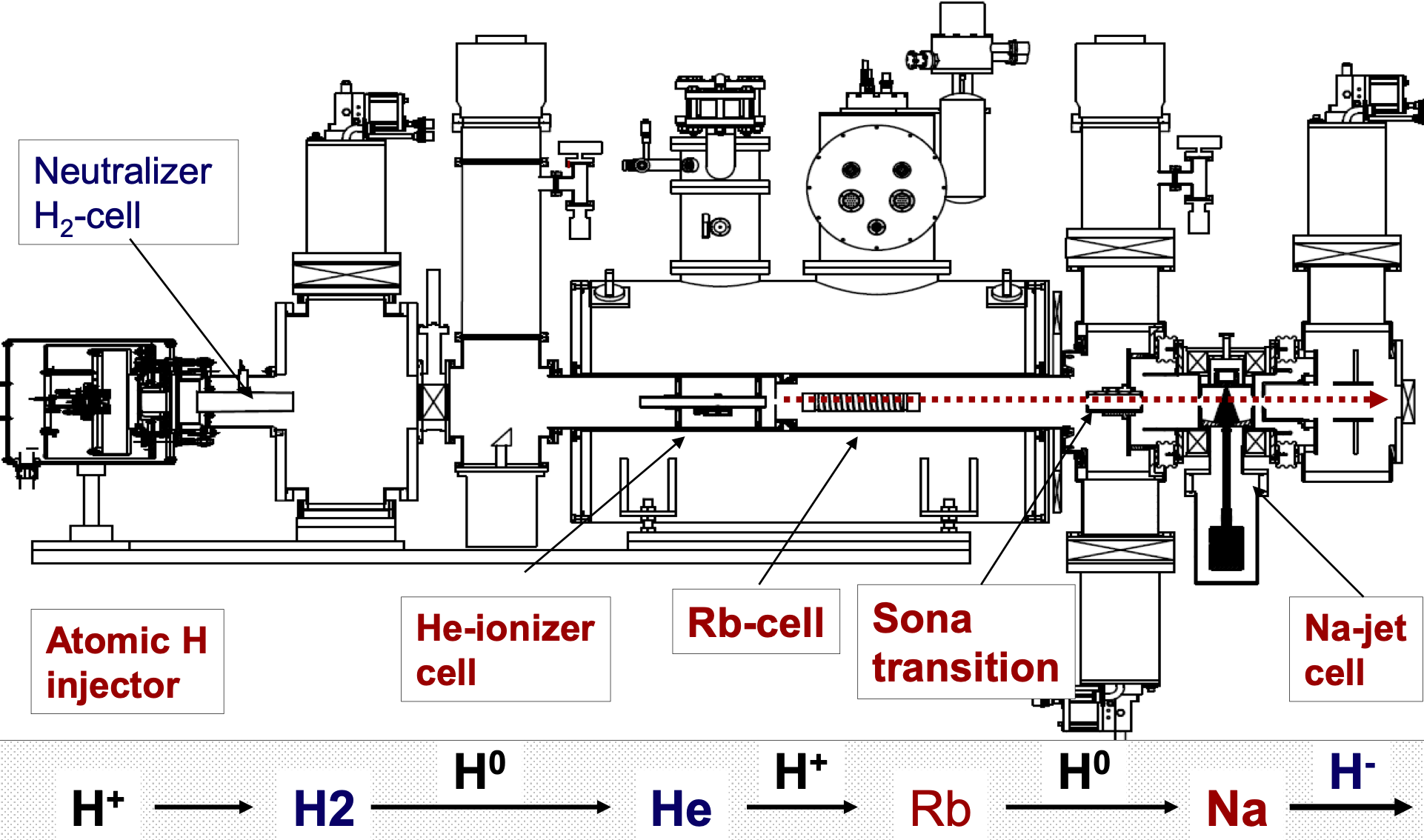}
    \caption{A Schematic Diagram of OPPIS, showing the  atomic hydrogen injector, the pulsed He-ionizer cell, the optically pumped Rb-vapor cell, the Sona Transition, and the Na-jet ionizer cell.}
    \label{fig:Oppis}
\end{figure*}

The Optically Pumped Polarized H- Ion Source (OPPIS) is a high-intensity polarized ion source developed to support the spin-physics program at the Relativistic Heavy Ion Collider (RHIC). Combined with charge-exchange injection, OPPIS enabled RHIC to achieve its maximum beam intensity and luminosity, limited primarily by beam–beam interactions.
OPPIS was first commissioned at Brookhaven National Laboratory (BNL) in 2000 \cite{Oppis1} and later upgraded for Run-2013 with the addition of a neutral hydrogen injector \cite{Oppis2}. The source operates through a multi-step process (Fig.\ref{fig:Oppis}): Proton Production: A 7 keV atomic hydrogen beam from an external source is decelerated to 3 keV and ionized in a helium gas-filled ionizer cell, located within a 3 T solenoidal magnetic field. Electron Polarization: Immediately downstream, in the Rb-cell, atomic hydrogen with polarized electrons is produced via electron transfer from optically pumped rubidium atoms excited by a circularly polarized laser ($\lambda = \SI{794.8}{nm}$) at ~85–95°C. Spin Transfer – SONA Transition: The polarized electrons’ spin is transferred to the protons through a magnetic field reversal region (the SONA transition), aided by field correction coils. H- Formation: In the Na-cell, operating at ~230°C within a 0.25 T field, an additional electron is captured, forming the polarized H- ions. These ions are extracted at 35 kV. Polarization Rotation: The longitudinal polarization is rotated to horizontal using a bending dipole, and then to vertical using a solenoidal magnet, matching RHIC injection requirements. Measurement: Polarization is measured after acceleration to 200 MeV using a polarimeter positioned at 12° and 16°.

The upgraded source consistently delivers polarized beam currents exceeding 4.0 mA. However, space-charge effects in the Low Energy Beam Transport (LEBT) line at 35 keV result in significant beam losses. Consequently, 1.4 mA reaches the RFQ, with 0.7 mA successfully accelerated to 200 MeV in the LINAC. Polarization measurements indicate up to 84\% polarization at 0.3 mA and 80\% at 0.6 mA.
The OPPIS reliably supported the 2024 RHIC polarized proton run \cite{Oppis3}, contributing significantly to achieve over 60\% polarization in RHIC’s colliding beams—one of the highest values to date for a high-energy polarized proton collider.

\subsection{Polarized Deuteron Beam}
\label{sec:pol-deuterons}

\begin{figure*}[htb]
    \centering
    \includegraphics[width=0.7\linewidth]{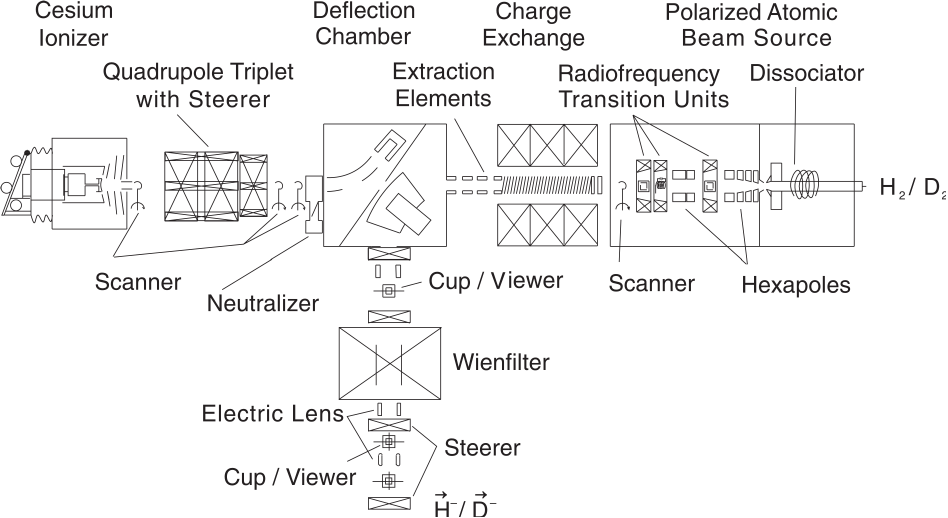}
    \caption{Scheme of the polarized ion source used at COSY\,\cite{PhysRevSTAB.9.050101}.}
    \label{fig:PPQ1B_ET}
\end{figure*}

The polarized colliding beams source at COSY\,\cite{HAEBERLI1982319,osti_6143540} comprises three major groups of components: the pulsed ground state atomic beam source \cite{Eversheim:1995am,Eversheim:1996qu,10.1063/1.1146665,Felden:2001vc}, the cesium beam source, and the charge exchange and extraction region.  The setup is shown schematically in Fig.\,\ref{fig:PPQ1B_ET}.

The ground state atomic beam source produces an intense pulsed polarized atomic hydrogen or deuterium beam. The gas molecules are dissociated in an inductively coupled RF discharge. The atoms are cooled to about \SI{30}{K} by passing an aluminum nozzle of \SI{20}{mm} in length and \SI{3}{mm} in diameter. A special admixture of small amounts of nitrogen and oxygen keeps a high degree of dissociation, reducing surface and volume recombination. The current output of the source depends sensitively on the relative fluxes of the gases and their timing to the dissociator radio frequency pulses. The cooled beams are focused by an optimized set of permanent hexapoles into the charge exchange region. By cooling down the supersonic atomic beam, the acceptance of the hexapole system and the dwell time in the charge exchange region are increased in proportion to the decrease of the beam velocity. Gas scattering in the vicinity of the nozzle partly reduces these beneficial effects. A peak intensity of \SI{7.5e16}{atoms\per \second} has been measured within a diameter of \SI{10}{mm} at the exit of the hexapole chamber.

The highly nuclear polarized atomic $\overrightarrow{D}^0$ beam meets inside the charge exchange region, the fast neutral $(\text{Cs}^{0}$ beam and charges are swapped according to the reaction $\overrightarrow{D}^{0} + \text{Cs}^{0} \rightarrow \overrightarrow{D}^- + \text{Cs}^+$. The negatively charged \({\overrightarrow{D}}^{-}\) ions are extracted from the charge exchange region by electric fields and are deflected magnetically by \SI{90}{\degree} into the beamline to the injector. The ions are transferred to the cyclotron, passing a Wien Filter to provide the proper spin alignment for injection into the cyclotron.

The fast neutral Cs\textsuperscript{0} beam for the charge exchange reaction is produced in a two-step process. Cesium vapor is thermally ionized on a hot porous tungsten surface at a beam potential of around \SI{45}{kV}. The beam is focused by a quadrupole triplet to the charge exchange region. Space charge compensation of the intense beam is improved by feeding \SI{1e-3}{mbar.l \per s} Argon to the beam tube following the extraction system. The long-term operation revealed that several components limited the effectiveness for routine experiments. To provide reliable operation for experiments at COSY, prototype parts,
mainly of the cesium beam section \cite{LEMAITRE1998345} of the source, were replaced through an improved design. Cesium sputtering and contamination are generally impeding long-term reliability. Therefore, pulsed operation of the cesium ionizer has been included in the source \cite{EGGERT2000514}. The cesium pulses reached peak intensities of over \SI{10}{mA} with reduced width around \SI{10}{ms}. For routine operation, cesium pulses with \SI{5}{mA} flat shape of \SI{20}{ms} width and a repetition rate of \SI{0.5}{Hz} are used \cite{Felden:2001vc}.

The neutralizer, a chamber filled with cesium vapor, is placed between the quadrupoles and the Cs deflector. The neutralizer comprises a cesium oven, a cell filled with cesium vapor and a magnetically driven flapper valve between the oven and the cell. The remaining Cs\textsuperscript{+} the beam is deflected in front of the solenoid to the Cs cup. Routinely, a neutralizer efficiency of over 90\% was measured.

The highly selective charge exchange ionization produces only little unpolarized background that would reduce the nuclear beam polarization. In the charge exchange solenoid, various beam properties can be adjusted. The transversal emittance can be traded for polarization by varying the solenoid's magnetic field. The magnitude of the electrical drift field inside the solenoid can be tuned to optimize the energy spread of the beam. A monotonous gradient in combination with a double buncher system in the injection beam line to the cyclotron led to an improved bunching factor.

The colliding beams type negative ion source can provide negative polarized hydrogen and deuteron beams without modification in comparable intensities. To prepare polarized deuterons with the desired combinations of vector and tensor polarization, the atomic beam part of the source needs to be equipped with new high-frequency transitions. These transition units are operated at the magnetic fields and radio frequencies to allow an exchange of occupation of the different hyperfine states in deuterium. A set of three installed devices, RFT1 to RFT3, allows a large number of combinations to be delivered to experiments. The unpolarized mode is used for normalization. Table.\,\ref{tab:polarization-modes} lists eight different combinations of vector and tensor-polarized deuteron beams that the source can generate\,\cite{PhysRevSTAB.9.050101}.

\begin{table}[htb]
\centering
\caption{The table lists eight configurations of  the polarized deuteron ion source, showing the ideal values of the vector and tensor polarizations and the relative beam intensities obtained by operating the three radio-frequency transitions(RFTs). Also shown are the measured vector and tensor polarizations of the deuteron beam with statistical errors. The measurements of $P^\text{LEP}_z$ were carried out at a momentum of \SI{539}{MeV/c},  using the Low Energy Polarimeter (LEP) during the ANKE run at COSY in November 2003\,\cite{PhysRevSTAB.9.050101}.}
\label{tab:polarization-modes}
\renewcommand{\arraystretch}{1.1}
\vspace{0.2cm}
\begin{tabular}{c|r|r|c|c|c|r}
\textbf{Mode} & \textbf{$P_z$} & \textbf{$P_{zz}$} & \textbf{RFT1} & \textbf{RFT2} & \textbf{RFT3} & \textbf{ $P^\text{LEP}_z$}\\
\hline
0 & 0      & 0      & Off & Off & Off & $0.000 \pm 0.010$ \\
1 & $-2/3$ & 0      & Off & Off & On  & $-0.572 \pm 0.011$ \\
2 & $+1/3$ & $+1$   & Off & On  & Off & $0.285 \pm 0.011$ \\
3 & $-1/3$ & $-1$   & Off & On  & On  & $-0.302 \pm 0.011$ \\
4 & $+1/2$ & $-1/2$ & On  & Off & Off & $0.395 \pm 0.014$ \\
5 & $-1$   & $+1$   & On  & Off & On  & $-0.758 \pm 0.015$ \\
6 & $+1$   & $+1$   & On  & On  & Off & $0.731 \pm 0.014$ \\
7 & $-1/2$ & $-1/2$ & On  & On  & On  & $-0.417 \pm 0.015$ \\
\hline
\end{tabular}
\end{table}

\subsection{Polarized He-3 Beam}
\label{sec:3He-beam}

A source of polarized neutrons will be central to the physics program at the Electron Ion Collider. The small magnetic moment of Deuterium poses challenges for spin manipulation in the RHIC and the planned EIC, but polarized He-3 is an attractive alternative. He-3 is predominantly found in the spatially symmetric S-wave state, with proton spins anti-aligned and the neutron carrying the nuclear spin, making it an effective spin-polarized neutron source\cite{PhysRevC.29.538}. The He-3 magnetic moment is larger than that of the deuteron and has been determined \cite{osti_1047667} to be compatible with the existing RHIC spin manipulation. The development of a polarized He-3 ion source at Brookhaven National Lab (BNL) was proposed in 2003~\cite{ICFA}, and has been pursued by a collaboration between MIT and BNL since 2013~\cite{Epstein:2013}. The concept is based around the existing Electron Beam Ion Source (EBIS) and it is scheduled to be installed during the maintenance period at the end of RHIC operations. 

\begin{figure*}[htb]
    \centering
    \includegraphics[width=0.7\linewidth]{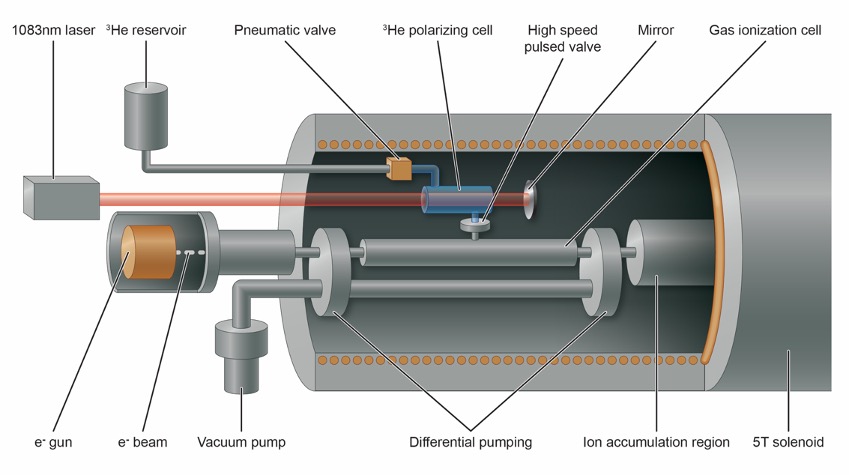}
    \caption{A Schematic diagram of the He-3 ion source gas purification and polarization scheme inside of the bore of EBIS. The glass cell containing the polarized He-3 is connected directly to the EBIS ionization region by a special, high-field pulsed valve.}
    \label{fig:schematic}
\end{figure*}

The Polarized He-3 source design \cite{ICFA,Epstein:2013} at BNL utilizes the Metastability Exchange Optical Pumping (MEOP) technique \cite{PhysRev.132.2561} to polarize He-3 gas inside of the 5T superconducting solenoid of the existing EBIS. A gas purification system and compact optical scheme has been developed to polarize and perform polarimetry on a glass cell of He-3 mounted directly to the EBIS drift tube, shown schematically in Fig.\ref{fig:schematic}. After polarization, He-3 is injected into the drift chamber for ionization by a high-field pulsed valve. The $^3 $He$^{++}$ ions are extracted from EBIS around 6 MeV and can be directed to either an absolute nuclear polarimeter or the Booster. The design goal for the source is to provide a minimum of 70\% polarization and a peak current near 4 mA. 

Previous studies at BNL have demonstrated the viability of the polarization scheme, with polarizations just under 90\% for a sealed sample of He-3\cite{MAXWELL2020161892} and up to 80\% with an ``open" cell connected to a gas purification system by a bellows valve \cite{ZELENSKI2023168494}. A new, minimal optical layout has been developed to fit the spatial constraints of the EBIS bore and polarizations around 60\% have been achieved\cite{Wuerfel24}. 

Currently, the new optical layout is being tested in a one-to-one copy of the EBIS solenoid while a final design for the gas system is being developed. The gas purification system and polarization and polarimetry optics will be installed during the EBIS shutdown at the end of 2025. There are plans~\cite{Epstein:2023} to make measurements of the extracted ion polarization in the absolute polarimeter during 2026. Additionally, the ions will be sent to the Booster and Alternate Gradient Synchrotron (AGS) during spring or summer 2026. An estimate of the polarization inside of the AGS will be made using the existing carbon Coulomb Nuclear Interference polarimeter \cite{Huang:2008zzf}, which has been used for many years to measure proton polarizations at RHIC. 

\subsection{Polarized Li-6 and Li-7 Beams}
\label{sec:pol-Li6-7}

The Medium Energy Physics (MEP) group at Argonne National Laboratory (ANL) in collaboration with the University of Kentucky has been developing highly polarized and high-current sources of lithium-6 and lithium-7 in support of the spin physics program. A primary scientific objective of nuclear physics research is to understand the spin structure of the nucleon in terms of fundamental quarks and gluons. Accordingly, highly polarized electron and ion beams are essential, as they allow probing dynamic spin-dependent phenomena that are inaccessible with unpolarized particles. Polarized light ions, specifically hydrogen (spin-1/2), deuterium (spin-1), and helium-3 (spin-1/2), have long been integral to the design of the U.S.-based Electron-Ion Collider (EIC). These polarized nuclei have also been extensively employed in fixed-target scattering experiments at various facilities. We propose expanding this list by including polarized lithium-6 (spin-1) and lithium-7 (spin-3/2). These isotopes offer unique spin structures and the potential to enhance the spin physics program significantly.

\begin{figure*}[htb]
\centering
\includegraphics[width=0.8\textwidth]{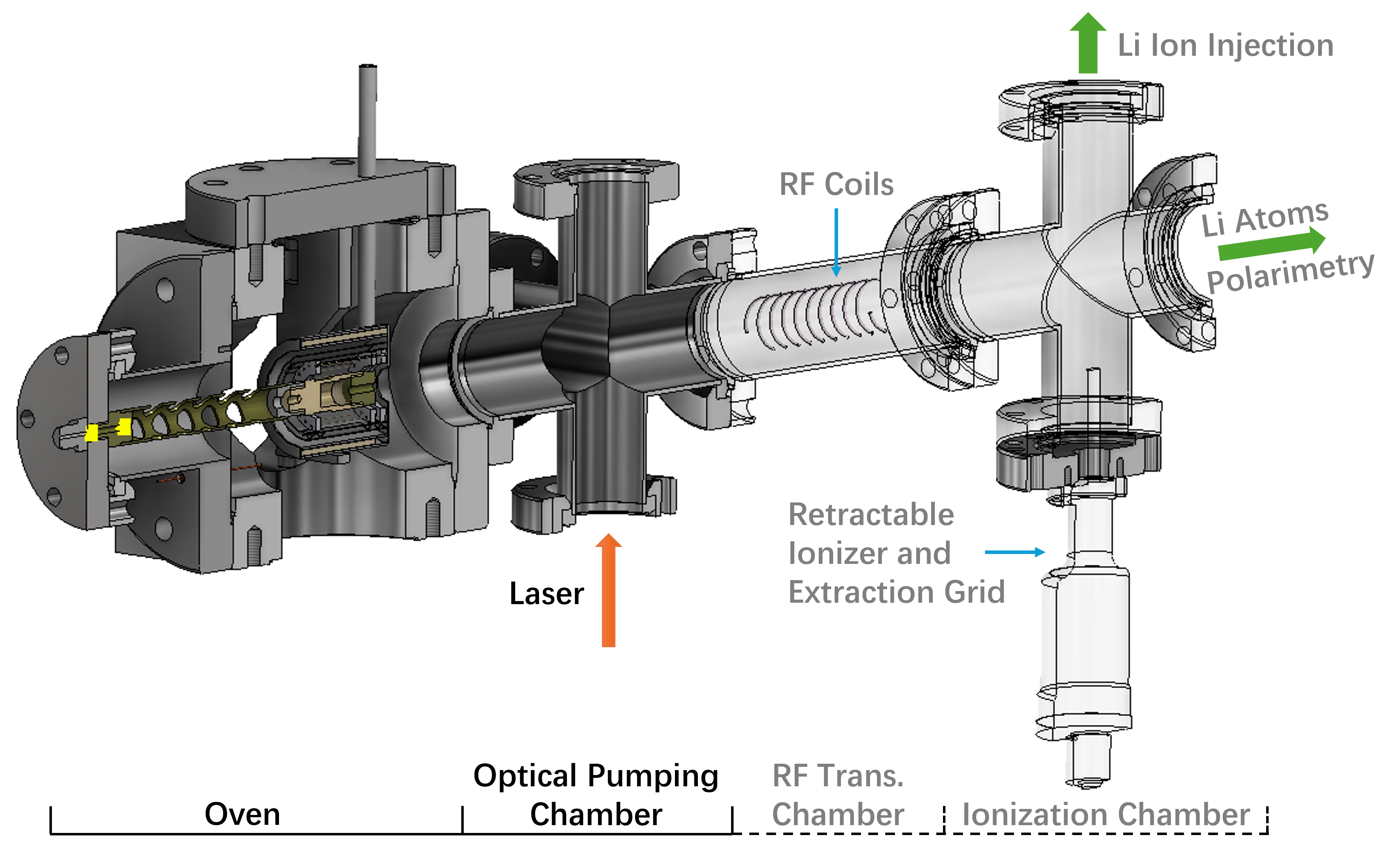}
 \caption{Schematic of the current polarization system for $^{6,7}$Li.}
\label{fig:pol_li_source}  
\end{figure*}

Figure \ref{fig:pol_li_source} shows the polarized source system at Argonne. It is still in the early stage of development, consisting of two major components: a vaporizing oven coupled with a de Laval nozzle, designed for generating a lithium atomic beam; a circularly polarized laser at around 671 nm, alongside a vacuum chamber with laser feedthrough, facilitates the optical pumping of lithium atoms. In addition, two other components are planned to be built: an RF transition chamber dedicated to the selection of specific nuclear magnetic sub-states for the polarized lithium atoms; an ionization chamber featuring a retractable ionizer and an extraction grid, designed to ionize and extract polarized lithium ions from the beam.

The polarized lithium sources will offer an essential tool for studying the spin properties of nucleons and light nuclei, advancing our understanding of their internal structure. By expanding the experimental capabilities to include polarized lithium ion beams, we expect to gain valuable new insights into quark-gluon dynamics within the nuclear medium and beyond.

\subsection{Other Polarized Beams}
\subsubsection{Polarized B-10 and B-11 Beams}

The large nuclear spin of $I = 3$ for Boron-10 makes it an attractive candidate for exotic gluon studies. However, the relatively complicated electronic structure of boron presents a challenge for optical pumping. We have identified a possible polarization scheme which has been presented as a poster at this meeting and which is being submitted for publication~\cite{Mil2025}.

Although beams of He-3 and alkali atoms have been optically 
pumped~\cite{ander1979prop,dreves1983production}, they have a comparatively simple electronic structure. Alkali atoms are hydrogen-like with a single valence electron and laser cooling of these species has been extensively studied.  Helium-3 possesses a metastable electronic state and established techniques for spin polarization have been developed using spin-exchange. 
Boron, on the other hand, is a group III element, possessing a $p$-orbital electronic ground state. Explorations into laser cooling and optical pumping of group-III elements have just begun in the past few years, with a beam of indium being successfully spin-polarized and laser-cooled~\cite{nicholson}.  
Given that boron has a similar electronic level structure, the techniques used for indium can directly translate to boron.

We note that both Boron-10 and Boron-11 are available commercially in ultra-high purity as crystalline solids. For thin film applications, they are available as rod, pellets, pieces, granules and sputtering targets and as either an ingot or powder.

\begin{figure}[htb]
\centering
\includegraphics[width=0.9\columnwidth]{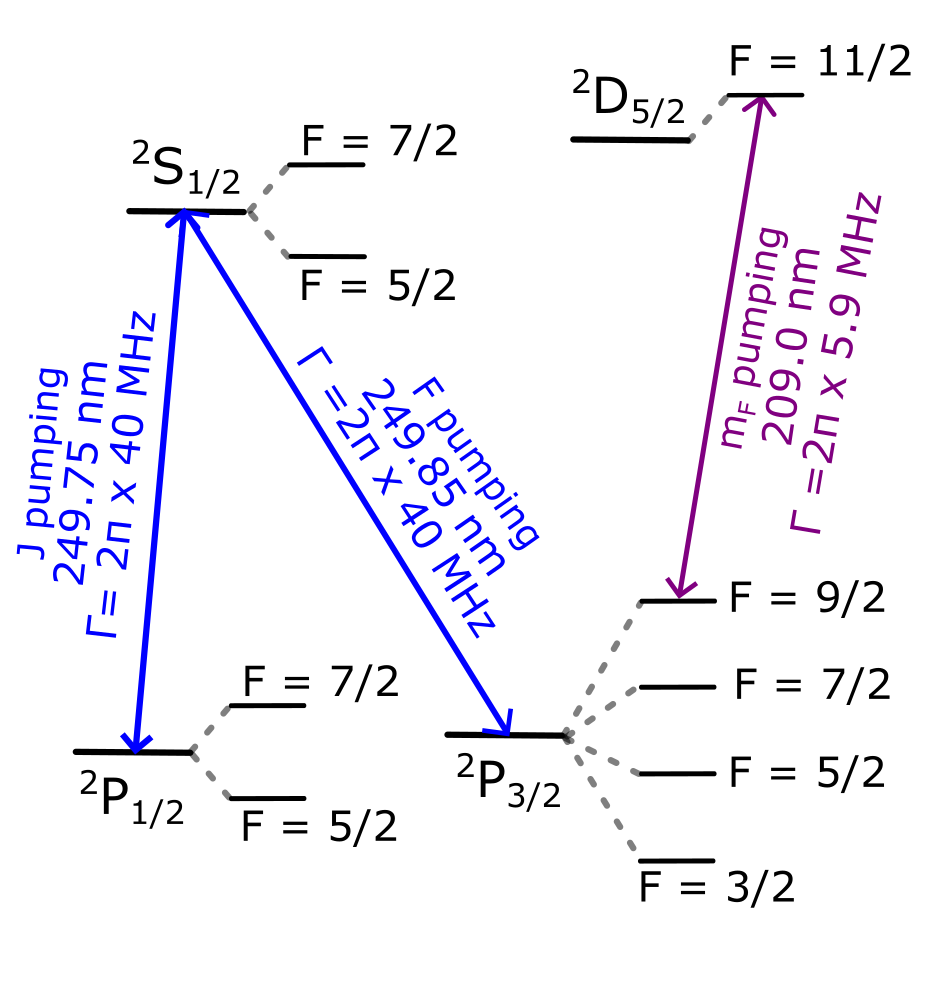}
\caption{$^{10}\text{B}$ energy levels. Transition rates $\Gamma$ are reported in Ref.~\protect\cite{carlsson1994lifetimes}. Transition wavelengths are determined from Ref.~\protect\cite{NIST_ASD}. Transitions are characterized in terms of $J$, $F$, and $m_{F}$ pumping steps. The $\ket{g}$ and $\ket{e}$ states are addressed by the 209\,nm transition.} 
\label{fig_1}
\end{figure}

\sloppy
All optical pumping schemes rely on the hyperfine interaction $H = A_{HF} I \cdot J$, which couples the electronic spin $J$ to the nuclear spin $I$. Before pumping, the atoms generally occupy a thermal distribution over all hyperfine sublevels $m_{F}$ $\{ -m_{F}, -m_{F} + 1, \dots, m_{F} \}$.  Using circular polarized light with angular momentum to drive electronic transitions, the atoms are pumped to a stretch state, $\ket{F, m_{F} = F}$ where the nucleus is fully polarized.  To effectively pump atoms, one wants to drive atoms to a closed transition where the excited state can only decay to a single, polarized ground state.  In boron, this condition can be achieved with the  $\ket{2s^{2}\,2p\,{}^{2}P_{3/2},F=9/2,m_{F}=9/2} \equiv \ket{g}$ to $\ket{2s\,2p^{2}\,{}^{2}D_{5/2},F'=11/2,m_{F'}=11/2} \equiv \ket{e}$ transition. Due to selection rules, atoms in the $\ket{e}$ state can only decay to $\ket{g}$ and both states are fully nuclear polarized. We note that the state $2s\,2p^{2}\,{}^{2}D_{J}$ lies energetically below $2s^{2}\,3d\,{}^{2}D_{J}$ for boron~\cite{NIST_ASD}, in contrast to In, Al, and Ga, and is the excited state used in  this scheme.

The energy levels of boron-10 are depicted in Fig.~\ref{fig_1}. Atoms start in an incoherent, thermal mixture in the $\ket{^{2}P_{1/2}}$ and $\ket{^{2}P_{3/2}}$ states. To transfer atoms to the target $\ket{g}$ state, the $J$, $F$, and $m_{F}$ quantum numbers must all be modified. First, atoms need to be transferred to the metastable $\ket{^{2}P_{3/2} }$ state. This is achieved using two `J pumping' lasers to drive both $\ket{^{2}P_{1/2},F= 5/2, 7/2}$ states to the $\ket{2s^2 3s\;^{2}S_{1/2},F'=7/2}$ state with a lifetime of $\tau \approx 4$ ns. To address all $m_{F}$ states, the lasers can be frequency modulated. Atoms rapidly decay and are distributed in the $F = 5/2, 7/2,$ and $9/2$ states in  the $J = 3/2$ manifold. Three additional lasers drive the `F pumping' transition $\ket{^{2}P_{3/2},F= 3/2, 5/2, 7/2}$ to $\ket{  ^{2}S_{1/2},F' =7/2}$  so that atoms accumulate in the $\ket{^{2} P_{3/2},F=9/2}$ manifold. All lasers should be turned on simultaneously to minimize de-pumping back to the $\ket{^{2}P_{1/2} }$ state. For the $J$ and $F$ pumping steps the polarization of the light is not critical, as the atoms are distributed among many $m_F$ states in $\ket{^{2}P_{3/2},F=9/2}$ following pumping. 

With atoms in the target $\ket{^{2}P_{3/2},F=9/2}$ manifold, we now perform optical pumping (`$m_{F}$ pumping') to transfer atoms to a single, polarized $m_{F}$ state. We use circularly polarized light with handedness $\sigma^{\pm}$ to transfer angular momentum so each absorbed photon drives a $\ket{m_{F}} \rightarrow \ket{m_{F'} = m_{F} \pm 1}$ transition. After absorbing many photons,  atoms are pumped to the $\ket{g}  \rightarrow \ket{e}$ cycling transition. Upon turning off the $m_{F}$ pumping light, any atoms in the $\ket{e}$ state quickly decay in $\tau =  27$ ns to the $\ket{g}$ state. 

In summary, optical pumping of boron-10 appears feasible, although there are technical challenges to achieve high nuclear spin polarization that need to be experimentally tested. High oven temperature is required for atomic beam operation and the hyperfine coupling for the $\ket{^{2}D_{5/2}}$ states will likely dictate the atomic beam divergence angle required for detection. Finally, achieving a large beam flux ideally approaching $10^{15}\text{cm}^{-2} \text{s}^{-1}$ will be of importance to successfully use this atomic beam in the EIC. Despite these technical uncertainties, experimentally testing this scheme appears absolutely worthwhile to open the door to potentially studying nuclear spin in an entirely new regime. The proposed scheme can also be used to produce polarized boron-11 nuclei, which can be used to enhance the pB fusion cross section.

\subsubsection{Polarized Na-23 Beam}

In the late 1970s a polarized $^{23}$Na (I=3/2) beam was developed~\cite{Na23a,Na23b} at the Max-Planck-Institut f\" ur Kernphysik at Heidelberg, Germany.  A polarized atomic beam was produced by an oven, Stern-Gerlach magnet and a system of hf transitions.  This beam was ionized to positive ions on a hot tungsten surface.  Ion currents up to 100 $\mu$A were extracted from the source and polarizations of order 50\% were measured.

\subsection{Development of New High-Flux Polarized Hydrogen (H, D, T) sources for Polarized Fusion}

Spin-polarized fusion (SPF) fuels can provide a significant boost towards the ignition of a burning plasma in a fusion reactor, increasing the $D + T \rightarrow  \alpha + n$ cross section by 1.5 and the Q (fusion-power/power-in) by 1.8 to 1.9 when the initial spins are parallel to the local field \cite{Baylor_2023,10.3389/fphy.2024.1355212}. 






The potential of SPF rests on two essential prerequisites: (a) the fuel-spin alignment, created at relatively low temperatures, must survive in a \SI{e8}{\kelvin} plasma for periods comparable to the $D$ and $T$ confinement times in a plasma ($\approx$ few s); (b) sources of fully polarized $D$ and $T$ must be able to keep up with the fueling demands of a power reactor ($\approx \num{e21}$ to $\num{e22}$ s$^{-1}$).

A SPF lifetime demonstration experiment has recently been funded by the US Department of Energy (DOE) Office of Fusion-Energy-Sciences (FES) \cite{Baylor_2023,10.3389/fphy.2024.1355212}. Since $D + T \rightarrow  \alpha + n$ and $D + ^3\text{He} \rightarrow \alpha + p$ are mirror reactions, polarized $^3\text{He}$ will be used as a surrogate for $T$, which avoids the complications of using tritium in a research tokamak. The experiment will utilize materials developed for Nuclear and Particle Physics (NP) and Medical Imaging. Pellets of polarized LiD and shells containing polarized $^3$He as pressurized gas will be injected into the DIII-D tokamak at the National Fusion Facility in San Diego. The resulting alpha and proton angular distributions and yields at the outer tokamak wall will be used to deduce {\it in-situ} polarization lifetimes \cite{Baylor_2023,Garcia_2023} The necessary equipment is under construction, and first measurements are expected by 2027 – 2028.  

We turn now to the remaining critical component -- a new method of producing intense polarized sources of hydrogen isotopes that can meet fueling demands. It is here that the interests of SPF and the EIC intersect. Many solid materials have been developed for NP experiments, such as Ammonia and Butanol, but the high electron-ionization energies in the heavier elements of these molecules (C, N, O) would be power sinks that could easily quench a plasma. Even the Lithium in the LiD pellets of the DIII-D demonstration experiment would become a problem if used on a massive scale. Polarized atomic (or diatomic) beam sources can select magnetic substates of interest (a Stern-Gerlach separation), but are limited in flux to a few times \SI{e17}{\per\second} by multiple scattering. 

In a new approach \cite{PhysRevLett.94.083005,Kannis:2020bcx}, molecules containing the hydrogen isotope(s) of interest are laser excited into a ro-vibrational state. The hyperfine interaction (HFI) with the molecular electrons splits the magnetic substates, and a natural (and calculable) beating with the rotational angular momentum moves spin to the nuclei. If the beating is stopped by a small magnetic field just when all angular momentum is carried by the nuclei, the molecules can be isolated and dissociated to yield the fully polarized nuclei of interest. Since the polarization process happens in each molecule, previous limitations are avoided and fluxes of $\SI{e22}{\per \second}$ should be possible. 

\begin{figure}[htb]
    \centering
    \includegraphics[width=0.85\columnwidth]{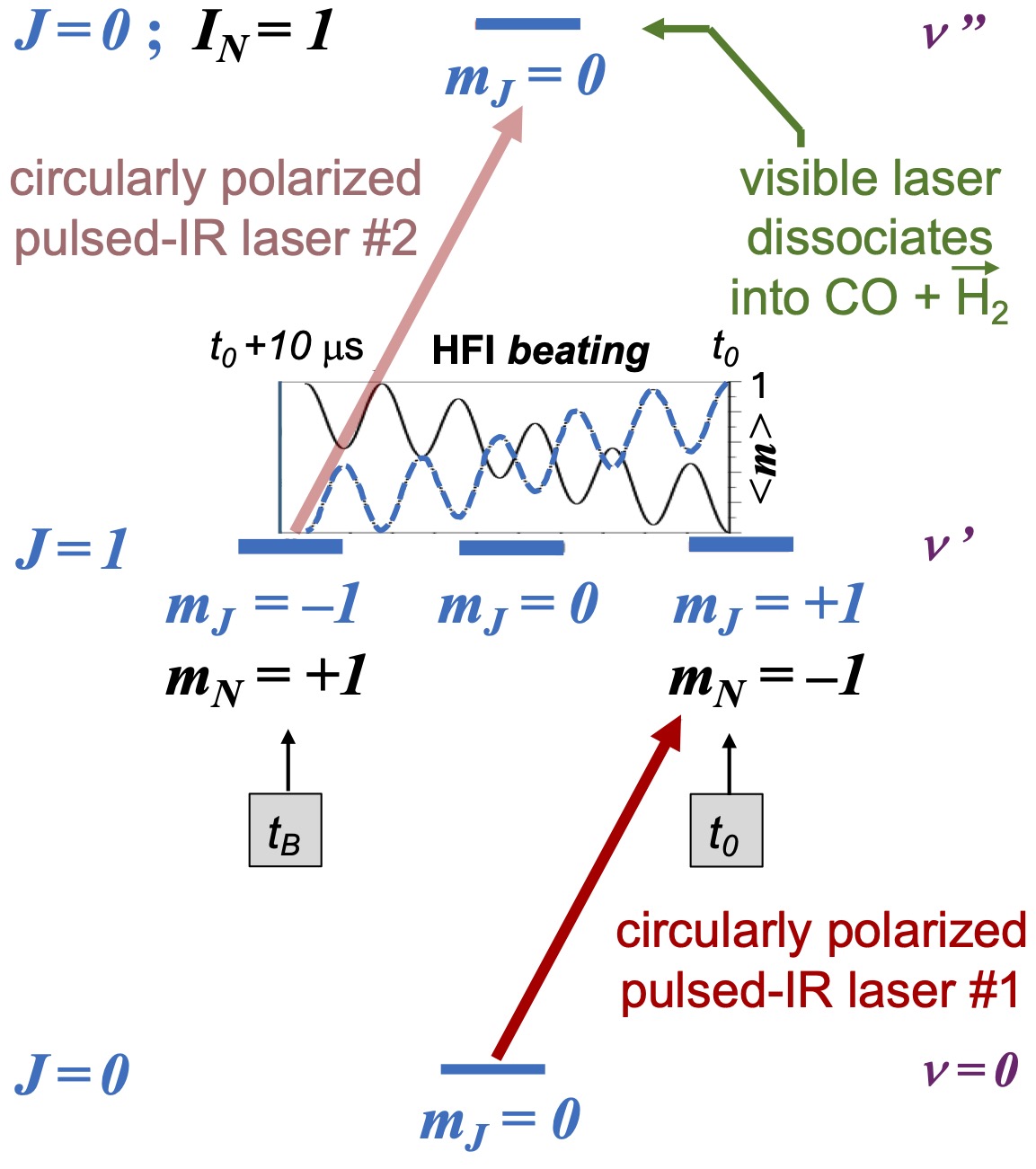}
    \caption{Schematic of a process using the successive laser excitation of molecules, with HFI beating between molecular rotation and nuclear spin, to produce high fluxes of fully polarized hydrogen.}    \label{fig:sandorfi}
\end{figure}

There are several versions of this general scheme, utilizing a variety of molecules. Following ref. \cite{Kannis:2020bcx}, we sketch a production of polarized H$_2$ that begins with formaldehyde, CH$_2$O. The process is shown schematically in Fig.\,\ref{fig:sandorfi}. 
Gaseous CH$_2$O is cooled to its ground state with $J \, (\text{rotational}) = 0$, $m_J =0$, $\nu \, (\text{vibrational}) = 0$ by expansion through a nozzle. The molecules exit the nozzle with a rather well-defined velocity of about $\SI{2e3}{\meter \per \second}$. The molecular beam passes through circularly polarized Infra-Red (IR) laser light, shining at right angles to the beam, which preserves the angular momentum projection along the beam direction. Through Raman scattering (Stimulated Raman Adiabatic Passage, or STIRAP) the molecules are pumped into the first HFI-resolved ro-vibrational state with $J =1$ and $m_J =+1$. (The vibrational energy is generally much larger than that of rotation. A rotational band built on a vibrational state is used to bring the transition into the frequencies of available lasers.) As the system evolves, the total angular momentum beats between rotation (blue dashed curve in the inset of Fig.\,\ref{fig:sandorfi}) and nuclear spins (solid black curve) and the population of the state with full nuclear polarization increases. (The beating curves always sum to unity.) After about \SI{10}{\micro \second}, the proton spins in the H$_2$ of the molecule are aligned and carry all the angular momentum. A small field ($\approx \si{\milli \tesla}$) applied at this point ($t_B$ in the figure), either by a fast current-switched coil or by a fixed magnet positioned at the appropriate distance from the expansion nozzle, stops the HFI beating. Within this small field, a second circularly polarized IR laser pumps these molecules up to the $J =0$, $\nu'' \,\text{2nd vibrational state}$. Only molecules with 100\% nuclear polarization reach this point. Molecules in this $\nu''$ state are then exclusively photo-dissociated with a high-resolution visible laser, tuned to break the CO-H$_2$ bond. Since the H$_2$ is the lightest fragment, it carries almost all the kinetic energy and is naturally separated. The same process can be applied to fully tritiated formaldehyde, CT$_2$O, to produce polarized $T_2$. For SPF, polarized beams would be captured on a cold surface and transferred to solid pellets for injection.

Each of these steps have been demonstrated with low power high-resolution lasers. In the last five years, a revolution in laser technology has yielded high-power tunable lasers of sufficiently narrow bandwidth that, if used for this process, could yield fluxes of $\approx \SI{e22}{\per \second}$. Similar methods can be used to produce beams of polarized $DT$ for SPF, or polarized $D_2$ for either SPF or for the EIC, although the steps are somewhat complicated by the higher spins involved \cite{Kannis:2020bcx}. (Alternatively, a potentially simpler method of directly pumping $DT$, with subsequent repumping to enhance the polarization, has also been suggested recently \cite{kannis2025productionspinpolarizedmolecularbeams}.) Such intense beams could fuel a fusion power reactor and open new windows at the EIC. Nonetheless, as with any new technique that jumps previous boundaries by many orders of magnitude, focused research and development will be necessary to demonstrate the predicted flux scaling. 


\section{Hadron Polarimetry}



\subsection{Absolute hadron beam polarimetry at EIC}
\label{sec:HJET}

Various methods can be used to measure the polarization of a stored beam. For protons scattered from an unpolarized carbon nucleus, the degree of polarization in a storage ring can be deduced from the left-right count rate asymmetry,
\begin{equation}
    \varepsilon_\text{beam} = \frac{\text{L} - \text{R}}  {\text{L} + \text{R}} = A_y \cdot P
    \label{eq:abs-had-pol-1}
\end{equation}
where L and R are the counts recorded in the left and right detectors, $A_y$ is the analyzing power, which is a measure of the polarization sensitivity of the scattering process, and $P$ the absolute value of the beam polarization. The problem, however, is that at the high energies at the AGS, RHIC and EIC, there are no scattering processes available for which the analyzing power $A_y$ is known with sufficient accuracy to perform an absolute and highly accurate measurement with an uncertainty of a few percent\,\cite{Haeberli:2005tj}. 

The interference of electromagnetic and strong interaction at small scattering angles provides sizable analyzing power for elastic proton-proton (and proton-nucleus) scattering (see, e.g., Fig.\,6 in\,\cite{Poblaguev2020-pm} and Sec.\,\ref{sec:theo-background-hadron-pol}). This analyzing power, which constitutes the basis of the RHIC high-energy (absolute) polarimeters, is derived from the same electromagnetic amplitude that generates the anomalous magnetic moment of the proton. Experiment E704 at Fermilab used \SI{200}{GeV/c} polarized protons from hyperon decay to detect the asymmetry in the scattering from a hydrogen target\,\cite{PhysRevD.48.3026}. The largest analyzing power $A_y$ was about 0.04 with large statistical errors. A calculation of the analyzing power agreed with these measurements, but the calculations are subject to uncertainties in the amplitudes of the strong interaction. Therefore, an accurate calibration of the reaction is required.

\begin{figure}[htb]
    \centering
    \includegraphics[width=0.85\linewidth]{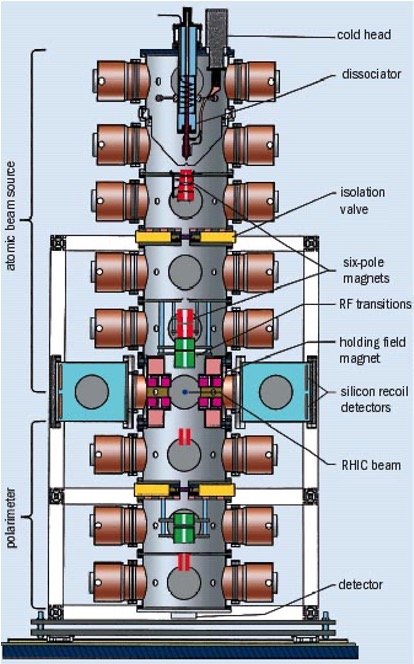}
    \caption{Schematic layout of the HJET polarimeter, taken from Ref.\,\cite{Haeberli:2005tj}, showing the atomic beam source, the scattering chamber, and the Breit–Rabi polarimeter at RHIC.}
    \label{fig:HJET}
\end{figure}

The idea for high-precision beam polarization calibration for the hadron beams in the EIC follows the concept developed for RHIC\,\cite{Haeberli:2005tj}. The stored beam passes through a beam of polarized hydrogen atoms (see Fig.\,\ref{fig:HJET}) with known nuclear polarization, and one measures the left-right ratio in the number of scattered particles [see Eq.\,(\ref{eq:abs-had-pol-1})]. The sign of the target polarization is periodically reversed to compensate for asymmetries caused by differences in detector geometry or efficiency in the left and right directions. This yields the target asymmetry $\varepsilon_\text{target} = A_y \cdot Q$, where $Q$ denotes the target polarization. A measurement of the corresponding asymmetry with beam particles determines $\varepsilon_\text{beam}$ [see Eq.\,\eqref{eq:abs-had-pol-1}], and since in elastic proton-proton scattering the analyzing power $A_y$ is the same regardless of which proton is polarized, the absolute beam polarization is given by
\begin{equation}
    P = \frac{\varepsilon_\text{beam}}{\varepsilon_\text{target}} \cdot Q \,.
\end{equation}

A similar approach to absolute hadron polarimetry as applied in $\vec p \vec p$ elastic scattering using a polarized hydrogen jet target, also making use of the elastic scattering of identical particles, shall be applied for the other polarized ion species anticipated for use at the EIC (see Chap.\,\ref{chap:polionsources}). It is important to note in this context that absolute beam polarization determination requires targets that maintain their position relative to the beam, thus are stable in time and position, to reliably determine the related beam spin asymmetries. Thus while pC scattering, as discussed in Sec.\,\ref{sec:pC-polarimetry}, provides the statistical precision for a fast determination of the relative polarization and of the polarization profile across the beam, due to the electromagnetic interaction with the bunch charges of the RHIC beam, the C targets presently in use lack the position stability to provide the required precision for an absolute polarization determination.

One of the science pillars that the EIC is expected to deliver, in addition to luminosity and $ep$ collision energy, is proton beam polarizations greater than 70\% with a relative uncertainty of 1\% or less. Achieving this goal will require modifications of the present atomic beam source and Breit-Rabi polarimeter (see Ref.\,\cite{rathmann2025eliminatingbeaminduceddepolarizingeffects} for further details. Currently, the accelerator chain leading to the EIC does not meet these EIC goals, with the AGS responsible for most of the polarization loss from source to maximum energy in the RHIC/EIC, as discussed in section \,\ref{sec:pol-intensity-optimization}. Thus, there is a need to further reduce the polarization losses in the AGS. One way of achieving this would be by enabling $\vec p \vec p$ and other identical particle scattering reactions in the AGS for the purpose of absolute polarimetry, successfully applied for $\vec p \vec p$ in RHIC. However, this is a difficult task due to the low areal density of polarized gas targets and the small number of bunches currently used in AGS to fill RHIC/EIC, thus the use of storage cells seems to be indicated to enhance the target thickness\,\cite{PGT_Review_Steffens_Haeberli_2003}.

In the  subsequent sections, pC polarimetry at EIC (Sec.\,\ref{sec:pC-polarimetry}), an absolute beam polarimeter for $^3$He beams (Sec.\,\ref{sec:cryogenic-3He-ABS}), the theoretical background of hadron polarimetry (Sec.\,\ref{sec:theo-background-hadron-pol}), the use of Siberian snakes in the EIC (Sec.\,\ref{sec:snakes}), and  lessons learned for the EIC from accelerating polarized
protons in the AGS (Sec.\,\ref{sec:learned-from-AGS-for-EIC}) will be discussed.

\subsection{\lowercase{p}C polarimetry at EIC}
\label{sec:pC-polarimetry}

The fast polarimeter at RHIC is based on the Coulomb--Nuclear Interference (CNI) process~\cite{PhysRevD.59.114010, PhysRevLett.89.052302} using a carbon target. Each measurement requires about 30~s to achieve a 2\% statistical error with a fully loaded beam of 110 bunches, each containing $2\times 10^{11}$ protons. The polarimeter uses a thin carbon ribbon target (\SI{10}{\micro\meter} wide, with an areal density of \SI{4}{\micro\gram\per\square\centi\meter}) together with silicon strip detectors. The large surface area of the ultrathin ribbon enables sufficient heat dissipation, keeping the target temperature below the sublimation temperature ($\sim\SI{2000}{\kelvin}$). The relatively narrow targets also allow the measurement of polarization profiles, which can arise when particles with larger betatron oscillation amplitudes experience greater polarization losses. Since each polarimeter target has a limited beam lifetime, multiple ribbons (six per holder) are installed to allow sequential use~\cite{Huang:2006cs}. Carbon events are selected within a time-energy window, with kinetic energy range $400 < T < \SI{900}{\kilo\electronvolt}$, optimized to suppress background.

Due to target orientation and thickness variations, the calibration of the $p$C polarimeter with the polarized hydrogen jet~\cite{OKADA2006450, PhysRevD.79.094014} cannot be performed as a one-time procedure. The polarized hydrogen jet must run in parallel, providing a polarization accuracy of $\pm 3\%$ for an 8~h store. Typically, four polarization measurements are taken during a store, at 0, 3, 6, and 8~h. Each set consists of two measurements with horizontal and vertical targets, respectively. In addition to absolute polarization, these data yield polarization profile information. Repeated measurements within a store also allow the observation of possible polarization decay with time.

In the EIC operation scenario, the total beam intensity will increase by a factor of three and the beam energy by about 10\%. In addition, the vertical beam size will be reduced through pre-cooling at injection. The bunch spacing will shrink from \SI{108}{ns} to about \SI{10}{ns}. These changes generate three potential problems.

The first issue is target heating. The increased intensity and energy will raise the target temperature if other beam parameters remain unchanged. To mitigate this, the vertical beta function will be enlarged to increase the beam size at the polarimeter, thereby reducing the heating effect. Preliminary simulations indicate that the carbon target equilibrium temperature will rise but remain below the sublimation threshold. By increasing the beta function at the polarimeter location (\SI{30}{m} $\rightarrow$ \SI{240}{m}), the beam size can be significantly enlarged. Target heating can thus be kept at a level similar to RHIC operation, even with higher intensity and smaller emittance. This mitigation, however, requires a longer target and a modified target chamber design.

The second issue concerns event selection. With the bunch spacing reduced to \SI{10}{ns}, carbon events from different bunches can no longer be separated. Since adjacent bunches may have opposite spin orientations (up or down), the measured asymmetry becomes diluted. For the polarimeter to function reliably under these conditions, a new detection technique will be required.

The third issue concerns beam stability. With short bunch spacing in the HSR, higher-order modes from pumping ports or other openings must be carefully evaluated. RF-induced heating is expected to be more severe in the HSR than in RHIC, making RF shielding of these ports essential. CST simulations of wakefields, impedances, and beam-induced resistive-wall (RW) losses for the HSR polarimeter have been performed using both RHIC and EIC beam parameters. To simplify the setup, only one target holder was included in the simulations. Two materials for the target holder were studied: aluminum and alumina (a dielectric). For RHIC beam parameters, the beam-induced losses are comparable for both materials. However, for the EIC proton beam, the aluminum target holder produces strong wakefield oscillations and resonances in the impedance analysis, whereas the use of alumina significantly reduces both the wakefield amplitude and the impedances.

Up to now, the temperatures of the polarimeter carbon targets during interaction with the proton beam have not been directly measured. Previous measurement efforts were inconclusive~\cite{huang:ibic14-mopd01}. Although simulations indicate that the equilibrium temperature remains below the sublimation point for both RHIC and EIC conditions, it is essential to benchmark these results with experimental data at RHIC. To this end, a dedicated optical light collection system was implemented at IP12 to capture and analyze the emitted light across the visible and near-infrared spectrum. This study explores the feasibility of using light emission as a diagnostic tool for determining target temperatures. A dedicated experiment with the proton beam is planned for the final RHIC operation in 2025. This work is critical for evaluating the viability of carbon fiber targets in the EIC under increased beam intensity.

At the EIC, the hadron polarimetry system will be located at IP4 and will include the pC polarimeter, the polarized hydrogen jet, and potentially deuteron and $^3$He jets. Consolidating these devices at a single location minimizes spin rotations between them. The  Hjet will be relocated for refurbishment and modifications. Planned upgrades include double-layer silicon detectors, a Breit–Rabi polarimeter with a quadrupole mass analyzer, an improved target chamber and magnetic field, as well as an upgraded slow-control system with integration into the EPICS database. These upgrades are essential to meet the EIC requirement of absolute beam polarization measurements with an accuracy of $\Delta P / P \approx 1\%$.

In summary, the basic polarimetry methods have been established. A new target chamber with an additional lock chamber will be implemented to accommodate extra carbon targets. The lock chamber functions as a vacuum transfer system, allowing new targets to be loaded or exchanged without breaking the vacuum of the main target chamber. This design ensures continuous operation and prevents time-consuming vacuum reconditioning after each target change. RF shielding is required for pump ports, view ports, and related openings. Longer carbon targets are needed, and the aluminum target holder must be replaced with alumina. The choice of readout and detector systems for the carbon targets is still under evaluation. The demands at the EIC are significantly higher than at previous facilities, and dedicated designs are in progress to meet these challenges. All polarimeters are planned to be operational at the start of polarized beam commissioning.

\subsection{Absolute Hadron Polarimeter based on Cryogenic $^3$He ABS}
\label{sec:cryogenic-3He-ABS}

In order to support the physics program envisioned for the Electron Ion Collider (EIC), the development of a polarized $^3 $He$^{++}$ beam utilizing the Electron Beam Ion Source (EBIS) is well underway \cite{Zelenski2023-pb,Maxwell2020-lb,Musgrave2020-cm,Maxwell2016-jd}. This necessitates accurate polarization measurements of the $^3 $He$^{++}$ beam. One approach involves measuring the polarization via elastic scattering of the $^3 $He$^{++}$ beam on an unpolarized $^4$He gas target in the CNI region \cite{Raparia2024-pb,Atoian2020-jt}. Absolute polarization measurement using $^3 $He--$^4$He scattering in the CNI region has been well studied \cite{Hardy1970-jv}, with precise analyzing powers available in the kinematic range of $4~\text{MeV} \leq E_{^3\text{He-beam}} \leq 18~\text{MeV} $ \cite{Spiger1967-aw, Boykin1972-mt}. However, for monitoring polarization along the Booster, the Alternating Gradient Synchrotron (AGS), or the high-energy Relativistic Heavy Ion Collider (RHIC) beamlines, precise analyzing powers are unavailable. At these kinematic regimes in which analyzing powers are unavailable, the most robust method involves elastic scattering of polarized $^3$He over a $^3$He target of known-polarization, also in the CNI region. The proposed $^3$He--$^3$He polarimeter follows the design of the established H-jet polarimeter, which has provided reliable absolute proton beam polarimetry at RHIC using $p$--$p$ scattering in the CNI region with a polarized hydrogen atomic beam source (ABS) target \cite{Eyser2007-it, Poblaguev2020-pm, Zelenski2005-bl}. The $^3$He--$^3$He polarimeter would utilize a polarized $^3$He ABS, as described in \cite{MohanMurthy2025-qk}. While alternative schemes, such as $p$--$^3$He scattering, have been proposed \cite{Poblaguev2024-zz,Kopeliovich2024-rh}, $^3 $He--$^3$He scattering remains the cleanest and simplest polarimetry technique \cite{Poblaguev2022-lv}. 

Conventional polarization methods like spin-exchange optical pumping (SEOP) \cite{Parnell2009-zi} and metastability-exchange optical pumping (MEOP) \cite{Maxwell2020-lb} are limited to polarizations below $90\% $ \cite{Gentile2017-ng}, whereas cryogenic $^3$He beams polarized via magnetic field gradients can exceed $99\%$ polarization with $\sim1\%$ precision \cite{Ahmed2019-oo}. Thermal $^3$He ABS systems operating above $37$ K with hexapole magnets have achieved $55\%$ polarization with a flux of $10^{14}$ atoms/mm$^2$/s \cite{Jardine2001-de}, while the first cryogenic polarized $^3$He ABS developed at Los Alamos demonstrated up to $95\%$ polarization with a flux of $10^{14}$ atoms/cm$^2$/s \cite{Esler2007-bl, Eckel2012-la}. This cryogenic ABS has been comprehensively upgraded at MIT \cite{MohanMurthy2025-qk}, and the details of the polarization measurement schemes for both the polarized $^3$He ABS target and the $^3$He beam at the EIC are presented in \emph{ref.} \cite{MohanMurthy2025-hs}.

\begin{figure}
\includegraphics[width=\columnwidth]{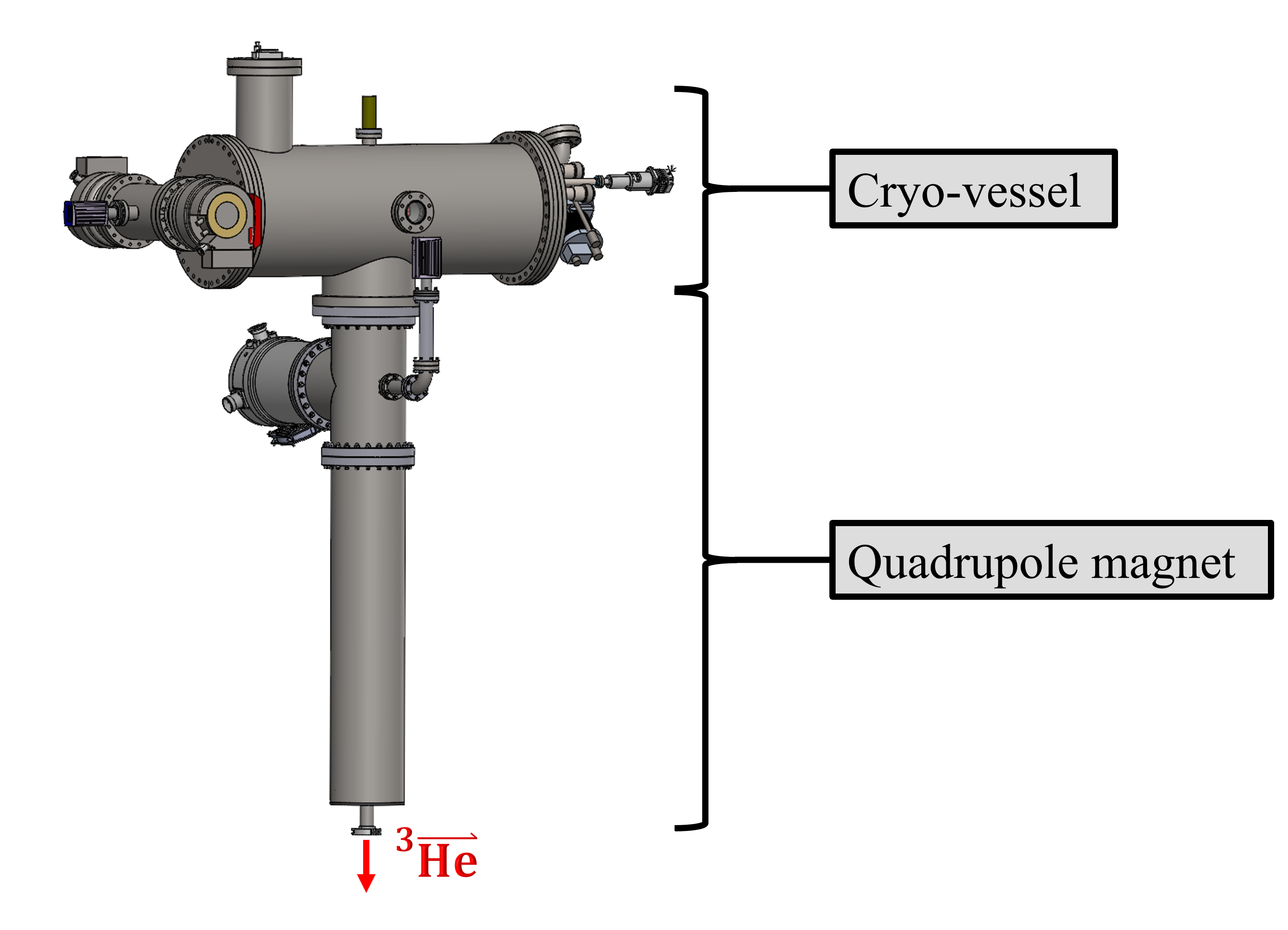}
\caption[]{An engineering diagram depicting the primary components of the polarized $^3$He ABS, including the three turbo pumps and two vacuum gauges. The cold head inlet can be seen on the flange to the right. Courtesy: \emph{Ref.}~\cite{MohanMurthy2025-qk}.}
\label{fig1-1}
\end{figure}

The polarized $^3$He ABS generates a finely collimated, highly polarized beam of neutral $^3$He atoms. It consists of two main components: (i) a cryogenic chamber with a multi-channel plate actuated nozzle cooled to $1~$K, and (ii) a quadrupole magnet section for beam polarization. These two components are shown in Fig.\,\ref{fig1-1}. A gas handling system supplies and regulates $^{{3,4}}$He gases to the ABS. A characterization section with two residual gas analyzers (RGAs) is used to map the beam profile but it is not operationally a part of the ABS. Successive cooling stages achieves $1~$K at the nozzle through which a beam of $^3$He atoms is produced effusively. A cold head capable of achieving a temperature of $4~$K is thermally coupled to a large copper pot, which also achieves a temperature of $4~$K. The feed gases of $^{3,4}$He thermalize with the $ 4~$K pot. A second smaller copper pot is fed the $^4$He gas at $4~$K, and vacuum pumped on, evaporatively achieving a temperature of $\sim1~$K. The $^3$He gas is then allowed to thermalize with the $1~$K pot, before being fed to the the quartz multi channel plate based nozzle. The nozzle is held at $ 1~$K by being thermally braided to the $1~$K pot. The $^3$He gas is in a molecular flow regime at the exit of the nozzle. The actuated nozzle has two degrees of motion to control its pitch and yaw. The beam produced at the nozzle can therefore be aligned along the central axis of the quadrupole magnet section. The $^3$He beam is passed through two skimmers constraining its angular spread before entering the quadrupole magnet section. In the quadrupole magnet section, four sets of permanent neodymium magnets, with a surface field of $\sim1~$T, generate a strong magnetic field gradient of $\sim1~$T/cm, over the entire $\sim1~$m pole length. One of the spin states of the $^3$He atoms is defocused and vacuum pumped away, while the other is focused, and therefore the beam emerging from the quadrupole magnet section is polarized. To prevent depolarization at the exit of the quadrupole magnet, four shimming magnets adiabatically transport the polarized $^3$He atoms to the holding field. The molecular beam is then supplied through a gate valve for further use. This method has previously demonstrated a flux of $ \sim10^{14}~ $atoms/cm$^2$/s with $ \sim95\% $ polarization \cite{Esler2007-bl, Eckel2012-la}, and further design and operational details are available in \emph{ref.}~\cite{MohanMurthy2025-qk}.

At the EIC, beams are transversely polarized and only converted to longitudinal polarization at the interaction points between the hadron and electron beams \cite{Poblaguev2024-zz,Nunes2022-vt}. Accordingly, the spin orientation of polarized $^3$He atoms from the ABS, initially exhibiting a spatially varying quadrupolar pattern, is aligned into a uniform transverse direction using spin evolution coils. The spin evolution coils are based around a solenoid, whose axis aligns with the axis of the ABS.

If a polarized beam undergoes elastic scattering off a polarized target, measuring the (beam) left-right asymmetry of both the scattered beam and target particles provides a means to determine beam polarization using the relationship
\begin{eqnarray}
p_{\text{beam}} = \frac{\epsilon_{\text{beam}}}{A^{\text{beam}}_N} = \frac{\epsilon_{\text{beam}}}{\epsilon_{\text{target}}}p_{\text{target}},~\label{eq-c-3}
\end{eqnarray}
where, $\epsilon_{\{\text{beam,target}\}}$ are the two asymmetries, and $p_{\text{target}}$ is the polarization of the cryogenic $^3$He ABS target. The analyzing powers, $A^{\text{beam}}_N=A^{\text{target}}_N$, and the corresponding kinematics have been studied in \emph{ref.} \cite{MohanMurthy2025-hs}. However, nuclear fragments from the scattering process must be vetoed to ensure sensitivity to elastic scattering events alone \cite{Poblaguev2022-lv}. The beam polarization measurement scheme uses two pairs of strip detectors to measure the two asymmetries in Eq.\,\eqref{eq-c-3}. The analyzing powers however vary for different scattering processes. The analyzing powers are studied in \emph{ref.} \cite{MohanMurthy2025-hs,Buttimore2014-ws} for the relevant cases of $^3$He--$^3$He, p--$^3$He, and C--$^3$He scattering at EIC. Since both ABS and beam polarization are $\mathcal{O}(1)$, the measured asymmetries are comparable in magnitude to analyzing powers. Consequently, the precision of the beam polarization measurement is primarily limited by the statistical uncertainty of the asymmetry measurements, with a secondary contribution from the uncertainty in the $^3$He ABS polarization. The fact that the analyzing powers exceed $10^{-3}$ across all relevant kinematics at the EIC for this polarimeter \cite{MohanMurthy2025-hs} enables fast and accurate beam polarization measurements. 

Given a $^3$He ABS flux of $\sim10^{14}$ atoms/cm$^2$/s \cite{Esler2007-bl, Eckel2012-la} and mean velocity of $\sim84~$m/s, the target density is $\sim10^{10}$ atoms/cm$^3$, lower than the H-jet based $p$--$p$ polarimeter by nearly two orders of magnitude \cite{Poblaguev2020-pm}. Employing a storage cell to act as an internal gas target, for polarized $^3$He, increases the target thickness to $\sim10^{12}$ atoms/cm$^2$, achieving $\sim1\%$ precision within $\sim1~$hour at peak EIC $^3$He$^{++}$ beam current of $\sim1~$A \cite{Zelenski2023-pb}. Spin-relaxation times of the order days have been demonstrated for polarized $^3$He in Cesium coated storage cells \cite{Heil1995-as}. But, in the context of accelerator based $^3$He storage cells, an adequately long relaxation time $\mathcal{O}(1~\text{minute}$) \cite{DeSchepper1998-vb} has been achieved. The $^3$He ABS target detectors are positioned to minimize background and can use silicon strip detectors, while the detectors measuring the asymmetry of scattered beam particles require radiation-hard diamond strip detectors \cite{Narayan2016-ha}. Using charge-to-digital (QDC) conversion of the signals from the scattered-beam detectors enables fragment rejection, allowing these detectors to also be used as veto counters.

\subsection{Theoretical background of hadron polarimetry}
\label{sec:theo-background-hadron-pol}
	
High energy hadron polarimetry will need measurements of the transverse analyzing power $A_\mathrm{N}$ of a polarized beam of hadrons \cite{PhysRevD.59.114010} with energy $E$, spin $j$, mass $m$, charge $Ze$, and magnetic moment $\mu$, scattering on a stationary target of mass $M$ and charge $\tilde Z e$, the spin of the target playing only a minor rôle \cite{PhysRevLett.123.162001}. In magneton units $\mu_p/(2em_p)$, based upon proton values, the non-flip and single spin-flip elastic amplitudes are \cite{10.1063/1.3122170}
	\begin{align*}
		f & = \frac{i + \rho}{\exp{i\delta_C}}
		-
		\frac{t_C}{t} \frac{\exp b_et}{\exp b_ht}
		\approx
		i + \varrho - \frac{t_C}{t} \,, \quad \text{and} \\
		g & = \frac{\sqrt{-t}}{2m_p}\left[\frac{R_S+iI_S}{\exp i\delta_C}
		-
		\frac{\kappa\,t_C}{t}\cdot\frac{\exp{b_m t}}{\exp b_h t}\right]
	\end{align*}
	where, in
	\(
	A_\mathrm N = 2\,\mathrm{Im}\,(f^\star g)\,/\,(|f|^2 + |g|^2)
	\),
	amplitudes divided by $\sigma_\mathrm{tot}(E)\exp(b_h t+i\delta_C)$ do not change the value of the analyzing power. The slope of \( \ln(d\sigma/dt) \) is $ 2\,b_h $\,, as amplitudes are squared. The slope of the target electric form factor multiplied by the incident electric form factor is $b_e$, and multiplied by the incident magnetic form factor, is $b_m$. Absorptive corrections to the Coulomb amplitude are small for light ions \cite{Poblaguev:2025zaj}. The Coulomb phase with Euler's $\gamma \approx 0.5772$ is
	\(
	\delta_C=-Z\tilde{Z}\alpha\left(\ln|b_ht+b_et|+\gamma\right)
	\),
	a few percent as $ \alpha = 1/137.036 $. The electromagnetic and hadronic non-flip amplitudes are equal in size near
	\begin{equation*}
		-t_C = \frac{\,8\,\pi\,|\,Z\,\tilde Z\,|}{137.036\,\sigma_\mathrm{tot}}
		\approx
		\frac{ |\,Z\,\tilde Z\,| }{14\,\sigma_{\mathrm{tot}}\mathrm{[mb]}}\ \si{(\giga\electronvolt/c)^2}\,, 
	\end{equation*}
	and where the anomaly\,\cite[Eq.\,(10.6.23), p.\,456]{Weinberg_1995} is given by  \cite{Buttimore:2003ct,Buttimore:2004vi}
	\begin{equation*}
		\kappa = \frac{\mu}{2jZ} - \frac{m_p}{m}\,.
	\end{equation*}
	
	The hadronic single spin-flip amplitude is $R_S + iI_S$, when normalized by the total cross section and a spin-flip kinematic factor. As $(b_e-b_h)t$, $\rho$, $\delta$ and absorption corrections for light ions \cite{PhysRevD.64.034004} are all negligible in the Coulomb Nuclear Interference (CNI) region,
	$ \varrho = \rho + \delta_C + (b_h - b_e)t_C $
	due to a first order expansion.
	The analyzing power for an incident polarized hadron is approximately
	\begin{align*}
		\frac{m_p}{ \sqrt{-t}^{\ } }\,A_\mathrm N
		\ = \ 
		\frac{ \left(\kappa - I_S\right)\frac{t_C}{t} + \varrho\,I_S - R_S }
		{ \frac{t_C^2}{t^2} - 2\left(\varrho + \frac{\kappa^2t_C}{4 m_p^2 }\right)\frac{t_C}{t} + 1 + \varrho^2}
	\end{align*}
	where the singular spin-flip electromagnetic term has been included in the denominator, omitting the small spin-flip hadronic terms. For pp elastic scattering, \(\varrho + \frac14 \kappa^2t_C/m_p^2 \) is zero near $E = \SI{200}{\giga\electronvolt}$ and a similar cancellation is expected for other elastic processes. As in the non-flip case, \( |I_S| > |R_S|\), so neglecting \( \varrho\,I_S - R_S \) and $\varrho^2$, the analyzing power may be approximated by an expression reaching a maximum size at $x = \sqrt{3} = t/t_C$,
	\begin{align*}
		A_N = \frac{\left(\kappa - I_S\right)\sqrt{-t_C}}{m_p}
		\cdot
		\frac{x^{3/2}}{x^2 + 1} \quad \mathrm{where}\ x = \frac{t}{t_C}\ .
	\end{align*}
	The level of the polarization of $m$ at each energy $E$ may be found from a comparison of the process involving $m$'s unknown polarization, namely,
	\(
	m\!\!\uparrow m'\; \longrightarrow\; m\;m'
	\)
	with a process where the polarization of $m$ (in a gas-jet say) is known accurately, viz.,
	\(
	m'\; m\!\!\uparrow\; \longrightarrow\; m'\; m\,.
	\)
	Analyzing powers would need to be compared at the same invariant $s$ values, that is, at $m\,E' = m'\,E$, where $E'$ is the energy of the mass $m'$ in the second reaction.
    
    In practice, the particle $m'$ should be the same as particle $m$, for statistical and kinematic reasons. Once the polarization of $m$ is known, the value of the imaginary part of the hadronic single spin-flip amplitude, $I_S$, for the reaction $m\!\!\uparrow\! M \longrightarrow m\,M$ can be estimated at each energy $E$ and polarimetry is assured at a particular level of accuracy.
	
	\subsection*{Hadron Kinematics}
	The analyzing power at higher energies is expected to largely result from events whose recoil scattering angle lies between $\theta_\mathrm{el}$ and $\theta_\mathrm{in}$ \cite[Eq.\,(4.16), p.\,80]{byckling1973particle}. The angle $\theta_\mathrm{el}$ refers to the elastic reaction, 
	\(
	m\!\!\uparrow\! M \longrightarrow m\,M
	\),
	and angle $\theta_\mathrm{in}$ to the inelastic process,
	\(
	m\!\!\uparrow\! M \longrightarrow m\!+\! \delta\;\, M
	\), so
	\begin{align*}
		\cos\theta_\mathrm{el} & = (1 + 2M/T)^{-\frac12}\,W/P, \quad \text{and}\\
		\cos\theta_\mathrm{in} & = (1 + 2M/T)^{-\frac12} 
		\left[
		\,W +\left(m+\tfrac\delta 2\right)\,\delta/T\,\right]/P
	\end{align*}
	where $P = \sqrt{E^2 - m^2}$ is the incident momentum and $\delta$ is the lowest excitation or break-up of the polarized incident mass $m$, that is, $m\!+\!\delta$ amounts to the ejected mass\,\cite{Poblaguev:2020qbw}. The target mass $M$ recoils unchanged with relativistic kinetic energy $T$ and $W = E + M$\,\cite{Buttimore:2013rez}. Examples of excitation or break-up are: $p\longrightarrow p + \pi\,$, in the case of an incident proton, or He-3$\,\longrightarrow p + d$, for an incident He-3 ion.
    
    It is quite possible that events scattered between the recoil angles of $\theta_\mathrm{in}$ and $\theta_\mathrm{inel}$ would also contribute significantly to an asymmetry, with recoil $\theta_\mathrm{inel}$ more generally given by
	\begin{align*}
		\cos\theta_\mathrm{inel}
		& =
		\left( 1 + 2\,\widetilde M / T \right)^{-\frac12}
		(\, W + k / T\; )
		/
		P, \\
		k & = \left( m + \tfrac\delta 2 \right) \delta
		+
		\left( E - \tfrac\Delta 2 \right) \Delta,
	\end{align*}
	this angle referring to inelastic scattering that includes a recoil mass of
	\( \widetilde M = M\! + \!\Delta \),
	the first excited state or break-up of the target mass $M$. The recoil angle for an incident mass $m$ remaining unchanged would have $\delta = 0$ here. As an example, consider a Carbon-12 target and its first nuclear energy level of $\Delta = 4.442$ MeV above its ground state.
    
    Measurements in the extended recoil angle region from $\theta_\mathrm{in}$ to $\theta_\mathrm{inel}$ should reveal the appropriateness of this contribution to the analyzing power. Though the elastic scattering angle can reach \SI{90}{\degree}, the inelastic angle has a maximum value of $\theta_\mathrm{inel}^\mathrm{max}$ at the particular recoil kinetic energy of $T_\mathrm{inel}$, where
	\begin{align*}
		T_\mathrm{inel} & = k\, \widetilde M
		/
		\left( W \widetilde M - k \right),
		\quad \text{and}\\
		\theta_\mathrm{inel}^\mathrm{max}
		& =
		\arccos\left( \frac{\sqrt{k \left(2\,W\widetilde M - k\right)}}{\widetilde M P} \right)
	\end{align*}
	The time taken for the recoiling particle of mass $\widetilde M$ to reach a detector distant $d$ away from the interaction point is
	\begin{align*}
		\frac dc \left( T + \widetilde M \,\right)
		/
		\sqrt{T \left( T + 2\,\widetilde M\;\right)}
		\approx \frac{ \widetilde M \, d }{c}/ \sqrt{-t_C}\,.
	\end{align*}
	and, as $t_C$ does not depend strongly on a particular process, the time taken is approximately proportional to the recoiling mass. The abort gap provides an opportunity to detect recoils from a bunch adjacent to the gap without the difficulty of pile-up at EIC.
	
	The above analysis relates to particles with spins of $\frac12$ and $1$. The higher spins of Lithium and Boron may require an analysis involving a study of the scattering amplitudes for all masses and spins \cite{Arkani-Hamed:2017jhn}.

\subsection{Snakes in AGS and EIC}
\label{sec:snakes}
\subsubsection*{The AGS partial snakes}

The Alternating Gradient Synchrotron (AGS) has two helical dipoles that are used for protons \cite{PSnake}. In the nominal configuration, there is a small spin-tune gap for $\nu_y$ and a modest tune jump \cite{PhysRevSTAB.17.081001} or skew quadrupoles for resonance correction \cite{Schoefer:2021xra} are used for horizontal resonance crossings. For helions, with the same magnetic field as protons, the spin-tune gap is larger and allows both $\nu_x$ and $\nu_y$ to fit inside the spin-tune gap \cite{Hock_phd}. Li$^7$ will have a smaller spin-tune gap than protons but will be sufficient for polarization transmission. A modest tune jump or skew quads for horizontal resonances will be required. Deuteron and Li$^6$ will only cross imperfection resonances and do not need a snake. The summary of relevant parameters for the different polarized species and their configuration in AGS is given in Tab.~\ref{tab:species-of-interest}.

\begin{table}[htb]
	\centering
	\caption{Polarized species of interest, their relevant parameters, $m$ is mass, $\mu/\mu_N$ is the nuclear magneton ratio, $S$ is the spin, $G$ is the anomalous magnetic $g$-factor, the resonance spacing given as $mc^2/G$, $\chi_c,\chi_w$ are the cold and warm snake rotations, and the size of the spin-tune gap ($\nu_s$~gap), and rotations by the existing AGS cold and warm snakes.}
	\label{tab:species-of-interest}
    \renewcommand{\arraystretch}{1.1}
    \vspace{0.2cm}
	\begin{tabular}{l|c|c|c|c|c}
		Parameter & $p$ & $d$ & $h$ & Li$^6$ & Li$^7$ \\
		\hline
		$m$ (MeV/c$^2$)     & 938.27 & 1875.61   & 2808.92 & 5601.52 & 6533.83 \\
		$\mu/\mu_N$         & 2.7928& 0.8574   &-2.1275 & 0.8220 & 3.2564\\
		$S$                   & $\frac{1}{2}$ &  1  & $\frac{1}{2}$ & 1 & $\frac{3}{2}$ \\
		$G$                   & 1.793 &-0.143   &-4.184 & -0.178 & 1.532\\
		$mc^2/G$  (GeV)     & 0.523 & 13.116     & 0.671    & 31.466    & 4.264 \\
		$\chi/\chi_{proton}$&1      & 0.0399 & 1.5572 & 0.0497 & 0.3668\\
		$\chi_c,\chi_w$ (\%)& 14, 9 & -- & 21.8, 14 & - &5.2, 3.3 \\
		$\nu_s$ gap         &  0.061& -- & 0.094 & -- & 0.025 \\\hline
	\end{tabular}
\end{table}

\subsection*{The HSR snakes}
The HSR will have six snakes, each consisting of four helical dipoles \cite{PTITSIN1997126}. For protons and helions, they will act as full-snakes where they rotate the spin $\pi$. The precession angle, $\phi$, requires attention to maximize polarization transmission, especially for helions in the $G\gamma>700$ range. Due to the larger scaling of snake rotation, helions can also satisfy any angle from $-\pi$ to $\pi$ \cite{Hock:2024ojr}. Preliminary analysis of Li$^7$ indicate polarization can be preserved with partial snakes for a modest spin-tune gap. This will require a jump quadrupole system for intrinsic resonances. Deuteron and Li$^6$ will have a solenoid that operates as a partial snake and avoid imperfection resonances. They will need a jump quadrupole system for vertical intrinsic resonances. The summary of these different configurations is given in Tab.~\ref{tab:snake-config}.

\begin{table}[htb]
	\centering
	\caption{Snake configuration (I being the outer and inner helix currents, $\phi$ the precession axis, $BL_\text{sol}$ being the solenoid strength, $\xi$ the total rotation at the exit of the fourth helix, and the spin tune $\nu_s$) for the six sets of helical dipoles for protons, helions, and Li$^7$ and the solenoid for $d$ and Li$^6$.}
	\label{tab:snake-config}
    \renewcommand{\arraystretch}{1.1}
    \vspace{0.2cm}
	\begin{tabular}{l|c|c|c|c|c}
		Parameter & $p$ & $d$ & $h$ & Li$^6$& Li$^7$ \\
		\hline
		$I_\text{out}$ (A) &100 & -- & 62 & -- & 174 (238)\\
		$I_\text{in}$ (A) &322 & -- & 206 & -- & 231 (322)\\
		$\phi$                  &$\pm45^\circ$ &&$\pm45^\circ$ && 0\\
		$BL_\text{sol}$ (Tm)  & --     & 15  &  --    &  15 &   -- \\
		$\chi~(\%\pi)$                  & 100 & 0.45 & 100  &0.42& 17 (31)\\
		$\nu_s$  & 1.5 & 0.01 -- 0.99& 1.5 & 0.01 -- 0.99& 0.2 - 0.8\\\hline
	\end{tabular}
\end{table}

\subsection{What should be learned for the EIC from accelerating polarized protons in the AGS}
\label{sec:learned-from-AGS-for-EIC}

Accelerating polarized protons to 22~GeV/$c$ in the AGS during 1983–1988 required overcoming 45 depolarizing spin resonances without Siberian snakes. This effort, led by researchers from the University of Michigan and Brookhaven National Laboratory following the cancellation of Isabelle, culminated in the first successful runs in 1988. Polarization correction was achieved using 95 harmonic correction dipoles and 10 fast pulsed quadrupoles, yielding a final polarization of about 42\%. However, the process demanded weeks of beam time and manual tuning. Each resonance had to be addressed individually, with polarization measurements taking 5–30 minutes per setting, making the setup operationally fragile and time-consuming \citep{khiari1989}.

Partial Siberian snakes were later introduced and proved decisive: imperfection resonances were effectively eliminated, and intrinsic resonances could be crossed by choosing a suitable vertical betatron tune. This made acceleration markedly more stable and reproducible.

These lessons remain relevant for the EIC. The planned RCS injector for polarized electrons traverses a range of energies and spin tunes where many weak imperfection resonances may appear, even as the design mitigates strong resonances through increased tune and super-periodicity. Experience shows that residual resonances can be influenced by misalignments and time-dependent effects (e.g., eddy currents, ground motion), and that spin dynamics are sensitive to small vertical deflections that standard orbit diagnostics may not fully reveal. Robust mitigation -- precise alignment, active orbit control, harmonic corrections, careful ramp optimization, and model-based spin tracking with dedicated polarization diagnostics --reduces operational risk.

Tuning a synchrotron via repeated polarization measurements, as done historically at the AGS, is operationally demanding and in contemporary operations is typically complemented by model-based spin tracking and dedicated diagnostics, with the expectation that optimization supported by AI/ML techniques will be significantly faster and more effective than was possible at the AGS. For polarized protons, designs incorporating full or partial Siberian snakes together with strong orbit control have repeatedly delivered reliable acceleration. For polarized electrons, linac-based injectors (e.g., CEBAF, SLC) have demonstrated long-term stability and high polarization while avoiding resonance management during acceleration.

\section{Spin manipulation in EIC}

\subsection{Aspects of polarization optimization}

\subsubsection{Polarization and intensity optimization in the injector chain for the EIC}
\label{sec:pol-intensity-optimization}

\begin{figure}[htb]
    \centering
    \includegraphics[width=0.75\linewidth]{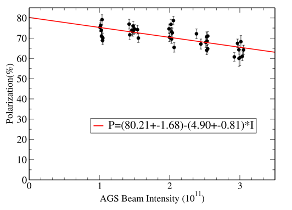}
    \caption{Proton beam polarization at AGS top energy E = 23 GeV versus intensity per bunch.  The decrease is mediated by intensity dependent increases in emittance and depolarizing intrinsic resonances.}
    \label{fig:AGSPolInt}
\end{figure}

The RHIC injector complex will be used as the hadron injector complex for the EIC. The injectors will be kept largely in their current state, except for upgrades required to accommodate new polarized species like helium-3 (see Section \ref{chap:polionsources}).  Assuming intensity transmission through the EIC HSR that is similar to that of RHIC and lossless transmission of polarization (owing to the installation of additional snakes), the AGS will have to provide bunches of $3\times10^{11}$ protons with a polarization of 70\% \cite{EIC_CDR}.  At current optimized performance, the AGS provides a polarization of 65\% at that intensity. We describe improvements aimed at increasing that number to 70-75\% to meet the EIC requirements and provide margin against any potential polarization losses in the HSR.  Figure \ref{fig:AGSPolInt} summarizes  current performance of the AGS.

The remaining depolarization mechanisms in the injectors are so-called intrinsic resonances, which produce polarization loss proportional to the beam emittance \cite{SYLee}.  Polarization improvements can therefore be divided into two types: correction of depolarizing resonance terms directly and reductions in the beam emittance (at a particular intensity).

One known source of depolarization are horizontal resonances driven by the AGS partial snakes. Since 2011 these have been mitigated with tune jump, which increases the resonance crossing speed and reduces depolarization \cite{PRAB_tune_jump,Schoefer2012_TuneJump}. A new system using skew quadrupoles is being commissioned to complete correct the resonances and gain a 5-10\% (relative) increase in polarization\,\cite{Schoefer_SkewQuadPRAB,IPAC24_SkewQuads}.  Further improvements could be realized by a full realignment of the AGS magnets to reduce sources of depolarization driven by misalignments.

The beam emittance increases with intensity due to a number of factors including collective effects like space charge. Space charge forces can be reduced by lowering the peak current of the accelerated bunches. This has previously been accomplished with multiple harmonics of RF to defocus the central part of the (typically Gaussian) bunch. Further reduction of the peak current is being explored using a technique to longitudinally split a bunch in the Booster, accelerate the pair in the AGS and merge them back at top AGS energy.  Since space charge defocusing is strongest at low energy, this produces lower peak forces at low energy in the AGS.

An additional, more invasive upgrade would involve upgrading the Booster main magnet and Booster to AGS transfer line equipment in order to increase the transfer energy from 1.7 GeV to 2.8 GeV.  This would reduce the peak space charge tune shifts by a factor of 3. Measurements are planned to assess the need for this upgrade during the RHIC to EIC transition period. The increased energy would also increase the number of spin resonances crossed in the Booster, the impact of which also needs to be measured experimentally.

\subsection{Ion spin rotators in the Hadron Storage Ring}

To transform vertical beam polarization in the arcs into longitudinal polarization at the experimental detector, the HSR will reuse existing RHIC spin rotators, which are based on helical dipole magnets.
Each spin rotator consists of four helical dipoles with a maximum field of 4 T \cite{RHIC_Polar_Manual}. Each magnet is $2.4$ meters long and has a single helical period.
The rotator is characterized by two magnetic field values, $B_1$ and $B_2$, for the outer and inner helices, respectively.
The internal configuration of the rotator is quite similar to that of the RHIC (or HSR) Snakes. However, unlike the Snakes, the orientation of the magnetic field at the center of the helical magnets is horizontal.

\begin{figure*}[htb]
\centering
\includegraphics*[width=0.7\textwidth, angle=-0]{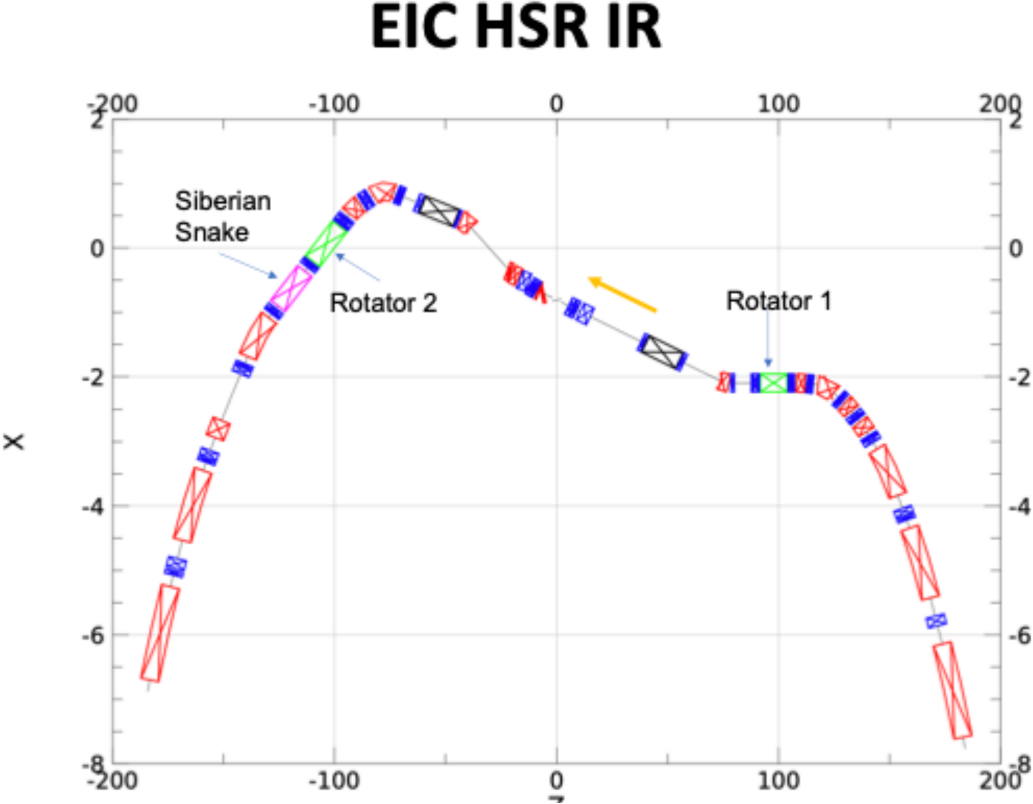}
\caption{Spin rotators in the EIC Interaction Region layout.}
\label{fig-SpinHRot-01}
\end{figure*}

To accommodate the higher proton beam current in the HSR, the RHIC spin rotators will be upgraded.
The vacuum pipe will be replaced with a new one coated with copper and amorphous carbon layers to mitigate resistive heating and electron cloud effects.
A new Beam Position Monitor will also be installed at the center of each rotator.

The EIC interaction region is much more complex than that of RHIC \cite{EIC_Hadron_SpRot}. The spin rotators cannot be placed symmetrically, as they are in RHIC.
They are also located farther from the interaction point, at $\SI{97.8}{m}$ and $\SI{107.8}{m}$ (see Figure~\ref{fig-SpinHRot-01}), compared to RHIC.
Additionally, the bending angles from the IP to the rotators are not symmetrical, $-17$ mrad and $61.35$~mrad, respectively.
As a result, the net bending angle between the HSR spin rotators is not zero. One impact of this asymmetry is a spin tune shift of approximately 0.01 that occurs when ramping the rotator magnet fields to their design settings at the beginning of the store.
To compensate for this shift, a slight adjustment of the Snake magnets will be required during the rotator magnet ramp.

At certain energies, the spin rotator cannot achieve longitudinal polarization at the interaction point because it would require fields exceeding the 4 T limit of the rotator magnets.
This occurs in the energy ranges of $143-151$~GeV and $226-232$~GeV, which do not include the EIC main operating energies (41~GeV, 100~GeV, and 275~GeV).
It should also be noted that, unlike in RHIC, the fields in the EIC helices will need to reverse polarity at different energies to cover as many energy points as possible.

For polarized $^3{\rm He}$ operation the required rotator magnet fields are smaller than for protons.

\subsection{Existing spin manipulators at RHIC}
In polarized proton collision experiments, it is highly advantageous to flip  the spin of each bunch of protons during the stores to reduce the systematic errors.  An artificial resonance can be introduced to flipper the spin. The traditional spin flipping technique uses a single RF spin rotator that rotates the spin around an axis in the horizontal plane. The spin rotator can be implemented as a dipole or a solenoid running with certain rf frequency.
  It is done as following: ramping the frequency of the spin rotator tune $\nu_{osc}$ across the spin tune $\nu_{sp}$ adiabatically and the spin can be flipped following the Froissart-Stora formula~\cite{FROISSART1960297}:
  \begin{eqnarray}
{{P_f}\over{P_i}} =2\exp[-{{\pi}\over {2}} {{|\epsilon|^2} \over {\alpha}}]-1,
\label{stora}
\end{eqnarray}
where  $\alpha={{\Delta \nu_{osc}}\over {2\pi N}}$, and $\Delta \nu_{osc}$ is the range of the rf spin rotator tune sweep range, $N$ is the number of turns the sweep covers. 
As long as the spin tune is covered by the sweeping range, a resonance will be crossed. With proper sweeping speed, the spin can be flipped. 

  It should be noted that such a single spin rotator generates  two spin resonances, one at $\nu_{sp}=\nu_{osc}$, and one at $\nu_{sp}=1-\nu_{osc}$ or so-called ``mirror'' resonance. As long as the spin tune is sufficiently far away from half integer, say at 0.47, then the two spin resonances are sufficiently far from each other and each one can be treated as an isolated resonance. 
  This is the case for  low energies when Siberian Snakes are not needed and the spin tune is not at or near half integer.
   In high energy polarized proton colliders such as RHIC, the spin tune is very close to half integer. The two spin resonances overlap and their interference makes the full spin flip impossible with such a single rf spin rotator. To reach full spin flip, the ``mirror'' resonance has to be eliminated~\cite{PhysRevSTAB.11.091001}.
In addition, it is critical to eliminate any global  vertical  betatron oscillations driven by the AC dipole to achieve full spin flip~\cite{PhysRevSTAB.12.099001}. The final spin flipper consists of five AC dipoles with horizontal magnetic field and four DC dipoles with vertical magnetic field, which not only eliminates the ``mirror'' resonance, but also forms two closed vertical orbital bumps and eliminates the global  vertical oscillations outside the spin flipper~\cite{Bai:2010zzd}.

 Besides eliminating the ``mirror'' resonance and any global vertical betatron oscillation driven by AC dipoles, the reduction of the spin tune spread is also critical for achieving full spin flip.
  In RHIC, the reduction comes from suppressing the local dispersion slope difference between the two Siberian Snakes. For a normal lattice, this difference  is about 0.045 at 255GeV, which corresponds to  0.007 spin tune spread for a beam with a momentum spread of 0.001.
 Such a small $\Delta D'$ lattice was achieved by using $\gamma_{tr}$ jump quads\,\cite{PhysRevAccelBeams.22.061002}.

\begin{figure}[ptb]
\centering
\includegraphics[width=3.4in, angle=-0]{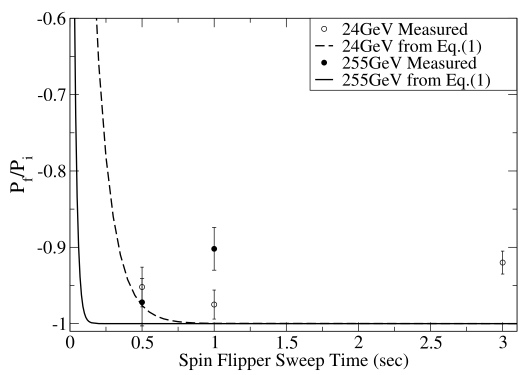}
\caption{The average final to initial polarization ratio  at 24GeV and 255GeV. The solid line is the polarization flip ratio from Eq.\,(\ref{stora}) for the resonance strength 0.00024 and the filled points are  the averaged spin flip efficiencies for three different sweep times at 24GeV.  The dashed line and open points are for 255GeV and  the resonance strength 0.00057.}
\label{f_flip_all}
\end{figure}

With the 0.005 tune sweep range and the given spin flipper strength, a 99\% spin flip efficiency is predicted for a  sweep time of 0.6 sec or slower at 24GeV from Eq.~(\ref{stora}) and numerical simulations~\cite{PhysRevAccelBeams.27.071002}.  The final to initial polarization ratio from Eq.~(\ref{stora}) for the given spin flipper strength at injection and 255GeV are plotted in Fig.~\ref{f_flip_all} as dashed line and solid line. But this is an over-simplified model. In reality, the synchrotron motion and residual spin tune spread  can have an impact on the final spin flip efficiency.  The measured spin flip efficiencies for three different sweep times are also shown in Fig.\,\ref{f_flip_all}~\cite{PhysRevLett.120.264804}. The best spin flip efficiency is about 97\% for both energies.  

In principle, the spin tune can be measured with a similar idea as the  betatron tune measurement: measuring the spin response to a driven spin coherence. Such a method  can also be  nondestructive. A coherent spin precession around the vertical direction can be adiabatically induced by driving the ac spin rotator at a drive tune near the spin tune.
 There are two advantages to this technique.  First,  it is an adiabatic spin manipulation and can preserve the beam polarization. Second, this is a relatively fast measurement.
 Hence, this technique is ideal for measuring the spin  tune at the store energy of a high-energy polarized synchrotron, such as RHIC or a future polarized electron ion collider.
 Driven coherent spin motion has been used to measure the spin tune in RHIC at 24 and 255~GeV.
The results show\,\cite{PhysRevLett.122.204803} that the spin tune can be measured by driven spin coherence when the tune separation is small enough.  For it to work, the drive tune needs to be close to the spin tune, which requires a  small spin tune spread. In RHIC, where a pair of Siberian snakes are used, the small spin tune spread was achieved by the reduction of the dispersion slope difference at  the two Siberian snakes\,\cite{PhysRevAccelBeams.22.061002, Ptitsyn:2010zz}.

\subsection{Novel spin tools for storage rings}

\subsubsection{Spin tune determination and feedback}
\label{sec:spin-tune-determination}

Polarized electron-hadron collisions at EIC necessite advanced techniques for polarized beam manipulation. The extensive R\&D program over the last decade by the Jülich Electric Dipole moment Investigations (JEDI) collaboration at COSY could provide valuable experience.

A key development was the online, continuous spin-tune determination, especially crucial for experiments utilizing in-plane polarized beams like it is proposed for investigating the Electric Dipole Moment (EDM) of charged particles in storage rings \cite{abusaif2021}. Precise knowledge of the precession frequency is essential for either maintaining the polarization direction or for applying resonant spin manipulation techniques. The JEDI collaboration has demonstrated methods for measuring this precession frequency through Fourier analysis of time-stamped events from a polarimeter/target setup that continuously probes a fraction of the circulating beam. At COSY this was achieved by inducing a controlled excitation around a betatron resonance, such that outermost particles are directed onto the target, allowing for the measurement of polarization asymmetries without significantly perturbing the entire beam. The analysis of the scattering events provides information on both the vertical and radial (in-plane) polarization components. The frequency at which the in-plane asymmetry oscillates yields the spin tune, while the phase obtained from the Fourier analysis indicates the direction of the spin relative to a precession starting longitudinally \cite{Bagdasarian:2014ega, JEDI:2015vwa}.

Furthermore, achieving long spin coherence times (in-plane polarization lifetimes) is paramount for such experimental programs. The JEDI collaboration has systematically investigated the contributing factors, achieving in-plane polarization lifetimes of up to 1000 seconds with 0.97-GeV/c deuterons \cite{JEDI:2016swi}. Studies have revealed the connection between zero chromaticity settings achieved through sextupole field corrections and extended polarization lifetimes, particularly when spin resonances are avoided \cite{JEDI:2017lbv}. The understanding and application of beam bunching, electron cooling \cite{Karanth:2020zat}, and sextupole optimization to minimize spin tune spread due to phase-space distribution might serve as valuable input for maintaining polarization of proton and ion beams at the EIC.

Finally, significant progress has been achieved in phase-lock feedback techniques to synchronize the spin precession with external radiofrequency devices used for spin manipulation. This was used to stabilize and manipulate the spin tune\,\cite{JEDI:2017bnp} or adjust the frequency of the rf devices to maintain resonance with the precessing spin\,\cite{JEDI:2025eec}. While this was used to phase-lock an rf Wien filter to the spin precession for EDM measurements over a complete cycle, also other applications might be of interest. Using the concept of fast switches\,\cite{pilotPRL} one could independently track the spin precession of a dedicated bunch to allow for real-time adjustments of the RF spin flipper frequency to maintain optimal resonance conditions for spin manipulation of the main beam. 

In summary, these developments underscore the importance of sophisticated beam instrumentation and control systems for realizing a scientific program with polarized beams at EIC.

\subsubsection{Lorentz-force free RF spin manipulator}
\label{sec:RF-Wien-filter}

The waveguide radio‑frequency (RF) Wien filter\,\cite{Slim2016116,SlimWFChaos} is a novel device designed as a resonant spin manipulator with minimal beam distortion. It has been commissioned in the framework of the electric dipole moment (EDM) studies conducted by the JEDI Collaboration at the Cooler Synchrotron COSY at the Forschungszentrum J\"ulich.
 

The waveguide RF Wien filter (RF WF) features a fundamentally different design from conventional models. Whereas a classical Wien filter uses crossed electric and magnetic subsystems to generate static fields to separate particles by velocity (or equivalently mass-to-charge ratio), the RF WF is a waveguide driven at the spin‐precession frequency and thus acts as a spin rotator rather than a mere velocity selector. Crucially, it must not induce beam distortions by generating RF electric and magnetic fields that are  mutually orthogonal and both perpendicular to the beam axis, with matched field amplitudes so that their Lorentz force cancels out.

The RF WF enforces the required orthogonality inherently by supporting the transverse electromagnetic (TEM) mode in its waveguide. Moreover, an RF circuit is integrated into the design to introduce a controlled impedance mismatch, establishing standing waves whose amplitudes and phases are precisely tuned so that the Lorentz force cancels\,\cite{SlimWFCircuit}. This concept has been demonstrated with both protons and deuterons, achieving induced beam–orbit oscillations as small as $\SI{430}{nm}$, only one order of magnitude above the quantum limit\,\cite{PhysRevAccelBeams.24.124601,PhysRevAccelBeams.26.014201}.

\begin{figure}[ht]
\centering
	\includegraphics[width=\columnwidth]{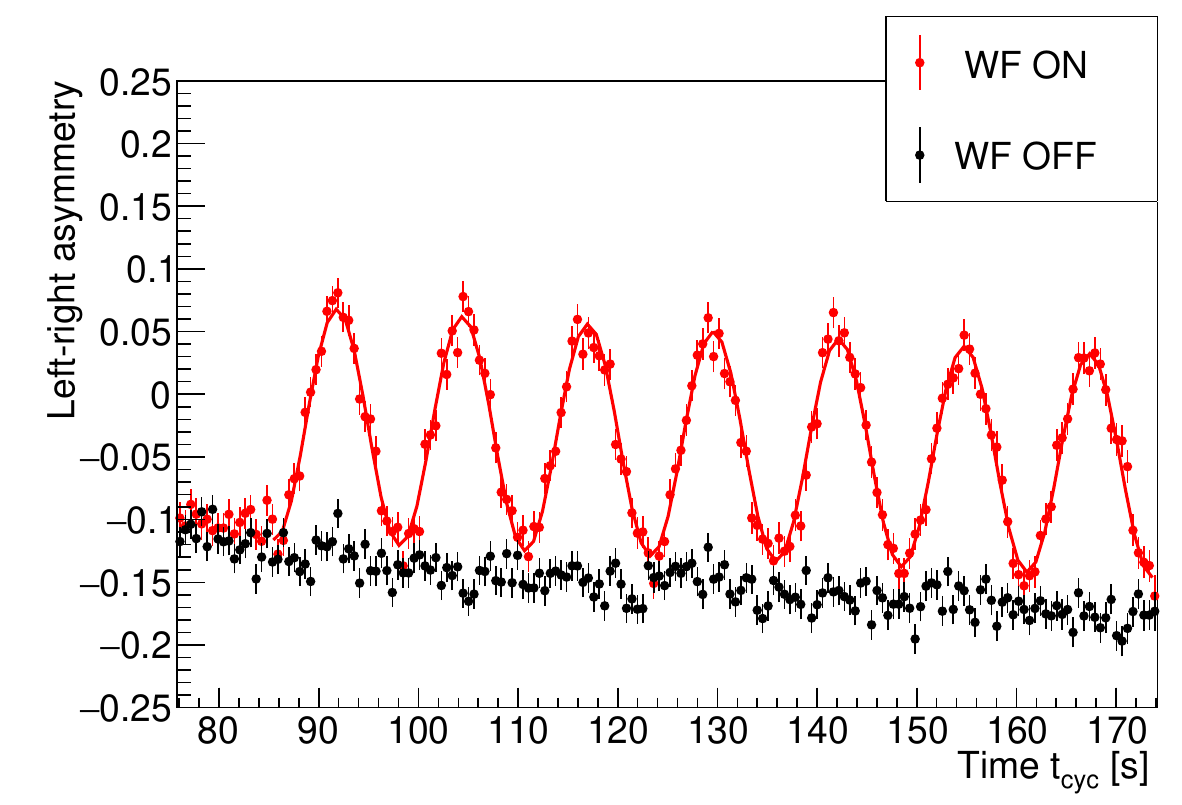}
\caption{\label{fig:drvOsc} Left-right asymmetry in the polarimeter, caused by the oscillating vertical polarization (red dots). The black dots are for a cycle with the RF WF completely turned off. (See also Ref.\,\protect\cite{pilotPRL} for further details.)}
\end{figure}
\begin{figure*}[t]
   \centering
    \includegraphics[width=0.6\textwidth]{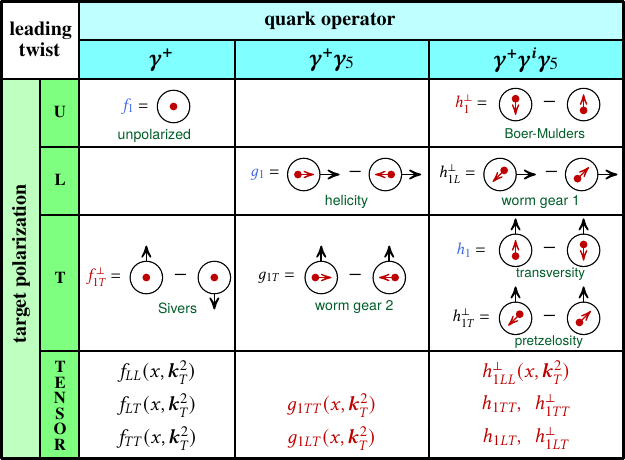}
    \caption{Spin-1/2 (first 3 rows) and spin-1 (last row) target transverse momentum dependent (TMD) distribution functions~\cite{Ninomiya:2017ggn}. }
    \label{fig:spin_one_tmds}
\end{figure*}

The RF Wien filter was successfully commissioned as a resonant spin rotator. By setting its magnetic field in the radial direction and starting from vertical polarization, driven spin oscillations could be induced with $\SI{1}{mW}$ of input RF power. Figure \ref{fig:drvOsc} shows a sample of these measurements.

Further studies were carried out in which the beam was split into two bunches (tests with four bunches were also successful). One serves as a comagnetometer—termed the “pilot” bunch—and the other as the “signal” bunch\,\cite{pilotPRL}. The spin tune, defined as the number of spin oscillations per turn, can only be unambiguously determined in an RF-free setup. To achieve this, the RF WF is gated off during the pilot bunch passage, allowing the feedback system to keep the filter precisely on resonance for the remainder of the cycle. An important implication of this work is that the EDM can be measured via a frequency-domain approach, and fully driven vertical spin oscillations were observed for the first time—all while maintaining the Lorentz force–free condition.


Driving a Lorentz‑force–free RF Wien filter at a fixed frequency provides a versatile spin‑manipulation tool for forthcoming facility such as the Electron–Ion Collider. Continuous spin flips becomes feasible without relying on the Froissart–Stora scan\,\cite{FROISSART1960297} employed by the RHIC RF‑dipole flipper\,\cite{RHIC_Spin_Flipper_PRL2018,RHIC_Spin_Flipper_PRL2019}. The RF WF parameters can instead be tuned to suppress the orbit distortions that RF dipoles and solenoids usually introduce. Consequently, the collider no longer needs to be filled with bunches that carry an alternating polarization pattern from the outset, it is then sufficient to inject a uniform polarization state and, at flattop, flip the vertical component on selected bunches—or whole bunch trains—via targeted RF gating.

Modifications to the RF WF design are required to meet the specifications of the Electron–Ion Collider (EIC). The spin‐flip rate of an RF Wien filter scales inversely with the square of the beam energy\,\cite[Eq.\,(13)]{SpinTuneMapping}. To achieve spin‐flip rates comparable to those observed in the pilot-bunch experiment\,\cite{pilotPRL} at the EIC’s design kinetic energy of \SI{275}{GeV} (corresponding to $\gamma \approx 295$ and $f_\text{rev}\approx\SI{78}{kHz}$), the RF field strengths must be increased by a factor of $\gamma^2 \approx \num{8.7e4}$. 

The precise implementation method remains an open question. Nonetheless, one can extend the length of the waveguide RF WF to \SI{10}{m}, operate the device at a sideband index of $K = \num{15000}$ i.e., frequency $f_\text{WF} \approx \SI{1.1}{GHz}$, and increase the driving RF power to a \SI{1}{MW}\,\cite{Carter:2013eaa}. Under these conditions, spin-flip periods on the order of one minute become feasible, which is acceptable for polarized proton beams.

Moreover, for selectively gating individual bunches with an RF WF at the EIC, the currently deployed RF switches\footnote{Developed by Barthel HF-Technik GmbH, 52072 Aachen, Germany \url{https://barthel-hf.com}} are capable of switching times of approximately $\SI{2}{ns}$. At the EIC, with $n_\text{b} \approx 1180$ bunches and a revolution frequency of $f_\text{rev}$, the bunch spacing is given by $(f_\text{rev} \cdot n_\text{b})^{-1} \approx \SI{11}{ns}$, which falls well within the operational limits of the existing switching technology. However, the challenge of managing the significantly higher RF power levels required for such operation remains an unresolved technical issue.

\section{Estimate of required Research and Development }
\subsection{Polarized ion source development}

\subsubsection{OPPIS reliability \& performance}

Over the last 25 years, the Optically Pumped Polarized Ion Source (OPPIS), described in Sec.\,\ref{sec:OPPIS}, has demonstrated reliable operation in supporting the polarized proton program at the Relativistic Heavy Ion Collider (RHIC). With its current performance, OPPIS is also poised to meet the requirements of the forthcoming Electron-Ion Collider (EIC). Nonetheless, there is scope for improvement in three critical areas: reliability, beam polarization, and output current. Targeted R\&D in these domains is needed to ensure system stability and performance for the EIC.

\subsubsection{Polarized deuteron ion source}

The implementation of a polarized deuteron ion source at BNL is discussed in Sec.\,\ref{sec:pol-deuterons}. With the COSY deuteron ion source no longer required in Germany, the system could be relocated to  BNL. Such a source could be accommodated at the BNL Tandem facility, though detailed studies where best to place it are still underway. For polarimetry measurements, vector polarization can be effectively monitored using conventional carbon targets throughout the accelerator chain leading to the EIC. However, tensor polarization presents additional challenges, requiring specialized polarimeter development such as Lamb-shift-based systems or BRP-type polarimeters specifically designed for deuteron ions. Additionally, resonant charge exchange should be seriously considered as an alternative, given its potential to deliver substantially higher efficiencies compared to conventional collision with Cs ions, making it an attractive option for optimizing the overall deuteron polarization program.

\subsubsection{Polarized $^3$He$^{++}$ ion source}

The development of a polarized $^3$He$^{++}$ ion source at BNL, described in Section\,\ref{sec:3He-beam}, represents a significant technical undertaking, integrating Metastability Exchange Optical Pumping (MEOP) within the spatial constraints of the existing \SI{5}{T} superconducting EBIS solenoid. The design required developing a specialized gas purification system, compact optical polarization scheme, and high-field pulsed valve system to inject polarized $^3$He atoms directly into the EBIS drift tube for ionization. Extensive R\&D efforts have systematically addressed the technical challenges, progressing from 90\% polarization in sealed samples to 80\% with open-cell gas handling systems, and developing a minimal optical layout achieving 60\% polarization within the tight geometric constraints. Current work involves testing the optical system in a full-scale EBIS solenoid replica while finalizing the gas handling design, with complete installation scheduled during the 2025 EBIS shutdown and commissioning planned for 2026 using both the absolute polarimeter and existing AGS  CNI carbon polarimeter. 

\subsubsection{Polarized Li-6/Li-7 ion source}

The development of polarized lithium-6 and lithium-7 ion sources at Argonne National Laboratory (Section\,\ref{sec:pol-Li6-7}) presents unique challenges in integrating atomic beam generation, optical pumping at \SI{671}{nm}, and selective nuclear state manipulation within a coordinated system. Key technical hurdles include preserving nuclear polarization during ionization and extraction processes, requiring development of retractable ionization systems and optimized extraction grids while maintaining precise coordination between thermal beam generation, laser-atom interactions, RF manipulation, and ion extraction throughout the source chain.

Laser heating of the lithium oven offers significant advantages over conventional thermal methods, providing precise temperature control, rapid response, and contamination-free operation without outgassing from heating elements. This approach enables spatial selectivity for optimal temperature gradients, eliminates electromagnetic interference affecting optical pumping systems, and maintains superior vacuum compatibility essential for high-quality atomic beam production, thereby advancing both precision and reliability of the polarized lithium source.

\subsection{Beam polarimeter systems for the EIC}

\subsubsection{HJET polarimeter} 
The hydrogen jet target (HJET) polarimeter (Section\,\ref{sec:HJET}) provides the only means for absolute beam polarization measurements with the stringent $\frac{\Delta P}{P} \le 1\%$ relative uncertainty demanded by the EIC physics program, representing a critical upgrade requiring comprehensive reassessment of systematic effects. The recoil detector system must handle approximately ten times higher bunch repetition frequencies than RHIC while maintaining adequate counting statistics, and the atomic beam source and Breit-Rabi polarimeter require optimization for enhanced precision at both injection and flattop energies. Measurement of the complete polarization vector $\vec{P} = (P_x, P_y, P_z)$ would enable better spin evolution diagnostics and minimization of unwanted beam components for higher experimental polarizations.

A critical modification involves upgrading the magnetic guide field from \SI{120}{mT} to approximately \SI{400}{mT} to prevent beam-induced depolarization by ensuring hyperfine transition frequencies remain safely separated from populated beam harmonics. This enhanced field system will reduces systematic uncertainties from beam-induced depolarizing effects by more than an order of magnitude compared to RHIC operation.

Additionally, the flexible system design permits easy conversion to a deuteron jet target (DJET), providing absolute polarimetry capabilities for deuteron beams in the EIC.

\subsubsection{pC polarimeter}

As described in Section\,\ref{sec:pC-polarimetry}, the transition from RHIC to EIC operation necessitates comprehensive modifications to the existing pC polarimetry infrastructure. The threefold beam intensity increase and reduced bunch spacing (\SI{108}{ns} $\rightarrow$ \SI{10}{ns}) create fundamental challenges requiring engineering solutions across multiple subsystems.

The target chamber requires redesign with an additional lock chamber for expanded target capacity and comprehensive RF shielding of all ports. Aluminum target holders must be replaced with alumina materials to eliminate wakefield oscillations and impedance resonances from CST simulations. Longer carbon ribbon targets are needed for the enlarged beta function geometry (\SI{30}{m} $\rightarrow$ \SI{240}{m}) required for thermal management. A critical gap -- direct temperature measurement of carbon targets during beam interaction -- may be addressed through optical light collection systems spanning visible and near-infrared spectra to ensure operation below the \SI{2000}{K} sublimation threshold.

The \SI{10}{ns} bunch spacing at the EIC requires implementation of advanced readout electronics and appropriate recoil detectors to maintain polarimetry precision while enabling bunch-by-bunch polarimetry and determination of transverse beam polarization profiles, representing a substantial enhancement over established RHIC methodologies.

\subsubsection{$^3$He atomic beam polarimeter}

The development of an absolute $^3$He$^{++}$ beam polarimeter for EIC (Section\,\ref{sec:3He-beam}) requires establishing a polarized $^3$He atomic beam source (ABS) system based on cryogenic magnetic focusing. The existing Los Alamos-MIT cryogenic ABS demonstrating $\sim95\%$ polarization at $\sim10^{14}$ atoms/cm$^2$/s flux must be adapted for EIC operational requirements through comprehensive engineering modifications.

Key technical developments include integration of the \SI{1}{\kelvin} cryogenic chamber with multi-channel plate nozzle and quadrupole magnet section into the EIC beamline infrastructure. The system requires precise alignment mechanisms for the actuated nozzle with pitch and yaw control, vacuum pumping systems for the defocused spin states, and shimming magnets for adiabatic transport to holding fields. Spin evolution coils based on solenoid geometry must be implemented to convert the spatially varying quadrupolar spin pattern from the ABS into uniform transverse polarization matching EIC beam requirements.

Target enhancement through a storage cell is essential to achieve the required $\sim 10^{12}$ atoms/cm$^2$ density for 1\% precision measurements within \SI{1}{\hour} at peak \SI{1}{\ampere} $^3$He$^{++}$ beam currents. The detector systems require development of radiation-hard diamond strip detectors for scattered beam particle detection with charge-to-digital conversion capability for nuclear fragment rejection, while silicon strip detectors will measure target asymmetries. Integration of residual gas analyzers for beam profile mapping and comprehensive gas handling systems for $^{3,4}$He supply and regulation complete the technical requirements for operational deployment.

\subsubsection{Li-6/Li-7 beam polarimeter}

While the lithium ion source development (Section\,\ref{sec:pol-Li6-7}) focuses on producing polarized $^{6,7}$Li$^+$ beams for collider operations, absolute polarimetry of these beams necessitates a complementary atomic beam target system. Unlike ion beam polarimetry which relies on nuclear elastic scattering with limited theoretical precision, atomic beam targets enable Li-Li elastic scattering in the Coulomb-Nuclear Interference (CNI) region, analogous to the established $pp$ scattering polarimetry at RHIC. The proposed polarimeter would utilize the same optical pumping and RF transition infrastructure from the ion source, replacing only the ionization chamber with a differentially pumped interaction region to provide absolute calibration capability through measurement of the known atomic target polarization.

Technical development requirements include adaptation of the existing vaporizing oven and nozzle system for continuous atomic beam operation, implementation of Stern-Gerlach magnetic focusing for beam collimation, and integration of Breit-Rabi polarimetry for absolute atomic polarization measurement. Detection systems must accommodate the unique kinematics of $^{6,7}$Li elastic scattering while providing sufficient rate capability for routine beam monitoring during EIC operations.

\subsection{Spin diagnostics tools, spin manipulators, detector systems}

\subsubsection{Spin tune determination and feedback}

Implementing continuous spin-tune determination (Section\,\ref{sec:spin-tune-determination}) at the EIC faces significant technical challenges due to the high bunch repetition rates and complex multi-species beam environment. Such a system would require real-time Fourier analysis of polarimeter data from thousands of circulating bunches while maintaining sufficient statistical precision for accurate spin precession frequency measurement. Advanced phase-lock feedback systems must synchronize with rapidly varying spin tunes across different beam energies and species, demanding sophisticated control algorithms and fast-response RF manipulation devices capable of tracking frequency drifts in real-time without introducing beam instabilities.

The integration of automated feedback correction presents additional challenges in coordinating chromaticity control, electron cooling systems, and sextupole optimization across the complex EIC lattice while maintaining stable polarization for extended store periods. Fast switching technology must selectively manipulate individual bunches within nanosecond timing windows while preserving main beam quality and avoiding cross-talk between different beam species. Despite these challenges, successful implementation would provide unprecedented polarization control capabilities at the EIC.

\subsubsection{RF Wien filter as non-invasive spin flipper}

The implementation of a Lorentz-force free RF Wien filter (Section\,\ref{sec:RF-Wien-filter}) at the EIC presents significant technical challenges due to extreme energy scaling requirements. The spin-flip rate scales inversely with the square of beam energy, necessitating RF field strengths increased by $\gamma^2 \approx \num{8.7e4}$ compared to COSY demonstrations to achieve comparable performance at 275 GeV. This requires extending the waveguide to approximately 10 m, operating at \SI{1.1}{GHz}, and managing \SI{1}{MW} RF power levels -- representing substantial engineering challenges in power handling, thermal management, and maintaining critical orthogonality between field components.

Despite these challenges, an RF Wien filter offers transformative benefits by enabling continuous, bunch-selective spin manipulation without beam distortions inherent in conventional RF dipole systems. Lorentz-force cancellation through precise impedance matching eliminates orbit perturbations, allowing spin flips of individual bunches or entire trains through targeted RF gating rather than pre-injection alternating patterns. With existing \SI{2}{ns} RF switching technology -- well within the \SI{11}{ns} bunch spacing -- the system provides unprecedented polarization control flexibility while preserving beam quality for the EIC physics program.

\subsection{Optimizing Polarized Beams with ML/AI}
\label{sec:AI/ML}

Polarized ion sources represent unique quantum devices: they generate intense, highly polarized beams whose spin orientation must be precisely controlled for efficient utilization. Unlike unpolarized beams—where optimization focuses on intensity and emittance—polarized beams demand simultaneous optimization of intensity, emittance, polarization, and spin orientation in all three spatial directions. Achieving this balance is critical for the EIC and its flagship detector ePIC, where scientific output depends directly on the stability and quality of these beams.

At BNL, we are testing Bayesian machine learning (ML) methods to optimize polarization, intensity, and emittance from the OPPIS source; transmission and emittance in the linac–booster and booster–AGS transfer lines; and beam intensity from the EBIS pre-injector. Early results indicate that ML and artificial intelligence (AI) can dramatically accelerate progress in these areas, enabling optimal use of polarized ion beams and maximizing their impact on the EIC physics program.
Key areas where ML/AI can contribute include:

\textbf{Ion Source Optimization}
\begin{itemize}
    \item Modeling plasma and beam dynamics: ML surrogate models can replace expensive particle-in-cell or Monte Carlo simulations, helping optimize plasma parameters for polarized sources (e.g., OPPIS, optical pumping schemes, SONA transitions).
    \item Real-time control: AI-driven feedback loops can dynamically adjust laser power, magnetic fields, or gas flows to stabilize polarization and intensity. 
\end{itemize}

\textbf{Polarimetry and Diagnostics}
\begin{itemize}
    \item Noise reduction: ML can improve signal-to-noise ratios in polarimeter detectors (e.g., H-jet, ³He), achieving higher statistical accuracy in shorter times.
    \item Automated calibration: AI can cross-calibrate different polarimetry methods, reducing systematic uncertainties.
    \item Anomaly detection: Trained models can identify drifts or unexpected artifacts in spin asymmetry measurements before they bias results.
\end{itemize}

\textbf{Spin Transport and Beam Dynamics}
\begin{itemize}
    \item Data-driven spin tracking: Neural networks can accelerate spin-tracking simulations through the EIC lattice, enabling exploration of more optics configurations for snakes and rotators.
    \item Control optimization: Reinforcement learning could tune RF solenoids, Wien filters, and Siberian snakes to maximize spin preservation during acceleration and storage.
    \item Uncertainty quantification: Bayesian ML tools can quantify how lattice imperfections or misalignments impact polarization.
\end{itemize}

\textbf{Accelerator Operations}
\begin{itemize}
    \item Predictive maintenance: AI can analyze signals from EBIS, the linac, booster , AGS and future EIC components to predict failures in ion sources, foils, RF source, or magnets.
    \item Beam scheduling: Intelligent algorithms can optimize run plans, balancing polarized beam R\&D with physics program demands.
    \item Adaptive operation: Real-time ML models can help operators adjust to fluctuations in ion intensity, polarization, and lifetime.
\end{itemize}

\textbf{Scientific Data Analysis}
\begin{itemize}
    \item Event classification: ML can separate rare spin-dependent scattering events from backgrounds in detectors.
    \item Global fits: AI can combine polarized DIS, SIDIS, DVCS and DVMP data to extract spin-dependent PDFs, TMDs, and GPDs more robustly.
    \item Discovery potential: Unsupervised ML can identify unexpected patterns in data that may hint at exotic gluonic states or novel QCD effects.
\end{itemize}

\textbf{Training and Workforce Development}
\begin{itemize}
    \item Student integration: AI/ML projects can be integrated into student training pipelines, attracting computational physicists to the polarized ion program.
    \item Shared frameworks: Common ML frameworks across laboratories and universities can reduce duplication of effort and accelerate technology transfer from atomic physics and medical imaging.
\end{itemize}

\subsection {Personnel Required}

The personnel estimates listed in Table\,\ref{tab:rd-personnel} were arrived at by considering personnel needed to \textit{i)} develop the OPPIS source, \textit{ii)} the RHIC/EIC polarized $^3$He ion source, \textit{iii)} the HERMES polarized targets, and \textit{iv)} the SLAC End Station A Spin Program, whose effort is briefly summarized in the Appendix \ref{sec:appendix}.

In summary, it is estimated that the R\&D effort to realize the polarized ion sources will require about 200 FTEs over about a decade, i.e. about 20 FTEs {\it p.a.}  Of these, about 80 FTEs of graduate student effort will be required.   With an average of 8 graduate students {\it p.a.}, this will necessitate the participation of order 5 university groups, each with a primary research focus in this area.   It is anticipated that the estimated 7 FTEs of scientists {\it p.a.} would be a mixture of postdocs and research scientists at laboratories and universities.  


\clearpage

\begin{sidewaystable}
\centering
\footnotesize
\adjustbox{margin=0cm 0cm 0cm 6cm}{
\begin{minipage}{\textwidth}
\centering
\captionof{table}{Personnel requirements for R\&D projects supporting the EIC polarization program, organized by technical area. Note that the three lines in the Personnel columns refer to the three (Prototyping, Testing and Installation) Stages on the left.}
\label{tab:rd-personnel}
\vspace{0.3cm}
\setlength{\tabcolsep}{4pt}
\begin{tabular}{|l|l|r|r|r|r|r|r|r|r|r|}
\hline
\multirow{3}{*}{\textbf{Item}} &  \multirow{3}{*}{\textbf{R\&D Project}} &  \multirow{3}{*}{\textbf{Duration (yr)}} & \multicolumn{3}{c|}{\textbf{Stages}} & \multicolumn{4}{c|}{\textbf{Personnel}} & \multirow{3}{*}{\textbf{Total FTE}} \\
\cline{4-10}
 &  &  & \textbf{Prototyping} & \textbf{Testing} & \textbf{Install./comm} & \textbf{PhD} & \textbf{Engineer} & \textbf{Scientist} & \textbf{Subtotal} & \\
& & & \textbf{(yr)} & \textbf{(yr)} & \textbf{iss. (yr)} & \textbf{FTE} & \textbf{FTE} & \textbf{FTE} & \textbf{FTE} & \\
\hline
\rowcolor{lightgray} \textbf{1} & \textbf{Ion Source Development} & & & & & & & & & \\
\hline
 \multirow{3}{*}{\textbf{1.1}} & 
 \multirow{3}{*}{\textbf{OPPIS reliability \& performance}} & 
 \multirow{3}{*}{\textbf{10.00}} 
& 0.00 & --  & --  & 0.00 & 0.00 & 0.00 & 0.00 & \multirow{3}{*}{\textbf{15.00}} \\
  & & &  -- & 5.00 & --  & 2.00 & 2.00 & 3.00 & 7.00 & \\
  & & &  -- & --  & 5.00 & 3.00 & 2.00 & 3.00 & 8.00 & \\
\hline
\multirow{3}{*}{\textbf{1.2}} & \multirow{3}{*}{\textbf{Polarized deuteron ion source}} & \multirow{3}{*}{\textbf{12.00}} & 4.00 & --  & --  & 3.00 & 3.00 & 3.00 & 9.00 & \multirow{3}{*}{\textbf{25.00}} \\
& & &  -- & 4.00 & --  & 3.00 & 3.00 & 3.00 & 9.00 & \\
& & &  -- &  -- & 4.00 & 3.00 & 2.00 & 2.00 & 7.00 & \\
\hline
 \multirow{3}{*}{\textbf{1.3}} & \multirow{3}{*}{\textbf{Polarized $^3$He$^{++}$ ion source}} & \multirow{3}{*}{\textbf{10.00}} & 3.00 & -- & -- & 2.00 & 2.00 & 1.00 & 5.00 & \multirow{3}{*}{\textbf{17.00}} \\
& & & -- & 4.00 & -- & 3.00 & 2.00 & 1.00 & 6.00 & \\
& & & -- & -- & 3.00 & 3.00 & 2.00 & 1.00 & 6.00 & \\
\hline
\multirow{3}{*}{\textbf{1.4}} & \multirow{3}{*}{\textbf{Polarized Li-6/Li-7 ion source}} & \multirow{3}{*}{\textbf{15.00}} & 5.00 & -- & -- & 4.00 & 2.00 & 3.00 & 9.00 & \multirow{3}{*}{\textbf{25.00}} \\
& & & -- & 5.00 & -- & 4.00 & 2.00 & 3.00 & 9.00 & \\
& & & -- & -- & 5.00 & 3.00 & 1.00 & 3.00 & 7.00 & \\
\hline
 \rowcolor{blue!8} & \textbf{Ion Source Subtotal} & \textbf{47.00} & \textbf{12.00} & \textbf{18.00} & \textbf{17.00} & \textbf{33.00} & \textbf{23.00} & \textbf{26.00} & \textbf{82.00} & \textbf{82.00} \\
\hline
\rowcolor{lightgray} \textbf{2} & \textbf{Beam Polarimetry Systems} & & & & & & & & & \\
\hline
\multirow{3}{*}{\textbf{2.1}} & \multirow{3}{*}{\textbf{HJET polarimeter upgrade}} & \multirow{3}{*}{\textbf{6.00}} & 0.00 & -- & -- & 0.00 & 0.00 & 0.00 & 0.00 & \multirow{3}{*}{\textbf{15.00}} \\
& & & -- & 3.00 & -- & 2.00 & 1.00 & 4.00 & 7.00 & \\
& & & -- & -- & 3.00 & 3.00 & 1.00 & 4.00 & 8.00 & \\
\hline
\multirow{3}{*}{\textbf{2.2}} & \multirow{3}{*}{\textbf{pC polarimeter upgrade}} & \multirow{3}{*}{\textbf{6.00}} & 0.00 & -- & -- & 0.00 & 0.00 & 0.00 & 0.00 & \multirow{3}{*}{\textbf{11.00}} \\
& & & -- & 3.00 & -- & 2.00 & 1.00 & 2.00 & 5.00 & \\
& & & -- & -- & 3.00 & 3.00 & 1.00 & 2.00 & 6.00 & \\
\hline
\multirow{3}{*}{\textbf{2.3}} & \multirow{3}{*}{\textbf{$^3$He atomic beam polarimeter}} & \multirow{3}{*}{\textbf{10.00}} & 2.00 & -- & -- & 2.00 & 2.00 & 2.00 & 6.00 & \multirow{3}{*}{\textbf{22.00}} \\
& & & -- & 4.00 & -- & 4.00 & 2.00 & 2.00 & 8.00 & \\
& & & -- & -- & 4.00 & 4.00 & 2.00 & 2.00 & 8.00 & \\
\hline
\multirow{3}{*}{\textbf{2.4}} & \multirow{3}{*}{\textbf{Li-6/Li-7 polarimeter}} & \multirow{3}{*}{\textbf{12.00}} & 4.00 & -- & -- & 4.00 & 2.00 & 3.00 & 9.00 & \multirow{3}{*}{\textbf{26.00}} \\
& & & -- & 4.00 & -- & 4.00 & 2.00 & 3.00 & 9.00 & \\
& & & -- & -- & 4.00 & 4.00 & 1.00 & 3.00 & 8.00 & \\
\hline
\rowcolor{blue!8} & \textbf{Beam Polarimetry Subtotal} & \textbf{34.00} & \textbf{6.00} & \textbf{14.00} & \textbf{14.00} & \textbf{32.00} & \textbf{15.00} & \textbf{27.00} & \textbf{74.00} & \textbf{74.00} \\
\hline
\rowcolor{lightgray} \textbf{3} & \textbf{Spin tools, detectors} & & & & & & & & & \\
\hline
\multirow{3}{*}{\textbf{3.1}} & \multirow{3}{*}{\textbf{Spin tune determination}} & \multirow{3}{*}{\textbf{6.00}} & 2.00 & -- & -- & 0.00 & 1.00 & 2.00 & 3.00 & \multirow{3}{*}{\textbf{13.00}} \\
& & & -- & 2.00 & -- & 2.00 & 1.00 & 2.00 & 5.00 & \\
& & & -- & -- & 2.00 & 2.00 & 1.00 & 2.00 & 5.00 & \\
\hline
\multirow{3}{*}{\textbf{3.2}} & \multirow{3}{*}{\textbf{RF Wien filter spin flipper}} & \multirow{3}{*}{\textbf{6.00}} & 3.00 & -- & -- & 2.00 & 1.00 & 1.00 & 4.00 & \multirow{3}{*}{\textbf{11.00}} \\
& & & -- & 3.00 & -- & 2.00 & 1.00 & 1.00 & 4.00 & \\
& & & -- & - & 2.00 & 1.00 & 1.00 & 1.00 & 3.00 & \\
\hline
\multirow{3}{*}{\textbf{3.3}} & \multirow{3}{*}{\textbf{Detector systems upgrade}} & \multirow{3}{*}{\textbf{6.00}} & 0.00 & -- & -- & 0.00 & 0.00 & 0.00 & 0.00 & \multirow{3}{*}{\textbf{11.00}} \\
& & & -- & 3.00 & -- & 2.00 & 1.00 & 2.00 & 5.00 & \\
& & & -- & - & 3.00 & 3.00 & 1.00 & 2.00 & 6.00 & \\
\hline
\rowcolor{blue!8} & \textbf{Spin Diagnostics Subtotal} & \textbf{18.00} & \textbf{5.00} & \textbf{8.00} & \textbf{7.00} & \textbf{14.00} & \textbf{8.00} & \textbf{13.00} & \textbf{35.00} & \textbf{35.00} \\
\hline
\rowcolor{blue!8}  & \textbf{GRAND TOTAL} & \textbf{99.00} & \textbf{23.00} & \textbf{40.00} & \textbf{38.00} & \textbf{79.00} & \textbf{46.00} & \textbf{66.00} & \textbf{191.00} & \textbf{191.00} \\
\hline
\end{tabular}
\end{minipage}
}
\label{tab:rd_project}
\end{sidewaystable}

\clearpage

\begin{acknowledgments}
The meeting was generously hosted by the Center for the Frontiers in Nuclear Science at Stony Brook University and we thank the Director Abhay Deshpande for his strong support.  In addition, we thank Collider-Accelerator Department Chair Wolfram Fischer for providing financial support for the speakers.

After the workshop, written contributions were invited and obtained from Grigor Atoian, Nigel Buttimore, Ian Cloet, Renee Fatemi, David Gaskell, Boxing Gou, Volker Hejny, Kiel Hock, Haixin Huang, Minxiang Li, James Maxwell, William Milner, Christoph Montag, Prajwal Murthy, Sergei Nagaitsev, Vadim Ptitsyn, Thomas Roser, Andrew Sandorfi, Vincent Schoefer, Jamal Slim, Noah Wuerfel, and Yaojie Zhai.  Subsequently, the writing of this paper was overseen by the EPIOS Steering Committee: Jaydeep Datta, Zein-Eddine Meziani, Richard Milner, Deepak Raparia and Frank Rathmann.  We thank R. Jaffe for critical reading and excellent feedback on earlier drafts of this paper.  All registrants of the Stony Brook meeting are co-authors of this paper.

We acknowledge support from the following institutions/agencies:

\begin{enumerate}
    \item DOE, Office of Science, Office of Nuclear Physics (USA) \\
    {\bf Contracts:} 
    DE-FG02-94ER40818 (R. Milner), 
    DE-SC0025511 (W. Korsch),
    DE-SC0018229, DE-SC0024846 (P. Mohan Murthy), 
    DE-FG02-05ER41372 (A. Deshpande),
    DE-SC0024602 (J. Jia),
    DE-SC0012704 (K. Hock)
   
    \item DOE, Office of Science, Office of Fusion Energy Sciences (USA) \\ 
    {\bf Contract:} 
    DE-SC0024682 (A. Sandorfi).    
    
    \item National Science Foundation (USA)

    \item Brookhaven National Laboratory  \\
    {\bf Contracts:} 
    460913 (P. MohanMurthy) \\
   N. W\"urfel's work is supported by Massachusetts Institute of Technology   under Contract No  435828  with the Brookhaven Science Associates, LLC  \\
   C. Ianzano's work is supported by Massachusetts Institute of Technology  under Contract No  449363 with the Brookhaven Science Associates, LLC
   
    \item We thank the Simons Foundation for generally supporting the Center for the Frontiers in Nuclear Science at Stony Brook University.

    \item The work of N. Buttimore was partially funded by the University of Dublin, Ireland.
    
    \item The work of V. Hejny and J. Slim was supported by the European Research Council (ERC) under the European Union’s Horizon 2020 program, Grant Agreement No. 694340 (“Search for electric dipole moments using storage rings”), and by the European Union’s Horizon 2020 research and innovation program under Grant Agreement No. 824093 (STRONG-2020).

    \item M. Li’s work was supported by the National Key Research and Development Program of China under Contracts No. 2023YFA1606800 and No. 2024YFA1611003.
   
    \item The work of W. Milner was supported by the MIT Department of Physics.
        
    \item The work by H. Mkrtchyan was supported by the High Education and Science Committee of RA under project 21AG-1C028.

    \item The work of P. MohanMurthy is supported by DOE grants DE-SC0019768, DE-SC0014448, DE-SC0018229, and DE-FG02-94ER40818. The $^3$He-ABS was built in part with the help of a sub-contract from UIUC to MIT from NSF grant PHY-1822502. In addition, this work is made possible by the generous support of MIT LNS internal funds, and funds from Prof.\ R.P.\ Redwine in particular, and  through Brookhaven National Laboratory via contract 460913. 

    \item C.-J. Naïm's work was supported by the Center for Frontiers in Nuclear Science and the Simons Foundation.

    \item N. Nikolaev acknowledges a support by the Russian Science Foundation grant No.\ 25-72-30005.

    \item Y. Zhai’s work was supported by the Natural Science Foundation of Beijing, China, under Contract No.\ JQ22002.

\end{enumerate}

\end{acknowledgments}

\section{Appendices}
\subsection{Summary of observables relevant to the 3D GPD and TMD Studies}

To obtain the most comprehensive and precise information on GPDs, the experimental strategy must combine a broad and complementary set of exclusive measurements, leveraging both Deeply Virtual Compton Scattering (DVCS) and Deeply Virtual Meson Production (DVMP) on proton and neutron (e.g., $^3$He) targets. DVCS provides clean access to the convolution of GPDs into Compton Form Factors (CFFs), which encode critical information about the spatial and momentum distributions of quarks in the nucleon. This access is realized through a suite of spin-dependent asymmetry measurements: the \textit{beam-spin asymmetry} (BSA) $\Delta\sigma_{LU}$, sensitive to the \textit{imaginary parts} of $\mathcal{H}, \tilde{\mathcal{H}}, \mathcal{E}$; the \textit{target-spin asymmetry} (TSA) $\Delta\sigma_{UL}$, and the \textit{double-spin asymmetry} (DSA) $\Delta\sigma_{LL}$, which probe their \textit{real parts}; and the \textit{transverse target-spin asymmetry} (TTSA) $\Delta\sigma_{UT}$, which offers direct sensitivity to the elusive GPD $E$, crucial for determining the quark orbital angular momentum via Ji’s sum rule~\cite{Ji:1996ek}. 

Table~\ref{table:QuarkGluonGPDObs} displays the different spin dependent measurements and their sensitivity to the relevant GPDs.
Table~\ref{table:QuarkGluonGPDObs} displays the different spin dependent measurements and their sensitivity to the relevant GPDs.
\begin{table*}[htb]
	\centering
	\caption{Overview of DVCS and DVMP observables and their sensitivity to Compton Form Factors (CFFs) and GPDs.}
	\label{table:QuarkGluonGPDObs}
	\resizebox{\textwidth}{!}{%
		\begin{tabular}{llllll}
			\hline
			\multicolumn{6}{c}{\textbf{Deep Virtual Compton Scattering (DVCS)}} \\
			\hline\hline
			\textbf{Observable} & \textbf{Beam} & \textbf{Target} & \textbf{CFF / amplitude} & \textbf{GPD sensitivity} & \textbf{Role} \\
			\hline
			Unpolarized ($d\sigma$) & No & No &
			mixed Re+Im; quadratic in CFFs &
			quarks (model-dep.), indirect gluons via evolution &
			BH dominated; weak for disentangling \\
			BSA ($\Delta\sigma_{LU}$) & Long. & No &
			$\mathrm{Im}[\mathcal{H}^q,\tilde{\mathcal{H}}^q,\mathcal{E}^q]$ &
			quarks (Im), gluons via $Q^2$ evolution &
			primary DVCS constraint \\
			LTSA ($\Delta\sigma_{UL}$) & No & Long. &
			$\mathrm{Im}[\mathcal{H}^q,\tilde{\mathcal{H}}^q,\mathcal{E}^q]$ &
			$\tilde H$, $E$ (quark) &
			helicity sensitivity \\
			DSA ($\Delta\sigma_{LL}$) & Long. & Long. &
			$\mathrm{Re}[\mathcal{H}^q,\tilde{\mathcal{H}}^q,\mathcal{E}^q]$ &
			quarks (Re); indirect gluon info &
			fixes real parts; cross-checks fits \\
			TTSA ($\Delta\sigma_{UT}$) & No & Trans. &
			$\mathrm{Im}[\mathcal{H}^q,\mathcal{E}^q]$ &
			$E^q$ (quark) &
			input to Ji sum rule / OAM \\
			\hline
			\multicolumn{6}{c}{\textbf{Deep Virtual Meson Production (DVMP, e.g.\ $J/\psi$, $\Upsilon$)}} \\
			\hline
			Unpolarized & No & No &
			amplitude $\propto H^g$ (LO) &
			gluon $H^g$ &
			clean small-$x$ gluon probe \\
			BSA ($\Delta\sigma_{LU}$) & Long. & No &
			$\mathrm{Im}[H^g]$ &
			gluon phase &
			phase info in exclusive channels \\
			LTSA ($\Delta\sigma_{UL}$) & No & Long. &
			$\mathrm{Im}[\tilde H^g]$ &
			polarized gluon $\tilde H^g$ &
			access to gluon helicity \\
			TTSA ($\Delta\sigma_{UT}$) & No & Trans. &
			$\mathrm{Im}[E^g]$ &
			gluon $E^g$ &
			link to gluon OAM (Ji) \\
			DSA ($\Delta\sigma_{LL}$) & Long. & Long. &
			Re/Im of $H^g$, $\tilde H^g$ &
			gluon helicity / long. spin &
			leading-twist gluon tests \\
			DSA ($\Delta\sigma_{LT}$) & Long. & Trans. &
			Re/Im of $E^g$, $\tilde H^g$ &
			gluon spin–orbit / transverse dyn. &
			constraints on subleading effects \\
			\hline\hline
		\end{tabular}%
	}
\end{table*}

DVMP complements this picture by offering flavor and chiral selectivity through the quantum numbers of the final-state meson~\cite{Goeke:2001tz,Diehl:2003ny}. For instance, production of vector mesons such as $\rho^0$, $\omega$, $\phi$ and $J/\psi$  primarily probes the unpolarized GPDs $H$ and $E$, while pseudoscalar mesons like $\pi^0$, $\eta$, and $K$ provide sensitivity to the polarized GPDs $\tilde{H}$ and $\tilde{E}$. Longitudinally polarized virtual photons, which dominate DVMP in the deep regime, enable this selectivity and allow for additional observables, such as longitudinal target-spin asymmetries and azimuthal modulations sensitive to interference terms~\cite{Belitsky:2005qn,Guidal:2004nd}.

\begin{table*}[htb]
	\caption{Mapping of SIDIS measurements to leading-twist transverse momentum-dependent distributions (TMDs). A comprehensive TMD program consists of measurements of asymmetries using polarized beams and targets, for both proton and neutron data, together with precision unpolarized data.}
	\label{table:TMD}
	\centering
	\begin{tabular}{lllll}
		\hline\hline
		\textbf{SIDIS Observable} & \textbf{Beam Pol.} & \textbf{Target Pol.} & \textbf{Sensitive TMD(s)} & \textbf{Role in the TMD program} \\
		\hline
		$A_{UT}^{\sin(\phi_h - \phi_S)}$  & No   & Transv. & $f_{1T}^{\perp}$ (Sivers) & $\vec{k}_T$ correl.\ with transv.\ target spin \\
		$A_{UT}^{\sin(\phi_h + \phi_S)}$  & No   & Transv. & $h_1$ (Transversity) $\otimes$ Collins FF & Transv.\ spin of quarks \\
		$A_{UT}^{\sin(3\phi_h - \phi_S)}$ & No   & Transv. & $h_{1T}^{\perp}$ (Pretzelosity) & Shape distortion of pol.\ quarks \\
		$A_{UL}^{\sin(\phi_h)}$           & Long.& Long.   & $h_{1L}^{\perp}$ (Worm-gear, chiral-odd) & Longit.\ target $\rightarrow$ transv.\ quark spin \\
		$A_{LT}^{\cos(\phi_h - \phi_S)}$  & Long.& Transv. & $g_{1T}$ (Worm-gear, chiral-even) & Transv.\ target $\rightarrow$ longit.\ quark spin \\
		$A_{UU}^{\cos(2\phi_h)}$          & No   & No      & $h_1^{\perp}$ (Boer--Mulders) $\otimes$ Collins FF & $\vec{k}_T$ correl.\ with transv.\ quark spin \\
		\hline\hline
	\end{tabular}
\end{table*}

The extraction of gluon GPDs follows two complementary experimental strategies grounded in distinct theoretical frameworks. The first involves Deeply DVCS over a wide range of $Q^2$~\cite{Belitsky:2001ns,Kumericki:2009uq}. In this case, gluon GPDs enter the process at next-to-leading order (NLO) via QCD evolution and influence the scaling behavior of the measured Compton Form Factors (CFFs). Although DVCS provides broad kinematic reach and is well-developed experimentally, its sensitivity to gluon GPDs is indirect and must be inferred through global fits that account for DGLAP/ERBL evolution of both quark and gluon contributions~\cite{Kumericki:2016ehc}.

A more direct probe is offered by exclusive electroproduction of heavy vector mesons such as $J/\psi$, $\psi'$, and $\Upsilon$, where the scattering amplitude at leading order is dominated by two-gluon exchange in the $t$-channel~\cite{Collins:1996fb,Diehl:2003ny}. Assuming QCD factorization holds, these processes provide access to the unpolarized gluon GPD $H^g(x, \xi, t)$ at leading twist. The large mass of the produced quarkonium ensures a hard scale and helps suppress higher-twist effects, allowing cleaner interpretation compared to DVCS~\cite{Ivanov:2004vd,Kumericki:2022dlx}. These reactions are particularly sensitive to gluon distributions at small skewness and low to moderate $x$.

While unpolarized cross sections are sufficient to access $H^g$, polarized beam and target configurations enable additional insights into spin-dependent gluon GPDs. Longitudinal target polarization provides sensitivity to the helicity-dependent gluon GPD $\tilde{H}^g$~\cite{Hatta:2005as}, while transverse polarization gives access to $E^g$, which is directly linked to the gluon contribution to nucleon spin via Ji’s sum rule~\cite{Ji:1996nm,Hatta:2012cs}. Beam polarization can further constrain the imaginary part of the amplitude, revealing the phase structure of gluon interactions.

A summary of these two approaches---including their theoretical foundation, kinematic requirements, and the role of polarization---is provided in Table~\ref{table:QuarkGluonGPDObs}.

By performing these measurements on both protons and neutrons—exploiting light nuclei like deuterium and $^3$He for effective neutron targets—and across a wide kinematic range in $x_B$, $Q^2$, $t$, and azimuthal angle $\phi$, one accesses different linear combinations of GPDs. The distinct electromagnetic form factors $F_1$ and $F_2$ for protons and neutrons further aid in \textit{flavor separation}, especially when asymmetries are combined across targets~\cite{Cates:2011pz}. Together, this global strategy—blending DVCS and DVMP channels, spin observables, and nucleon isospin diversity—enables a model-independent, multidimensional mapping of CFFs. This paves the way toward reconstructing the nucleon’s full 3D structure and rigorously quantifying the role of quark orbital angular momentum in the nucleon spin decomposition~\cite{Burkardt:2002hr,Ji:2004gf}.

To achieve a multidimensional picture of the nucleon in momentum space, the experimental strategy centers on precise and diverse measurements of SIDIS, where a high-energy lepton scatters off a nucleon and a specific hadron is detected in the final state~\cite{Airapetian:2004tw, Alexakhin:2005iw}. This process enables access to \textit{Transverse Momentum Dependent parton distributions} (TMDs), which describe the probability of finding a parton carrying a longitudinal momentum fraction $x$ and intrinsic transverse momentum $\vec{k}_T$ inside the nucleon~\cite{Collins:2011zzd,Bacchetta:2006tn}. 

The extraction of TMDs relies on measuring a variety of \textit{spin- and azimuthal-angle-dependent asymmetries} in SIDIS, each sensitive to different combinations of TMDs and fragmentation functions. For instance, the \textit{Sivers asymmetry}, observed in measurements with a transversely polarized target, probes the \textit{Sivers function} \(f_{1T}^{\perp}(x, k_T^2)\), which encodes a correlation between the transverse momentum of unpolarized partons and the transverse spin of the nucleon~\cite{Sivers:1989cc,Airapetian:2009ae}. This observable provide access to parton orbital angular momentum and initial-states QCD effects. The \textit{Collins asymmetry}, measured with transversely polarized targets and sensitive to the azimuthal modulation of the final-state hadron, accesses the \textit{transversity distribution} $h_1(x, k_T^2)$ in conjunction with the Collins fragmentation function, revealing the transverse polarization of quarks inside transversely polarized nucleon~\cite{Collins:1992kk,Seidl:2008xc}.

Additional asymmetries, such as those involving longitudinally polarized electron beam and/or ion beams (targets), allow access to TMDs like the Boer-Mulders and worm-gear distributions $g_{1T}$ and $h_{1L}^{\perp}$~\cite{Boer:1997nt,Efremov:2006qm}. By performing these measurements across a wide kinematic range in $x$, $Q^2$, $z$, and transverse momentum $P_T$, and on both \textit{proton and neutron targets} (with effective neutrons accessed through polarized $^3$He or deuteron), one can disentangle the flavor and spin structure of the nucleon’s partonic content~\cite{Zhou:2009rp, Belostotski:2007yf}. Furthermore, TMD evolution with $Q^2$ can be studied through measurements at different beam energies and scales, providing vital information on non-perturbative QCD dynamics and the scale dependence of TMDs~\cite{Aybat:2011zv}. Table~\ref{table:TMD} shows the sensitivity of different measurements to different TMDs. Altogether, this comprehensive program of multidimensional SIDIS measurements—spanning azimuthal, polarization, and hadron-flavor dependence—enables a full momentum-space tomographic imaging of the nucleon and advances our understanding of the spin-orbit correlations and dynamics that underlie the nucleon’s internal structure. 

Altogether, this comprehensive program of multidimensional SIDIS measurements—spanning azimuthal, polarization, and hadron-flavor dependence—enables a momentum-space tomographic imaging of the nucleon and deepens our understanding of the spin-orbit correlations and partonic dynamics that underlie nucleon structure~\cite{Boussarie:2023fsy}.

\subsection{Consideration of Personnel Needed to Construct Previous Polarization Experiments in Nuclear Physics}
\label{sec:appendix}
To develop estimates of the personnel required to develop the EIC polarized ion capability, we consider the personnel required for the following: 

\medskip
\noindent
{\bf The OPPIS polarized proton source for RHIC} 

The OPPIS technique for polarized H$^-$ ion beam production was developed in the early 1980s at KEK (Japan), INR (Russia), LAMPF (USA), and TRIUMF (Canada).

This particular OPPIS has a long development history. In the 1980s, it was first developed at KEK.
In the 1990s, it was transferred to TRIUMF for further upgrades for RHIC in collaboration with INR.
In 2000, it was brought to BNL (Brookhaven National Laboratory) and underwent several upgrades.

The BNL OPPIS is the highest-intensity polarized H$^-$ ion source, providing polarized protons to RHIC (Relativistic Heavy Ion Collider). Over the years, approximately 30 physicists, 10 engineers, and 5 postdocs/students have contributed to its development and optimization.

\medskip
\noindent
{\bf The RHIC/EIC polarized $^3$He source development} 

The polarized $^3$He  ion source utilizes a technique based on metastability-exchange optical pumping (MEOP) within the 5T field of the existing Electron Beam Ion Source (EBIS).

The development of polarized $^3$He  for RHIC began in the first decade of the 21st century through a collaboration between MIT and BNL.

Approximately 12 physicists, 4 postdocs, and students have contributed to the project, which is still under development at BNL. The polarized $^3$He  source is expected to be ready for injection into the BNL Booster for initial testing in mid-2026.

\medskip\noindent
{\bf The HERMES Polarized Targets}
\begin{itemize}
    \item The HERMES H/D target was principally developed by Erlangen, Madison, Marburg, Munich, Liverpool, Ferrara and Boulder.  
    It is estimated that the effort took 40 graduate student years, 50 postdoc/senior years and technical support from the workshops at Heidelberg, Liverpool, Ferrara and Madison.
    \item The HERMES polarized He-3 target was principally developed by Caltech and MIT.  
    It is estimated that the effort took 10 graduate student years, 12 postdoc/senior years and two years of engineering support. 

\end{itemize}

\medskip
\noindent
{\bf The SLAC End Station A Spin Program}
\begin{itemize}
\item The SLAC nucleon spin physics program started in the late 80s and extended to the late 90s. The series of experiments comprising this program where SLAC E142 (polarized $^3$He), E143 (Polarized ammonia and deuterated ammonia), E154 (polarized $^3$He) and E155 (polarized ammonia and deuterated ammonia).
\item In the E142 and E154 experiments the target development, construction, and operation was shared between two institutions: the University of Michigan 
and Princeton University.
In total, a lower estimate of the number of people focused on the development and operation of the target during the experiments was 7 for a period of about 5 years.
\item In E143 and E155 experiments, the target responsibilities were carried by mainly one group from the University of Virginia 
supported by another visiting expert from Bonn.
The target group consisted of three senior people 
and three graduate students.
In both cases, the work of these groups was also supported by at least one laboratory personnel.
\end{itemize}

\clearpage
\bibliographystyle{apsrev4-2}
\bibliography{references_cleaned}

\end{document}